\documentclass[aps,preprint,floatfix,superscriptaddress,balancelastpage,nofootinbib,amsmath,preprintnumbers]{revtex4-1}[12pt]
\pdfoutput=1

\usepackage{amsmath,amssymb,graphicx, graphics, color,verbatim}
\usepackage{rotating}
\usepackage{multirow}
\usepackage{latexsym}
\usepackage{cancel}
\usepackage{enumerate}
\usepackage[normalem]{ulem} 
\usepackage {slashed}
\usepackage{epsfig}
\usepackage{url}
\usepackage{hyperref}
\usepackage{booktabs}
\usepackage{pstricks}
\usepackage{fancyvrb}
\usepackage{amsthm}
\usepackage{bm}
\usepackage{epstopdf}
\usepackage{wrapfig}
\usepackage{sidecap}
\usepackage{bbold}
\usepackage{fancyhdr}
\usepackage{comment}
\usepackage{array}
\usepackage{titlesec}
\usepackage{lineno}
\usepackage[icelandic,english]{babel}
\usepackage[T1]{fontenc}

\usepackage{setspace}

\graphicspath{{figs/}}

\newcommand{\nc}{\newcommand}

\nc{\beq}{\begin{equation}}
\nc{\eeq}{\end{equation}}
\nc{\beqa}{\begin{eqnarray}}
\nc{\eeqa}{\end{eqnarray}}
\nc{\bea}{\begin{eqnarray}}
\nc{\eea}{\end{eqnarray}}
\nc{\ra}{\rightarrow}
\nc{\Tr}{{\rm Tr}}
\nc{\slsh}{\slash\hspace*{-0.22cm}}

\def\be{\begin{equation}}
\def\ee{\end{equation}}
\def\bea{\begin{eqnarray}}
\def\eea{\end{eqnarray}}
\def\bit{\begin{itemize}}
\def\eit{\end{itemize}}

\nc{\barray}{\begin{eqnarray}}
\nc{\earray}{\end{eqnarray}}
\nc{\barrayn}{\begin{eqnarray*}}
\nc{\earrayn}{\end{eqnarray*}}
\nc{\mc}{\mathcal}
\nc{\M}{\mathcal{M}}

\nc{\h}{$h$}
\nc{\infinity}{\infty}

\def\ben{\begin{enumerate}}
\def\een{\end{enumerate}}

\def\to{\rightarrow}

\newcommand{\ga}{\alpha}
\newcommand{\gb}{\beta}
\newcommand{\gam}{\gamma}
\newcommand{\gd}{\delta}
\newcommand{\eps}{\epsilon}

\newcommand{\gt}{\theta}

\newcommand{\gs}{\sigma}

\newcommand{\gD}{\Delta}

\newcommand{\gO}{\Omega}

\newcommand{\pt}[1]{\left( #1\right)}
\newcommand{\pq}[1]{\left[ #1 \right]}
\newcommand{\pg}[1]{\left\{ #1\right\}}
\newcommand{\de}{\mathrm{d}}

\renewcommand\thesection{\arabic{section}}
\renewcommand\thesubsection{\arabic{section}.\arabic{subsection}}
\renewcommand\thesubsubsection{\arabic{section}.\arabic{subsection}.\arabic{subsubsection}}
\makeatletter
\def\p@subsection{}
\def\p@subsubsection{}
\makeatother

\makeatletter
    \renewcommand\@make@capt@title[2]{%
     \@ifx@empty\float@link{\@firstofone}{\expandafter\href\expandafter{\float@link}}%
      {\textbf{#1}}\@caption@fignum@sep#2\quad}%
    \makeatother
   
\makeatletter 
\renewcommand{\fnum@figure}{\textbf{Figure~\thefigure}}
\makeatother

\newcommand{\mywidth}{0.48}

\begin{document} 

\begin{flushright}
YITP-SB-14-23 
\end{flushright}

\title{Strong Optimized Conservative {\it Fermi}-LAT Constraints on Dark Matter Models from the Inclusive Photon Spectrum}

\author{Andrea~Massari}
\thanks{Contact Author: andrea.massari@stonybrook.edu}
\affiliation{C.N.~Yang Institute for Theoretical Physics, Stony Brook University, Stony Brook, NY 11794-3840}

\author{Eder~Izaguirre}
\thanks{Contact Author: eizaguirre@perimeterinstitute.ca}
\affiliation{Perimeter Institute for Theoretical Physics, Waterloo, ON N2L 6B9, Canada}

\author{Rouven~Essig}
\thanks{Contact Author: rouven.essig@stonybrook.edu}
\affiliation{C.N.~Yang Institute for Theoretical Physics, Stony Brook University, Stony Brook, NY 11794-3840}
 
\author{Andrea~Albert}
\affiliation{SLAC - KIPAC, Stanford University, Stanford, CA 94305-4085}

\author{Elliott~Bloom}
\affiliation{SLAC - KIPAC, Stanford University, Stanford, CA 94305-4085}

\author{Germ\'{a}n~Arturo~G\'{o}mez-Vargas}
\affiliation{Instituto de Fis\'ica, Pontificia Universidad Cat\'olica de Chile, Avenida Vicu\~na Mackenna 4860, Santiago, Chile}
\affiliation{Istituto Nazionale di Fisica Nucleare, Sezione di Roma ``Tor Vergata'', I-00133 Roma, Italy}


\begin{abstract} 
We set conservative, robust constraints on the annihilation and decay of dark matter into various Standard Model final states under various assumptions about the distribution of the dark matter in the Milky Way halo.  
We use the inclusive photon spectrum observed by the {\it Fermi Gamma-ray Space Telescope} through its main instrument, the Large-Area Telescope (LAT).  
We use simulated data to first find the ``optimal'' regions of interest in the $\gamma$-ray sky, where the expected dark matter signal is largest compared with the expected astrophysical foregrounds.  
We then require the predicted dark matter signal to be less than the observed photon counts in the {\em a priori} optimal regions. This yields a very conservative constraint as we do not attempt to model or subtract astrophysical foregrounds.
The resulting limits are competitive with other existing limits, and, for some final states with cuspy dark-matter distributions in the Galactic Center region, disfavor the typical cross section required during freeze-out for a weakly interacting massive particle (WIMP) to obtain the observed relic abundance. 
\end{abstract}

\maketitle

\tableofcontents



\section{INTRODUCTION}
\label{sec:intro}

The {\it Fermi Gamma-ray Space Telescope} ({\it Fermi}), through its main instrument, the Large Area Telescope (LAT)~\cite{Atwood:2009ez}, has been surveying the $\gamma$-ray sky since August 2008 in the energy range from 20~MeV to above~300~GeV (with detected events up to $\sim 1$~TeV). 
In addition to $\gamma$ rays produced by known astrophysical sources, the {\it Fermi}-LAT can detect photons from postulated decay or annihilation of dark matter (DM) to Standard Model (SM) particles. 
The possibility that DM can annihilate is particularly motivated by the ``WIMP miracle'' \cite{Bertone:2004pz}. Here one hypothesizes the existence of weakly interacting massive particles (WIMPs) with few-GeV to few-TeV masses and weak-scale annihilation cross sections. These WIMPs would have been in thermal equilibrium with the SM sector in the early Universe and they generally produce the observed relic abundance of DM from thermal freeze-out. This suggests that WIMPs could still be annihilating today to SM particles. 
The annihilation could produce various SM particles, which can either radiate photons, further decay to other SM particles including photons, or inverse Compton scatter (ICS) off background light, producing high-energy $\gamma$ rays. 
Those photons that arrive at the {\it Fermi}-LAT could then be used to infer properties of the DM particles and their distribution around us.

Many WIMP searches have been performed using {\it Fermi}-LAT data. Analyses by the {\it Fermi}-LAT Collaboration and outside groups have searched for monochromatic $\gamma$-ray lines~\cite{Abdo:2010nc,Ackermann:2012qk,Ackermann:2013uma,Albert:2014hwa, Bringmann:2012vr, Weniger:2012tx, Su:2012ft} and continuum $\gamma$-ray excesses in the diffuse spectrum from different target regions {\it e.g.}, dwarf spheroidal galaxies~\cite{Ackermann:2011wa, Abdo:2010ex, Ackermann:2013yva, Mazziotta:2012ux,GeringerSameth:2011iw, Cholis:2012am,Geringer-Sameth:2014qqa}, clusters of galaxies~\cite{Ackermann:2010rg,Han:2012uw,MaciasRamirez:2012mk}, the Galactic halo~\cite{Ackermann:2012rg, Mazziotta:2012ux, Cirelli:2009dv, Papucci:2009gd}, the Inner Galaxy~\cite{Gomez-Vargas:2013bea, Cholis:2009gv,Goodenough:2009gk, Hooper:2010mq, Boyarsky:2010dr, Hooper:2011ti, Abazajian:2012pn, Gordon:2013vta, Abazajian:2014fta, Hooper:2013rwa, Daylan:2014rsa, Huang:2013pda, Macias:2013vya,Calore:2014xka}, the Smith cloud~\cite{Drlica-Wagner:2014yca, Nichols:2014qsa}, and the extragalactic $\gamma$-ray background~\cite{Abdo:2010dk,Fornasa:2011yb,Calore:2013yia,Cholis:2013ena}.
No undisputed signal of DM has been detected thus far, and the cross-section upper limits from these analyses for DM masses $m_{\text{DM}} \lesssim 10$~GeV are approaching the typical cross section required during freeze-out for a WIMP 
to obtain the observed relic abundance, namely $\langle \sigma v\rangle_{\rm relic} \sim 3\times 10^{-26}$~cm$^3$ s$^{-1}$. 

While DM is often thought of as being a stable particle, viable DM candidates only need to be stable on {\it cosmological} time-scales.
In particular, DM lifetimes of the order of the age of the Universe or longer ($\tau_{\rm DM}  > 10^{17}$~s) can typically evade cosmological and astrophysical bounds more easily than annihilating DM, such as constraints from Big Bang Nucleosynthesis~\cite{Hisano:2009rc}, the extragalactic $\gamma$-ray background \cite{CyrRacine:2009yn}, and re-ionization and the Cosmic-Microwave-Background~\cite{Belikov:2009qx, Galli:2009zc, Slatyer:2009yq, Huetsi:2009ex, Cirelli:2009bb,Madhavacheril:2013cna}. The more relaxed constraints on decaying DM are a result of the DM decay rate being linear with $\rho_{\rm DM}$, as opposed to quadratic with $\rho_{\rm DM}$ in the case of annihilation.

In this paper, we will provide conservative DM cross-section upper limits and decay-lifetime lower limits from the {\it Fermi}-LAT inclusive photon spectrum.  The inclusive spectrum is presumably dominated by astrophysical foregrounds in the Milky Way, though DM could contribute to it.  
We make no attempt at subtracting foregrounds and simply require that any putative DM signal contribute less than the observed flux.  
A similar idea has been used in other papers to derive conservative constraints~\cite{Ackermann:2012rg,Gomez-Vargas:2013bea, Papucci:2009gd}, where the DM signal is maximized until saturating the observed flux.  
The approach in this paper differs from such previous analyses in several ways, resulting in stronger constraints on DM. Firstly, we restrict our regions of interest (ROIs) to have a particular symmetric shape determined by only a few free parameters, and we optimize over these parameters. Secondly, we also optimize the energy range that we use for deriving the constraint. Thirdly, we optimize with respect to the constraint itself and not, for example, the signal-to-noise ratio, and last, we optimize our constraints on 10 simulated data sets, not on the measured data.
After finding the optimal ROI on simulated data, we use the real data from that same ROI to find the constraint.  
We derive constraints in this fashion for various DM-halo shapes and for various annihilation and decay final states. 
The resulting constraints, while being robust and conservative as no foregrounds have been subtracted, are competitive with other existing constraints and stronger than other conservative bounds obtained by ~\cite{Ackermann:2012rg,Gomez-Vargas:2013bea, Papucci:2009gd}.

The paper is organized as follows. In \S \ref{sec:signal} we discuss the calculation of the expected $\gamma$-ray flux from DM annihilation and decay.  In \S \ref{sec:methods} we discuss the event selection, method, simulated data sets, and ROI 
selection.  \S \ref{sec:results} discusses the resulting constraints, while our conclusions are in \S \ref{sec:conclusions}.  
In Appendix~\ref{sec:gamma-ray-excess} we use our method to calculate the limits on DM-annihilation models that have been invoked to explain an excess of $\gamma$ rays from the Galactic Center (GC) and Inner Galaxy region.  
Appendix~\ref{sec:results-roispectra} presents the optimal ROIs together with the corresponding count spectra for several DM channels.  
Appendix~\ref{subsec:SS-FB} discusses the effect on our results of source masking and choosing front-/back-converting events. Appendix~\ref{app:ICS} describes the astrophysical assumptions affecting the results that include contributions from ICS.  
Finally, Appendix~\ref{app:MC} provides more details on the simulated data sets that we use, and Appendix~\ref{subsec:MC-data-comp} compares the limits obtained from our simulated data sets with those derived from real data. 

\section{EXPECTED DARK MATTER SIGNAL}
\label{sec:signal}

Gamma rays from DM annihilation or decay to SM final states can be produced in two dominant ways. The first possibility, which we refer to as \emph{prompt}, is from either final-state radiation (FSR) produced by Bremsstrahlung by SM particles or from the decay of hadrons that arise in hadronic final states. 
The second possibility is from electrons and positrons (produced either directly or at the end of a cascade decay chain) that inverse Compton scatter off background ambient light, which primarily consists of starlight, the infrared background light, and the Cosmic Microwave Background (CMB). This ICS process boosts the energy of the background light to produce $\gamma$ rays. Unlike prompt radiation, ICS depends on various unknown astrophysical parameters discussed below. Although a sizable contribution to the energy lost by the electrons propagating through the Galaxy consists of synchrotron radiation due to acceleration by the Galactic magnetic field, we note that the synchrotron radiation does not make up a noticeable fraction of the $\gamma$ rays in the energy range under study, as we only consider DM particles with mass below 10~TeV~\cite{Cirelli:2012tf,Meade:2009iu}. We thus do not include it in this study. Moreover, the DM signal can receive additional sizeable contributions due to Galactic substructure, particularly for annihilations~\cite{Sanchez-Conde:2013yxa}, but we do not include this effect in our study. This makes our analysis more conservative and model independent in this regard.
We now outline the calculation of the DM-initiated $\gamma$-ray flux.

\subsection{Prompt radiation}
\label{subsec:Prompt}

The differential flux, d$\Phi_\gamma$/d$E_\gamma$, of \emph{prompt} photons coming from DM annihilation within the Milky Way halo is given by 
\bea\label{eq:dm-ann-flux}
\frac{\de \Phi_{\gam}}{\de E_\gam} = {1\over 8 \pi} \,{\langle \gs v\rangle \over m^2_{\rm DM}}\, {\de N_\gam \over \de E_\gam}\, r_\odot \, 
\rho^2_\odot \, J_{\rm ann}\,,
\eea
where $\langle \sigma v \rangle$ is the thermally averaged DM annihilation cross section, $m_{\rm DM}$ is the DM mass, and $\de N_\gam/\de E_\gam$ is the photon spectrum per annihilation. We assume $\rho_\odot = 0.4~{\rm GeV/cm^{3}}$ is the DM density at the Sun's location in the Galaxy~\cite{Catena:2009mf,Salucci:2010qr} \footnote{A range of values between 0.2 and 0.85 $\rm GeV/cm^3$ are possible at present though \cite{Prada:2004pi, Pato:2010yq,Catena:2009mf,Salucci:2010qr,Garbari:2012ff}. Note that a different value for the local DM density would shift up or down our predictions for DM annihilation and decay by a factor proportional to $\rho_0^2$ and $\rho_0$, respectively for annihilations and decays.}, and $r_\odot=8.5~{\rm kpc}$ is the distance between the Sun and the GC~\cite{Ghez:2008ms}.   
The ``$J$-factor'' is given by 
\bea\label{eq:J-ann}
J_{\rm ann} \equiv \int_{\rm ROI} \,\de b  \,\de \ell\, \de s  \,\frac{\cos b}{r_\odot} \, \pq{\rho\pt{r(s,b, \ell)}\over \rho_\odot}^2\,,
\eea 
which depends on the distribution of DM in the Milky Way halo, $\rho(r)$, where $r\equiv r(s,b,\ell)$ is the Galactocentric distance, given by $r=\sqrt{s^2 + r_{\odot}^2 - 2sr_{\odot} \cos\ell \cos b}$, where $\ell$ and $b$ are the Galactic longitude and latitude, respectively, and $s$ is the line-of-sight distance. The integral is over a particular ROI. For decays we can replace $\langle \sigma v\rangle \rho_\odot^2/2m_{\rm DM}^2$ with $\rho_\odot/\tau m_{\rm DM}$ in Eq.~(\ref{eq:dm-ann-flux}), where $\tau$ is the DM decay lifetime, with the $J$-factor
\bea\label{eq:J-dec}
J_{\rm dec} \equiv \int_{\rm ROI} \,\de b  \,\de \ell\, \de s  \,\frac{\cos b}{r_\odot} \, {\rho\pt{r(s,b,\ell)}\over \rho_\odot}\,.
\eea
Moreover, for decays the $\de N_\gam/dE_\gam$ should be interpreted as the photon spectrum for individual DM particle decays.

\begin{figure}[t!]
\includegraphics[width=0.48 \textwidth]{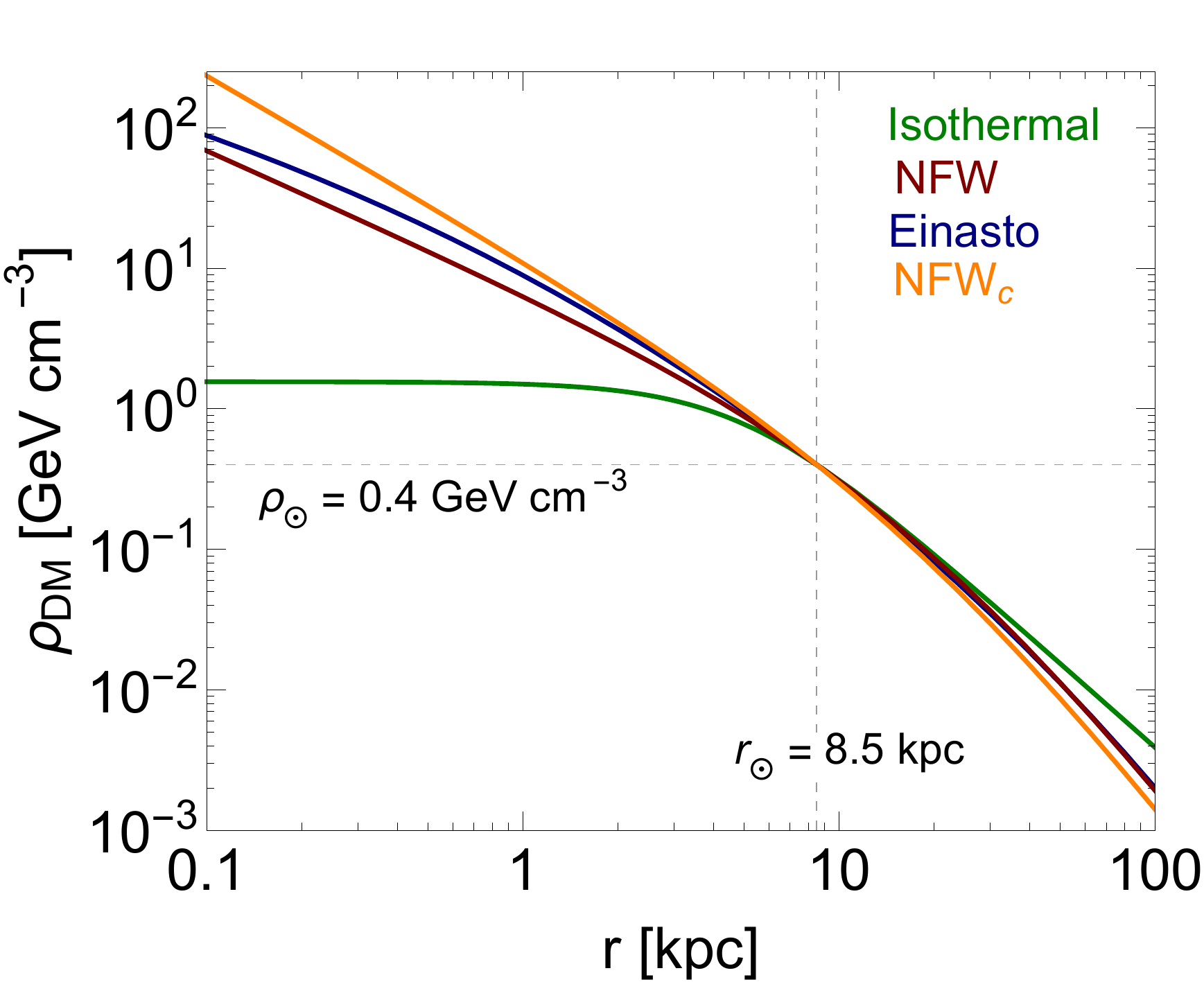}\;\;\;
\includegraphics[width=0.48 \textwidth, height=6.4cm]{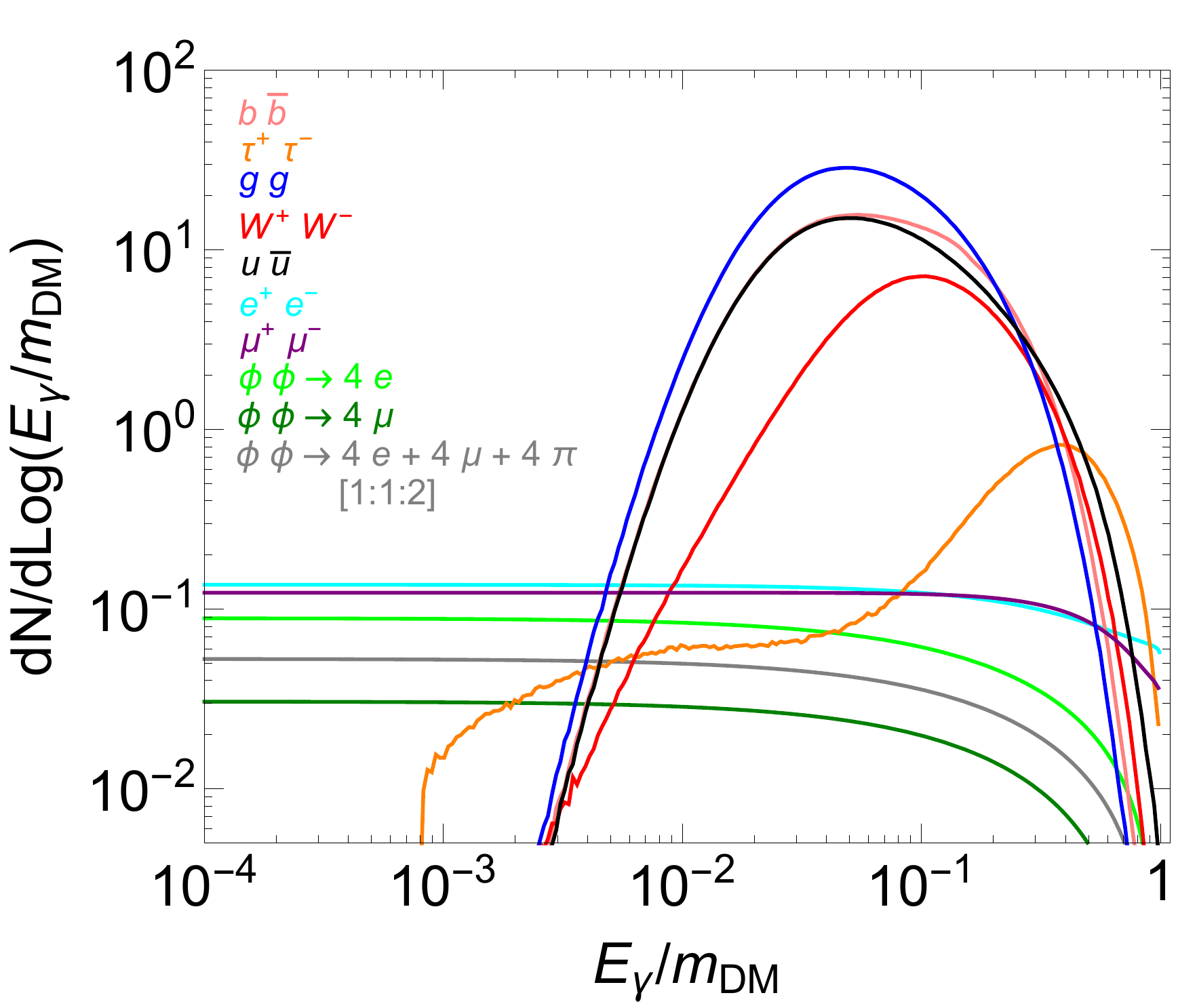}
\vskip -0.2cm
\caption{ 
{\bf Left:} 
Dark-matter density profiles versus distance from the Galactic Center (GC).  We use the 
Isothermal (green), NFW (red), Einasto (blue), and a ``contracted'' NFW (NFW$_c$, orange, with 
$\rho \propto 1/r^{1.3}$ for $r\to 0$) profile.  
{\bf Right:} 
Prompt $\gamma$-ray spectra produced in the annihilation of 1~TeV dark matter to $e^+e^-$, $\mu^+\mu^-$, $\tau^+\tau^-$, $b\bar{b}$, $W^+W^-$, $u \bar{u}$, $gg$ ($g=$ a gluon), and $\phi\phi$, where $\phi$ decays either only to $e^+e^-$ (with $m_\phi = 0.1$~GeV), 
or only to $\mu^+\mu^-$ (with $m_\phi = 0.9$~GeV), or to $e^+e^-$, $\mu^+\mu^-$, and $\pi^+\pi^-$ in the 
ratio $1:1:2$ (with $m_\phi = 0.9$~GeV).  
\label{fig:profiles}
}
\end{figure}

We consider four popular DM density profiles: the Navarro-Frenk-White (NFW)~\cite{Navarro:1995iw,Navarro:1996gj}, Einasto~\cite{Einasto,Navarro:2008kc}, Isothermal~\cite{Bahcall:1980fb},\footnote{Another popular parametrization of a ``cored'' profile is the Burkert profile~\cite{Burkert:1995yz}. Adopting the best-fit parameters in~\cite{2013JCAP...07..016N} yields a distribution that is very close to the Isothermal one for radii $\lesssim 10\, {\rm kpc}$. We will see that the optimal ROIs for the Isothermal profile for DM annihilation are contained with this region, and therefore the limits for the two cored distributions would be very similar. Thus we not explicitly consider the Burkert profile in our analysis.} and ``contracted'' NFW (${\rm NFW}_c$)~\cite{Prada:2004pi,Gustafsson:2006gr} with slope values taken from~\cite{Ackermann:2013uma}. 
\bea
\rho_{\mbox{\footnotesize Isothermal}}(r)&=&{\rho^{\rm Iso}_0  \over 1+\pt{r / r_{s,iso}}^2}\\
\rho_{\mbox{\footnotesize NFW}}(r)&=& \frac{\rho^{\rm NFW}_0}{{r/ r_s} \pt{1+{r/ r_s}}^2}\\ 
\rho_{\mbox{\footnotesize Einasto}}(r)&=&\rho^{\rm Ein}_0\, {\rm exp } \pg{-{(2 / \ga)}\pq{\pt{r / r_s}^\ga-1}}\\
\rho_{\mbox{\footnotesize NFW$_c$}}(r)&=& {\rho^{{\rm NFW}_c}_0 \over \pt{r / r_s}^{1.3} \pt{1+{r / r_s}}^{1.7}}\,.\label{contra}
\eea
We set $\ga=0.17$, $r_s=20\, {\rm kpc}$~\cite{Navarro:2008kc,Gustafsson:2006gr}, and $r_{s,iso}=5 \,{\rm kpc}$ \cite{Bahcall:1980fb}.  
The normalization $\rho (r_\odot)=\rho_\odot$ fixes $\rho^{\rm Iso}_0 \simeq 1.56$, 
$\rho^{\rm NFW}_0 \simeq 0.35$, $\rho^{\rm Ein}_0 \simeq 0.08$, and $\rho^{{\rm NFW}_c}_0 \simeq 0.24$ in units of GeV/cm$^3$. Our choice of the functional form and parameters in Eq.~(\ref{contra}) is a representative example of the possibility that, due to adiabatic contraction from the inclusion of baryonic matter, the DM profile might have a central slope steeper even than that of the NFW or Einasto profiles (although note that high-resolution observations of the rotation curves of dwarf and low-surface-brightness galaxies favor cored distributions~\cite{deBlok:2002tg,Simon:2004sr}). The four profiles are shown in Fig.~\ref{fig:profiles} ({\bf left}).

The (prompt) photon spectra, ${\de N_{\gam} / \de E_\gam}$ have been generated 
with \texttt{Pythia 8.165}~\cite{Sjostrand:2007gs} or are based on formulas 
in~\cite{Beacom:2004pe,Birkedal:2005ep,Mardon:2009rc,Essig:2009jx}.  They are the same as 
in \texttt{DMFIT} \cite{Jeltema:2008hf} after the latest update described in~\cite{Ackermann:2013yva}.  
We will consider the ten different final states $e^+e^-$, $\mu^+\mu^-$, $\tau^+\tau^-$, $b\bar{b}$, $W^+W^-$, $u \bar{u}$, 
$gg$ ($g=$ a gluon), and  $\phi\phi$, where $\phi$ decays either only to $e^+e^-$ (with $m_\phi = 0.1$~GeV), 
or only to $\mu^+\mu^-$ (with $m_\phi = 0.9$~GeV), or to $e^+e^-$, $\mu^+\mu^-$, and $\pi^+\pi^-$ in the 
ratio $1:1:2$ (with $m_\phi = 0.9$~GeV) (the latter ratio is motivated if $\phi$ is a dark photon that kinetically mixes 
with the SM hypercharge gauge boson). Other SM final states are of course possible but they would yield constraints very similar to the channels we consider in our analysis. The annihilation channels to $\phi\phi$  are motivated by DM models~\cite{ArkaniHamed:2008qn,Pospelov:2008jd} that attempt to explain the rising positron fraction measured by PAMELA~\cite{Adriani:2008zr}, {\it Fermi}~\cite{Abdo:2009zk}, and AMS-02~\cite{Aguilar:2013qda, Aguilar:2014mma}; the $\phi$ can also facilitate an inelastic transition between the DM ground state and an excited state~\cite{Finkbeiner:2007kk, ArkaniHamed:2008qn} to explain e.g.,~the 511 keV line anomaly~\cite{Knodlseder:2003sv}.
For DM decays, the $\phi$ channels can be viewed as ``simplified models'' that can capture how the constraints change when there is a cascade, e.g.,~\cite{Meade:2009iu}. We will sometimes refer to these scalar-mediated processes as ``eXciting Dark Matter'' (XDM). These spectra are shown in Fig.~\ref{fig:profiles} ({\bf right}) in the case of annihilating DM and $m_{\rm DM} = 1$~TeV. We do not consider other popular DM candidates like axions and gravitinos.

We note that the observed differential photon flux can also be written as 
\bea
\frac{\de \Phi_{\gam}}{\de E_\gam} \equiv {\de N_{\gam}\over  t_{\rm tot}\,  A_{\rm eff} \,\de E_\gam} \equiv \frac1{\mathcal{E} } {\de N_{\gam}\over \de E_\gam}\,,
\eea 
where we have now explicitly included $A_{\rm eff}$, the effective area (which is a function of energy), $t_{\rm tot}$, the LAT's total live time, and $\mathcal{E}$, the LAT's exposure. 
Given the photon spectra, the number of photons from a DM annihilation signal in a spatial region $\gO^i$, with $J$-factor 
$J_{\rm ann}^i$, and energy range $\pq{E_k,E_{k+1}}$ is given by 
\bea\label{anni}
N_{\gam}^{i,k}={1\over 8 \pi}\, {r_\odot \, \rho^2_\odot  \over m^2_{\rm DM}}\, {\langle \gs v\rangle}\,J_{\rm ann}^i \,\mathcal{E}^{i,k}
\, \int_{E_k}^{E_{k+1}}\, \de  E_\gam \, {\de N_{\gam} \over \de E_\gam}\,,
\eea
where $\mathcal{E}^{i,k}$ is the exposure averaged over $\gO^i$ and calculated at the midpoint of $\pq{E_k,E_{k+1}}$ (since the variation of the exposure over a single energy bin is very small).  
For decays the predicted counts are 
\bea
N_{\gam}^{i,k}={1\over 4 \pi} \, {r_\odot \rho_\odot \over m_{\rm DM}} \, {1\over \tau} \, J_{\rm dec}^i \,\mathcal{E}^{i,k} \, \int_{E_k}^{E_{k+1}}\, \de E_\gam \, {\de N_{\gam}  \over \de E_\gam}\,.
\eea
The approximately homogeneously distributed DM in the Universe could provide an extragalactic contribution to the observed photon flux. However, the observed $\gam$-ray spectrum will be different than that expected from Galactic DM interactions since the photons redshift as they propagate to us and there is a finite optical depth --- the result of interactions of the $\gamma$ rays with low-energy photons that compose the extragalactic background light (EBL). This yields the following expected extragalactic photon intensity for decaying DM~\cite{Cirelli:2010xx,Essig:2013goa} 
\begin{equation}
\frac{\de^2 \Phi_\gam}{\de E_\gam \de \Omega} = \frac{1}{4\pi}\,\frac{\Omega_{\rm DM}\,\rho_{c,0}}{\tau m_{\rm DM}} 
\,\int_0^\infty \de z\,\frac{e^{-\tau(E_\gam(z),z)}}{H(z)} \frac{\de N_\gam}{\de E_\gam}(E_\gam(z),z)\,.
\label{smoothedmap}
\end{equation}
Here, $\Omega_{\rm DM} \simeq 0.267$ is the present DM energy density, $\rho_{c,0} \simeq 4.7 \times 10^{-6}$~GeV/cm$^3$ 
is the critical density today, 
$E_\gam(z)=E_\gam(z+1)$ is the energy of the emitted photon, 
$H(z) = H_0\sqrt{\Omega_m(1+z)^3+\Omega_\Lambda}$, where $\Omega_m\simeq 0.317$ and  $\Omega_\Lambda \simeq 0.683$ are the total matter and cosmological-constant energy densities~\cite{Ade:2013zuv}, respectively, and we assume a flat Universe with $\Omega_m + \Omega_\Lambda = 1$.  The optical depth is given by $\tau \pt{E_\gam,z}$, and we use the parameterizations found in~\cite{Cirelli:2010xx}. We note that, for annihilating DM, the smooth extragalactic contribution is subleading compared to the Galactic one and we ignore it, whereas for decays it is a factor of order $\lesssim 1$ as large as its Galactic counterpart and we include it in our analysis.

\subsection{Inverse Compton Scattering}
\label{subsec:ICS}

We include the flux generated by ICS for the cases where DM annihilates/decays to $e^{+}e^{-}$, $\mu^{+}\mu^{-}$, $\tau^{+}\tau^{-}$, as well as $\phi\phi$ channels. 
In all cases we end up with high-energy electrons and positrons. 
These propagate within the Galaxy and can lose energy through ICS off starlight, infrared background light, or CMB photons, or via synchrotron radiation in the Galactic magnetic field. 
The ICS process ($e^\pm {}' \gamma' \rightarrow e^\pm \gamma$) can produce high-energy $\gamma$ rays that are observed by the LAT.  
The synchrotron-radiation contribution in the {\it Fermi}-LAT energy range is subdominant for the DM masses and models that we consider \cite{Meade:2009iu} and thus we neglect it when we derive our limits. 
However, it must be included when calculating the ICS $\gamma$-ray signal, since it determines how fast the electrons and positrons cool. 
In fact, the cooling time is strongly dependent on the Galactic magnetic field, whose values at different locations in our Galaxy are not known very accurately. This leads to large uncertainties in the ICS signal. 
Moreover, the calculation of the ICS signal requires additional assumptions; for example, we assume, as generally done, that the density of electrons and positrons after propagation follows a steady-state solution. 
However, phenomena such as the {\it Fermi} bubbles~\cite{Su:2010qj}, pointing to a dynamical event in the Milky Way's recent history, might make this assumption not fully justified.  
We also assume that the steady-state propagation of the electrons/positrons only occurs inside a cylindrical region of the Galaxy that has a maximum radius $R_h$ and half-height $z_h$. The steady-state diffusion equation is given by (e.g.,~\cite{Delahaye:2007fr})
\bea\label{eq:diffusion}
-D_{xx}\pt{E_e'} \nabla^2 {\de n_e \over \de E_e'} - { \partial \over \partial E_e'} \pq{b \pt{r,z,E_e'}\, {\de n_e \over \de E_e'}}
= 
\left\{ \begin{array}{l l}
    {1 \over 2} \, \pt{\rho (r,z) \over m_{\rm DM}}^2 \, \langle \gs v\rangle \, {\de N_e \over \de E_e'}, \,\, \mbox{annihil.}\\
   {1 \over 2} \, {2 \rho (r,z) \over m_{\rm DM}} \, \frac1{\tau}\, {\de N_e \over \de E_e'}, \,\, \mbox{decays}
  \end{array} \right.\,.
\eea
Here ${\de n_e / \de E_e'} \equiv {\de n_e (r,z,E_e')/ \de E_e'}$ is the energy-dependent differential electron+positron density at a given point in the Galaxy, $(r,z)$, where $r$ and $z$ are the cylindrical coordinates of the electron/positron in the Galaxy. The right-hand side of Eq.~(\ref{eq:diffusion}) is the source term and contains the DM density profile, $\rho(r,z)$ (a function of cylindrical coordinates) and the electron+positron energy spectrum, 
$\de N_e/\de E_e'$; also, there is a factor $1/2$ for Majorana fermions, otherwise $1$ for Dirac fermions. The first term on the left-hand side accounts for the spatial diffusion and is characterized by an energy-dependent coefficient,  
\be
D_{xx} \pt{E_e'}=D_0 \pt{E_e' \over E_0}^\gd\,.
\ee
The second term is the energy-dependent loss and is given by 
\be\label{bfactor}
b \pt{r,z,E_e'}\equiv -{\de E_e'\over \de t}={4 \gs_T\over 3m^2_e}E_e'^2 \pq{u_B(r,z) + \sum_{i=1}^3 u_{\gamma i}(r,z) \,R_i (E_e')}\,,
\ee
where $\gs_T={8\pi r_e^2/3}$, with $r_e={\ga_{\rm em}/m_e}$, is the Thomson cross section, and $u_B\pt{r,z}=B^2/2$ is the energy density of the Galactic magnetic field $B$, chosen to have the form~\cite{Strong:1998fr}  
\bea
B \equiv B(r,z) = B_0 \,e^{- [(r - R_\odot) / R_b + z / z_b]}\,, 
\eea
where $R_b = 10$~kpc and $z_b = 2$~kpc. 
The $u_{\gamma i}\pt{r,z}$ are the energy densities of the three relevant light components in the Galaxy, i.e.:~CMB, infrared light, and starlight. 
The factors $R_i(E_e')$ take into account relativistic corrections.
The $\gamma$-ray differential flux at energy $E_\gamma$, resulting from ICS off an electron is 
\bea
\frac{\de^2\Phi_{\gamma}}{\de E_\gamma \de \Omega} = \frac{\alpha_{\rm em}^2}{2}
\int \de s \iint \de E_\gamma' \, \de E_e' \,\frac{ f_{\rm IC}\, \pt{q,\eps} }{{{E_\gamma'}^2}{E_e'^2}}\, \frac{\de n_e}{\de E_e'} \pt{r, z, E_e'}\, \frac{\de u_\gamma}{\de E_\gamma'} \pt{r, z, E_\gam'} \,,
 \label{eq:Compton1}
\eea
where $s$ is the line-of-sight distance, and 
\begin{eqnarray}
f_{\rm IC}(q,\epsilon )& \equiv & 2q\log q+(1+2q)(1-q)+\frac{1}{2}\frac{(\epsilon q)^2}{1+\epsilon q}(1-q) \,,\\
q&\equiv &\frac{\epsilon}{\Gamma(1-\epsilon)} ,\qquad \epsilon\equiv\frac{E_\gamma}{E_e'}, \qquad \Gamma\equiv\frac{4E_\gamma' E_e'}{m_e^2}\,. 
\eea
We calculate the ICS contribution with \texttt{GALPROP V50}~\cite{Vladimirov:2011rn}. We use a version of \texttt{GALPROP V50} that was modified by the authors of~\cite{Cholis:2008wq} to include various DM annihilation and decay final states.
We fix $\delta = 0.33$, $E_0=4$~GeV, and take the cylindrical geometry to have a maximum radius $R_h =20$~kpc and a maximum half-height $z_h = 4$~kpc. As mentioned above, the greatest source of uncertainty is due to the Galactic magnetic field, $B$. To capture some of this uncertainty, we vary $B_0$ between $1-10~\mu$G, when showing our results in \S\ref{sec:results}. We fix the spatial diffusion coefficient parameter to be $D_0 = 4.797 \times10^{28}$~cm$^2$/s ($6.311 \times10^{28}$~cm$^2$/s) for $B_0=1~\mu$G ($10~\mu$G). (See Appendix~\ref{app:ICS} for sources for these values.) In Appendix~\ref{app:ICS} we show how our results are affected when varying $z_h$ and $R_h$, in addition to $D_0$ and $B_0$; we find that the largest effect on the results comes from the variation of $B_0$.

\section{DATA SETS AND METHODS}
\label{sec:methods}

We aim to set conservative, robust constraints on the annihilation and decay of DM into various SM final states.  
We consider the inclusive photon spectrum observed by the {\it Fermi}-LAT, and use simulated data to first find the ``optimal'' ROI in the 
$\gamma$-ray sky, i.e.~the one that yields the strongest constraint. We then require the DM signal to be less than the observed photon counts. 
We note that our approach does not allow us to search for the existence of a DM signal.

In this section we describe the event selection, how we use the simulated data sets in our analysis, the ROI choice, and 
how we construct optimal upper bounds on the DM annihilation cross section and lower bound on the DM decay lifetime. 
We also provide a detailed example of our procedure. 

\subsection{Event Selection}
\label{subsec:data}
The data set used for this study consists of $\sim 5.84$ years of {\it Fermi}-LAT data (from August~2008 until June~2014) in the energy range $1.5-750$~GeV. 
We select photons using the \texttt{P7REP\_CLEAN} event-class selection~\cite{Bregeon:2013qba}, to minimize contamination by residual cosmic rays. We also require the zenith angle to be smaller than $100^{\circ}$ to remove photons originating from the bright Earth's Limb. 
Details on the {\it Fermi}-LAT instrument and performance can be found in \cite{Atwood:2009ez,Ackermann:2012kna}.  
All data reduction and calculation of the exposure maps were performed using the {\it Fermi}-LAT \emph{ScienceTools}, version v9r34p1 \cite{ScienceTools}. As for the {\it Fermi}-LAT instrument response functions (IRFs), we use \texttt{P7REP\_CLEAN\_V15} for both MCs and data. As described in Appendix \ref{subsec:SS-FB}, the results shown in this paper are obtained after masking all known point sources identified in the 5-year {\it Fermi} catalog (3FGL)~\cite{TheFermi-LAT:2015hja}, using a PSF (point spread function)-like masking radius, except for those photons coming from within the inner $2^\circ\times2^\circ$ square at the GC.  Moreover, we include both front- and back-converting events.  In Appendix~\ref{subsec:SS-FB} we show that, although this choice is generally optimal, our results are not significantly affected if we mask only the brightest sources, or no sources at all, and if we include only front- or only back-converting events.

\subsection{Simulated (Monte Carlo) Data Sets}
\label{subsec:MC1}
For our study, we use 10 Monte Carlo (MC) data sets, each a statistically independent $\sim 5.84$-year representation of the $\gamma$-ray sky.  
The same event selection described above is applied to MC data. 
We use the simulated data sets to select ``optimized'' ROIs, independent of the real data, as described below in \S\ref{subsec:roichoice}.  
By finding optimal ROIs based on the MC simulations, we avoid the possibility of accidentally obtaining a strong constraint due to 
statistical fluctuations in the data.  
We describe the details of the simulated data in Appendix~\ref{app:MC}.  
Note that the MC simulations contain photons with an energy range of 0.5~GeV to 500~GeV (as opposed to 1.5~GeV to 750~GeV in the data). 
We account for this difference by extrapolating the MC data up to 750~GeV as described in Appendix~\ref{app:MC}.  

\subsection{ROI Choice}
\label{subsec:roichoice}

We take the ROI for annihilating DM to have the dumbbell shape as shown in Fig.~\ref{fig:ROI} ({\bf left}).  This shape depends on three parameters: the radius from the GC to the edge of the ROI, $R$, the width in latitude of the Galactic Plane (GP) that is to be excluded from the ROI, $2\Delta b$, and the width in longitude of the GC region that is to be included in the ROI, $2\Delta \ell$. The motivation for choosing such shape is that the DM distribution is approximately spherically symmetric (hence the choice of a circular region, parametrized by $R$), but the Galactic foregrounds are largest in the GP region, which we then remove. However, we include the GC in our ROI as this is where the DM signal peaks as well, dramatically so for cuspy profiles (since $N_{\gamma ,\rm DM}\propto\rho_{\rm DM}^2$). 

\begin{figure}[t!]
\includegraphics[width=0.48 \textwidth]{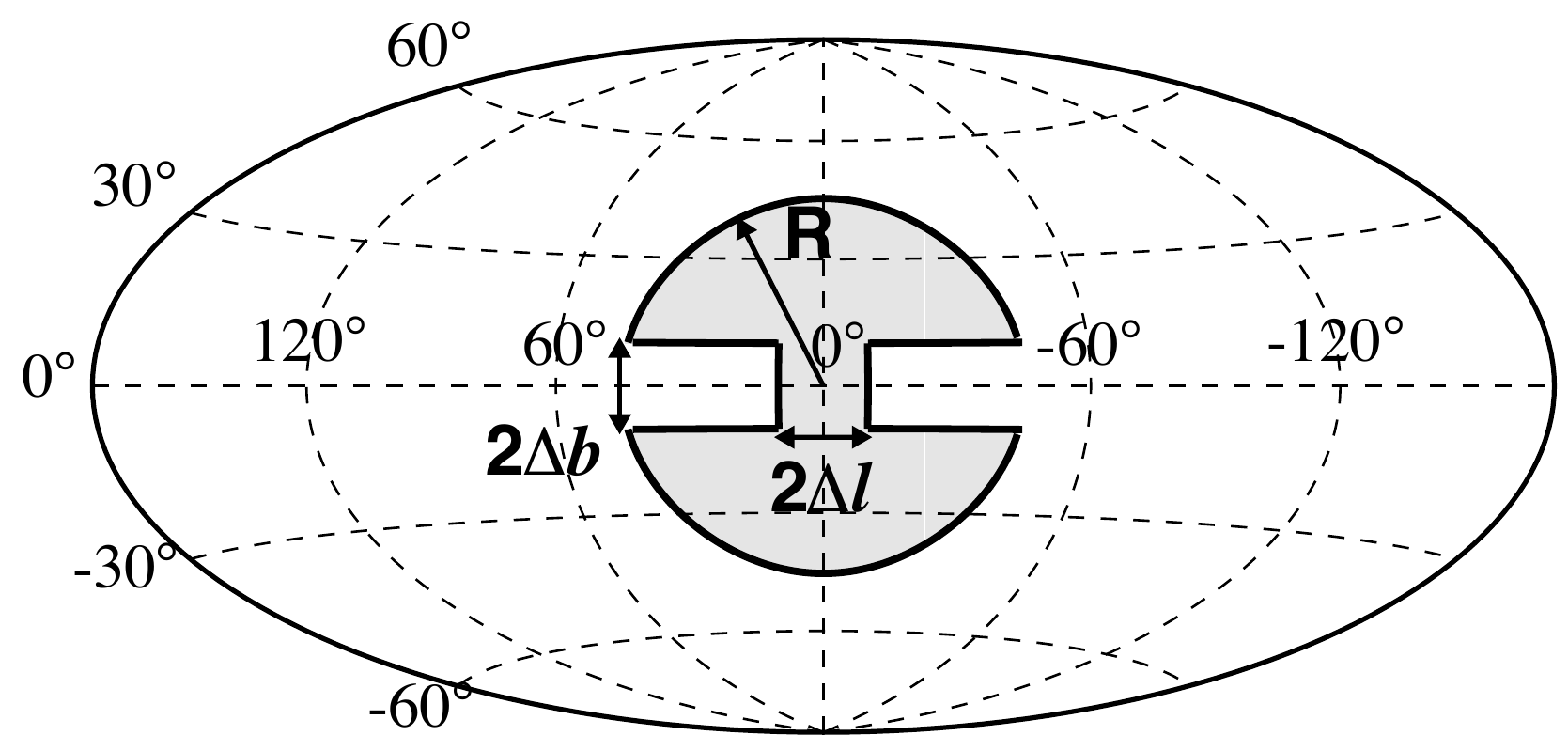}\;\;\;
\includegraphics[width=0.48 \textwidth]{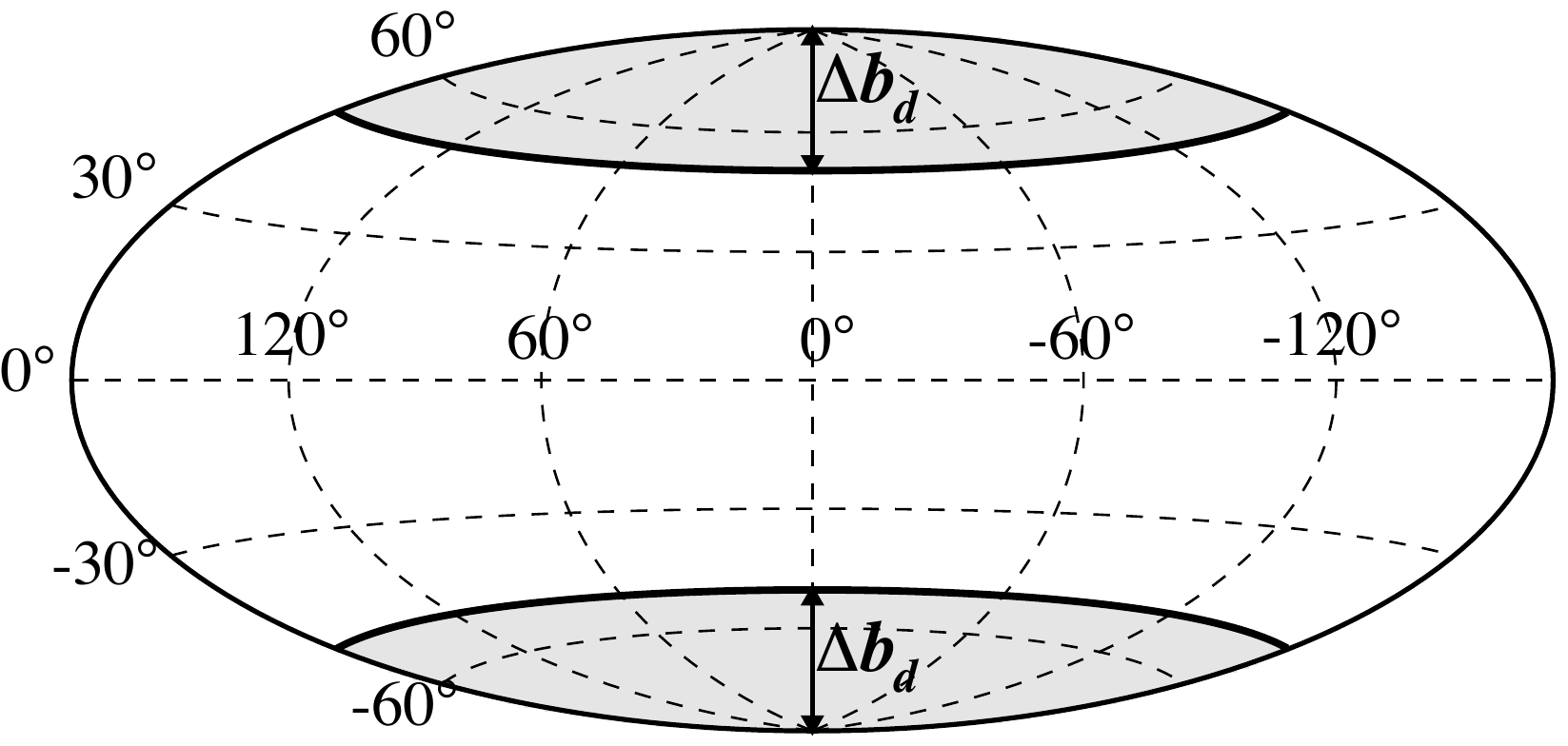}
\caption{{\bf Left:} 
The choice of ROI (shaded) in the $\gamma$-ray sky for dark-matter annihilation.  
The ROI depends on 3 parameters, as indicated.
{\bf Right:} The choice of ROI for dark-matter decays (shaded), which depends on one parameter, as indicated.
\label{fig:ROI}
}
\end{figure}

For decaying DM, our choice of ROI will consist of the two high-latitude regions shown in Fig.~\ref{fig:ROI} ({\bf right}), 
and depends on only one parameter: the width in latitude from the Galactic poles to the edge of the ROI, $\Delta b_d$. 
In contrast to annihilation, the decaying DM signal is expected to be much less concentrated in the GC, since $N_{\gamma, \rm DM}\propto \rho_{\rm DM}$. 

\subsection{Optimizing the ROIs and Energy Ranges using Simulated Data}
\label{subsec:optimization}

A particular DM model or {\it Theory Hypothesis}, $T_H=[m_{\rm DM},\rho,\text{annihilation/decay},\text{channel}]$, is characterized by the DM mass ($m_{\rm DM}$), the DM density profile ($\rho$), whether it is annihilating/decaying DM, and the annihilation/decay final state. 
Given any ROI and a photon energy range, [ROI, $\Delta E$], we obtain a constraint on either the DM annihilation cross section, 
$\langle \sigma v\rangle$, or decay lifetime, $\tau$, for a given $T_H$ by requiring that the number of DM events, 
$N_{\gamma, {\rm DM}}$, in [ROI, $\Delta E$] does not exceed the observed value, $N_{\gamma, O}$.  
More precisely, to set a limit with a confidence level (C.L.) of $1-\alpha$, we vary $\langle \sigma v\rangle$ or $\tau$ 
until the probability that $N_{\gamma, {\rm DM}}>N_{\gamma, O}$ is $\alpha$; in equations, 
the bound on $\langle \sigma v\rangle$ or $\tau$ is obtained by solving 
\bea\label{eq:limit}
\sum_{k=0}^{N_{\gamma, O}} {\rm Poisson}\pt{k\,\vert \,N_{\gamma, \rm DM}}=1-\ga, 
\eea
where as usual 
\bea
{\rm Poisson}(k \, \vert \lambda) = \frac{\lambda^k \,e^{-\lambda}}{k!}\,.
\eea
For each $T_H$, we find the optimal ROI and optimal photon energy range, [ROI, $\Delta E$]$_O$, which 
provides the best limit on $\langle \sigma v\rangle$ or $\tau$.  
If we simply scan over all [ROI, $\Delta E$] in the data, this would subject our constraints to statistical fluctuations.  
Instead, we use the 10 simulated data sets to find [ROI, $\Delta E$]$_O$ as follows.  
For the $i$-th ROI and energy range, [ROI, $\Delta E$]$_i$, and $j$-th simulated data set, we calculate the bound 
on the cross section or lifetime, $\langle \sigma v\rangle_{i,j}$ or $\tau_{i,j}$, as described above in Eq.~(\ref{eq:limit}). 
We then average the resulting expected limit across the 10 simulations, i.e.
\bea
\overline{\langle \sigma v\rangle}_{i} &=& \frac{1}{10}\,\sum_{j=1}^{10}\, \langle \sigma v\rangle_{i,j}\,,\\
\overline{\tau}_{i} &=& \frac{1}{10}\,\sum_{j=1}^{10} \,\tau_{i,j}\,.
\eea  
We then find [ROI, $\Delta E$]$_O$ by scanning over all [ROI, $\Delta E$]$_i$'s and selecting the one that yields the minimum $\overline{\langle \sigma v\rangle}_{i}$ (maximum $\overline{\tau}_i$), i.e.
\bea
\overline{\langle \sigma v\rangle} &=& \min_{i}\,{\overline{\langle \sigma v\rangle}_{i}}\,,\\
\overline{\tau} &=& \max_{i}\, { \overline{\tau}_{i}}\,.
\eea 
We then use [ROI, $\Delta E$]$_O$ on the real data to calculate the limits on $\langle \sigma v\rangle$ or $\tau$ for the given $T_H$. 

The ROIs used in our optimization are given in \S\ref{subsec:roichoice}.  
We bin each simulated data set into $0.18^\circ \times 0.18^\circ$ rectangular pixels in Galactic latitude and longitude 
and $N=127$ logarithmically-uniform energy bins between $1.5-750$~GeV.  
We then vary the ROI shape parameters described in \S\ref{subsec:roichoice} in steps of $0.5^\circ$ for $R$, steps $\sim 1^\circ$ for $\Delta b$ and $\Delta \ell$, and $2.5^\circ$ for $\Delta b_d$.  For each choice of ROI, we scan over all $(N-1)(N-2)/2  = 8064$ choices of adjacent bins in energy, assuming a minimum of three adjacent bins. In Appendix~\ref{sec:results-roispectra} we show a sample of the resulting optimized ROIs and energy ranges.

We note that for large enough $N_{\gamma, \rm DM}$ or $N_{\gamma, O}$, the statistical distributions are approximately Gaussian, and we would obtain a 95\% C.L.~bound by requiring $N_{\gamma, \rm DM} < N_{\gamma, O} + 1.64 \sqrt{N_{\gamma, O}} \simeq N_{\gamma, O}$.  Even our smallest optimal ROIs with the highest optimal energy ranges contain at least $\mathcal{O}(10)$ photons. Our method thus does not produce constraints that are susceptible to Poisson fluctuations of the number of events in [ROI, $\Delta E$]$_O$, and, as a consequence, our constraints are not expected to improve significantly with more data (some small improvements may arise from, e.g., a better rejection of backgrounds).  

We also note that since [ROI, $\Delta E$]$_O$ was selected using simulated data, other choices of [ROI, $\Delta E$] may  
provide a stronger constraint on the data.  Also, the simulated data is not a perfect representation of the data.  
Indeed, there are certain regions in the sky where the simulations do not model the data perfectly, and the ``expected'' limits using MC data may differ from the limit obtained on the real data (see Appendix~\ref{subsec:MC-data-comp}). One notable example is in the GC and in the Inner Galaxy region, which has led to claims of a $\gamma$-ray excess, see Appendix \ref{sec:gamma-ray-excess}.  
However, an imperfect modeling of the sky does not affect the validity of our constraints.  
We use the simulations as a tool to pick [ROI, $\Delta E$] in an unbiased way.  
Even if the simulations were a totally inaccurate representation of the real data, it would not \emph{invalidate} our limits, although other choices of [ROI, $\Delta E$] would provide stronger constraints.  

We note that for prompt radiation we include the effects of the {\it Fermi}-LAT's PSF, by performing its convolution with the J-factors, using the {\it Fermi}-LAT \emph{ScienceTools}. For the constraints that include prompt and ICS, however, convolving the PSF for the DM signal calculation is computationally intensive, so we do not account for these effects. To see by how much this could potentially affect our limits, we constrained the ROIs to have a shape which is safe w.r.t. the PSF containment radius at the lowest energies considered. If the ROI includes a portion of GC (i.e.~$\gD \ell >0^\circ$), then we require the width of this window to be at least $6^\circ$ (i.e.~$\gD \ell >3^\circ$); for the width of the top and bottom of the ROI shape (resembling crescents) we require that $R < \gD \ell$ (so the ROI is a circle), $R < \sqrt{{\gD b}^2+{\gD \ell}^2}$ (so the two crescents have no tips), and $R > 4^\circ+\gD b$ (so the two crescents are thick enough).  
The upper bounds thus obtained are only degraded by at most $\sim 20-40\%$ with respect to the unconstrained-ROI case. This is a small number; especially in view of the fact that the largest uncertainty for the DM ICS signal comes from the value of the magnitude of the local magnetic field, see Appendix~\ref{app:ICS}.  

We note that systematic effects of the PSF are not included in our analysis, as they are much smaller than the other sources of systematic uncertainty considered, such as in the ICS signal and DM density profile.

\subsection{Illustration of Procedure}
\label{subsec:illustration}

%
\begin{figure}[t!]
\includegraphics[width=0.48 \textwidth]{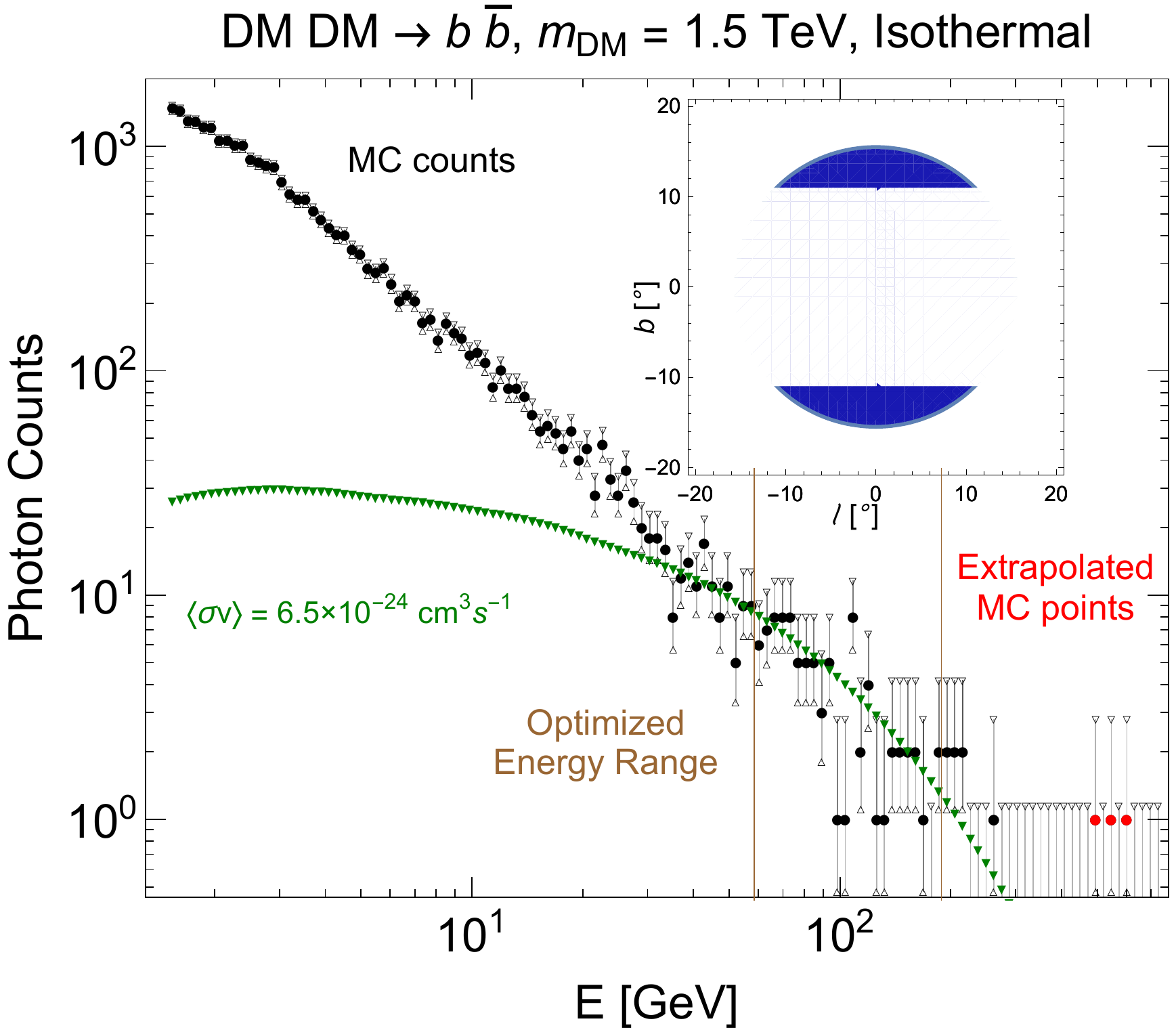}\;\;\;
\includegraphics[width=0.48 \textwidth]{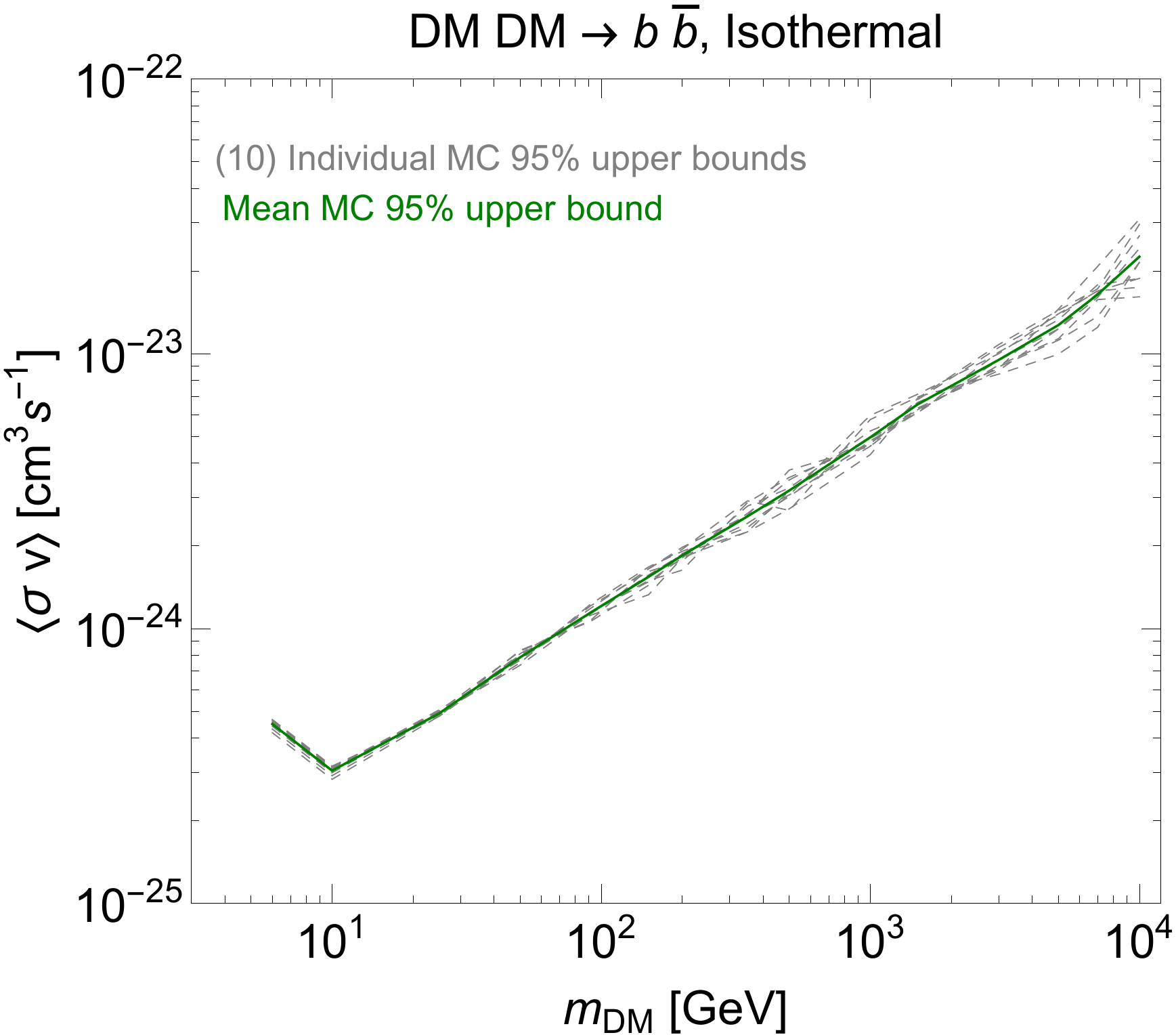}
\caption{{\bf Left:} Count spectrum from one of the MC data sets for the ROI shown in the inset.  
The green points show the spectrum for 1.5~TeV DM distributed according to the Isothermal profile, annihilating to $b\bar{b}$, with a cross section chosen such that the number of signal events in the energy range from 68~GeV to 142~GeV (vertical brown lines) is larger than the number of events in the MC data (at 95\% C.L.), as given by Eq.~(\ref{eq:limit}).  
Since the simulated data only contains photons up to 460~GeV, we extrapolate it to 750~GeV (red points), using a power-law fit to the photon spectrum above $\sim 6.2$~GeV. See Appendix~\ref{app:MC} for more details.
{\bf Right:} The best cross-section limit averaged over all ten MC data sets is shown with a green solid line, while the individual cross-section limits for each of the 10 MC data sets are shown with dashed gray lines. As explained in \S\ref{subsec:illustration}, the average cross-section limit is used as a figure of merit for our ROI/energy range optimization.
\label{fig:method-example}
}
\end{figure}
An illustration of our method is shown in Fig.~\ref{fig:method-example}.
The left plot shows the count spectrum from one of the MC data sets for the ROI shown in the inset. The green triangles show the spectrum for a 1.5~TeV DM annihilating to $b\bar{b}$, assuming isothermally distributed DM, with the cross section set at the 95\% C.L. upper limit. This limit is derived by requiring that the number of signal events in the optimal energy range from 68~GeV to 142~GeV (vertical brown lines) be larger than the number of events in the MC data as given by Eq.~(\ref{eq:limit}), where we set $\ga=0.95$. The number of events in this ROI and energy range will fluctuate from one MC data set to another, and we calculate the average cross-section limit for all ten MC data sets. We show the best average cross-section limit as a function of DM mass with a green solid line in Fig.~\ref{fig:method-example} ({\bf right}), together with the cross-section limit for the ten individual MC data sets (dashed gray lines). In Fig.~\ref{fig:method-example}, we masked all point sources and included both front- and back-converting events.  

We now have all the ingredients put in place for calculating constraints from the $\gamma$-ray sky observed by the {\it Fermi}-LAT. In the next section we give the 95\% C.L. bounds on the annihilation cross section (upper bound) and on the DM lifetime (lower bound) for annihilations and decays into various SM modes, respectively. 

\section{RESULTS AND DISCUSSION}
\label{sec:results}

In this section we give the results from the optimization procedure described in \S\ref{sec:methods}. We emphasize that the constraints obtained in this study are conservative and robust, since they do not depend on the modeling and subsequent subtraction of astrophysical foregrounds. In \S\ref{subsec:results-annihilation} (\S\ref{subsec:results-decays}) we discuss the constraints on annihilating (decaying) DM.  Additionally, in Appendix~\ref{sec:gamma-ray-excess} we use our method to derive bounds on models invoked to explain a putative $\gamma$-ray excess at the GC \cite{Goodenough:2009gk, Hooper:2010mq, Boyarsky:2010dr, Hooper:2011ti, Abazajian:2012pn, Gordon:2013vta, Abazajian:2014fta, Hooper:2013rwa, Daylan:2014rsa, Huang:2013pda, Macias:2013vya,Calore:2014xka}. The effect on our constraints due to different choices of source-masking, and due to the variation of ICS parameters is discussed in Appendix~\ref{subsec:SS-FB} and Appendix~\ref{app:ICS}, respectively.
\vspace{-2mm}
\subsection{Constraints on Dark Matter Annihilation}
\label{subsec:results-annihilation}

The constraints on the DM-annihilation cross section as a function of DM mass are presented in Fig.~\ref{fig:result-Ann} for annihilation to $e^+e^-$, $\mu^+\mu^-$, $\tau^+\tau^-$, and $\phi\phi$, where $\phi$ decays either only to $e^+e^-$ (with $m_\phi = 0.1$~GeV), or only to $\mu^+\mu^-$ (with $m_\phi = 0.9$~GeV), or to $e^+e^-$, $\mu^+\mu^-$, and $\pi^+\pi^-$ in the 
ratio $1:1:2$ (with $m_\phi = 0.9$~GeV). 
Fig.~\ref{fig:result-Ann-2} shows the results for the final states $b\bar{b}$, $W^+W^-,$\footnote{Note that limits for the $W^+W^-$ 
channel extend to $m_{DM} < m_W$.  In this region, the $W^+W^-$ final state is not produced on-shell, but instead 
the annihilation is to a three- or four-body final state consisting of leptons and/or quarks through off-shell $W^\pm$.  
(The expected cross-section in any concrete DM model for the off-shell process would be highly suppressed compared to the on-shell process.)\vspace{0mm}} $u \bar{u}$, and $gg$.
In all cases we present the results for four different assumptions about the DM profile $\rho(r)$ introduced in \S\ref{subsec:Prompt}. 
We note that each DM mass for each spatial distribution and final state choice has been separately optimized, and an 
optimal ROI, ROI$_{o,i}$, and photon energy range, $\Delta E_{o,i}$, were obtained to set the 95\% C.L.~constraint.  
In Appendix~\ref{sec:results-roispectra} we illustrate how the optimal ROI and energy range change for various DM 
density profiles and for different DM masses (see Figs.~\ref{fig:result-ROIs} and~\ref{fig:result-ROIs2}).

\begin{spacing}{1}
\begin{figure}[t!]
\begin{center}
\vskip -2mm
\includegraphics[width=\mywidth\textwidth]{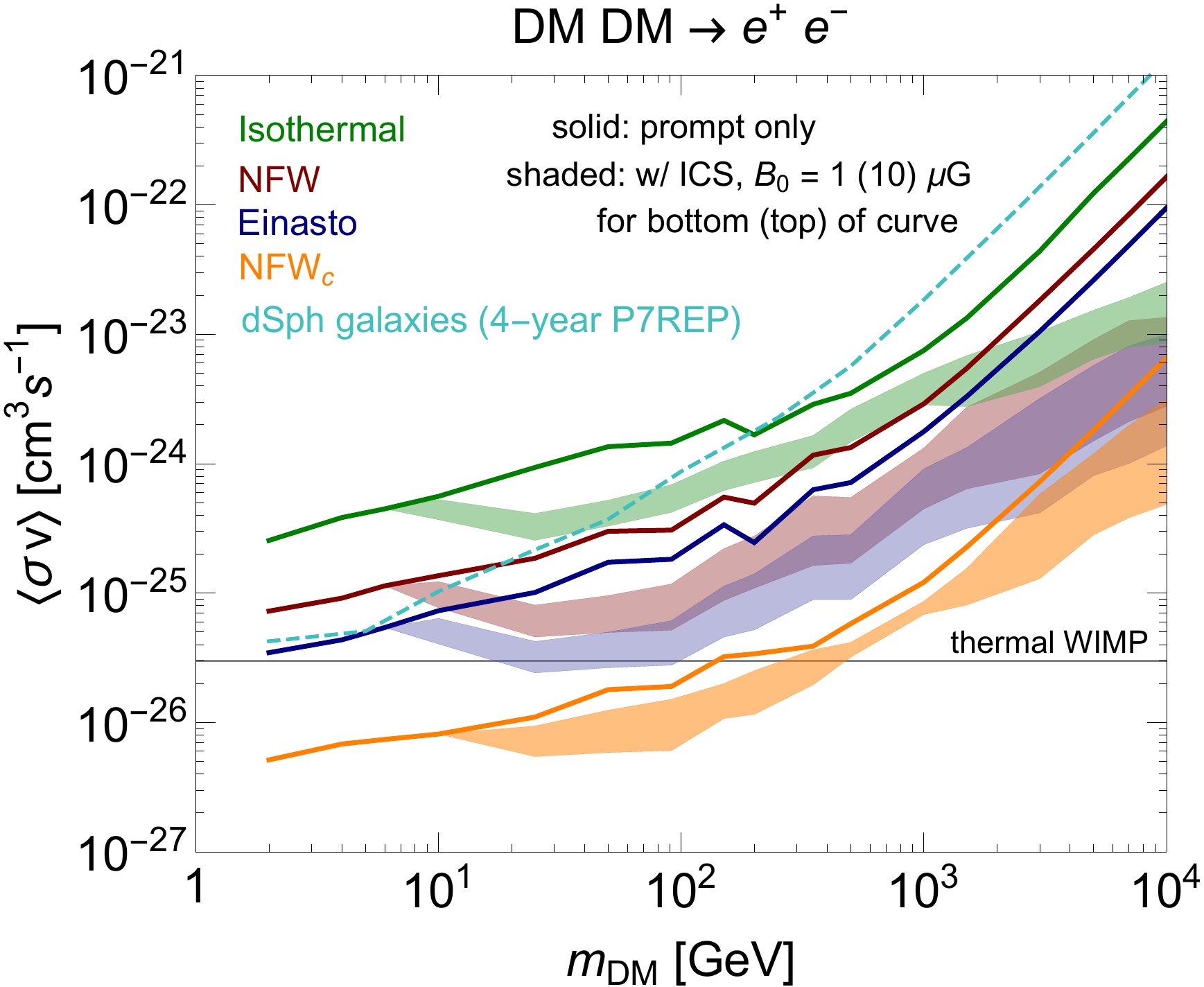}\;\;
\includegraphics[width=\mywidth\textwidth]{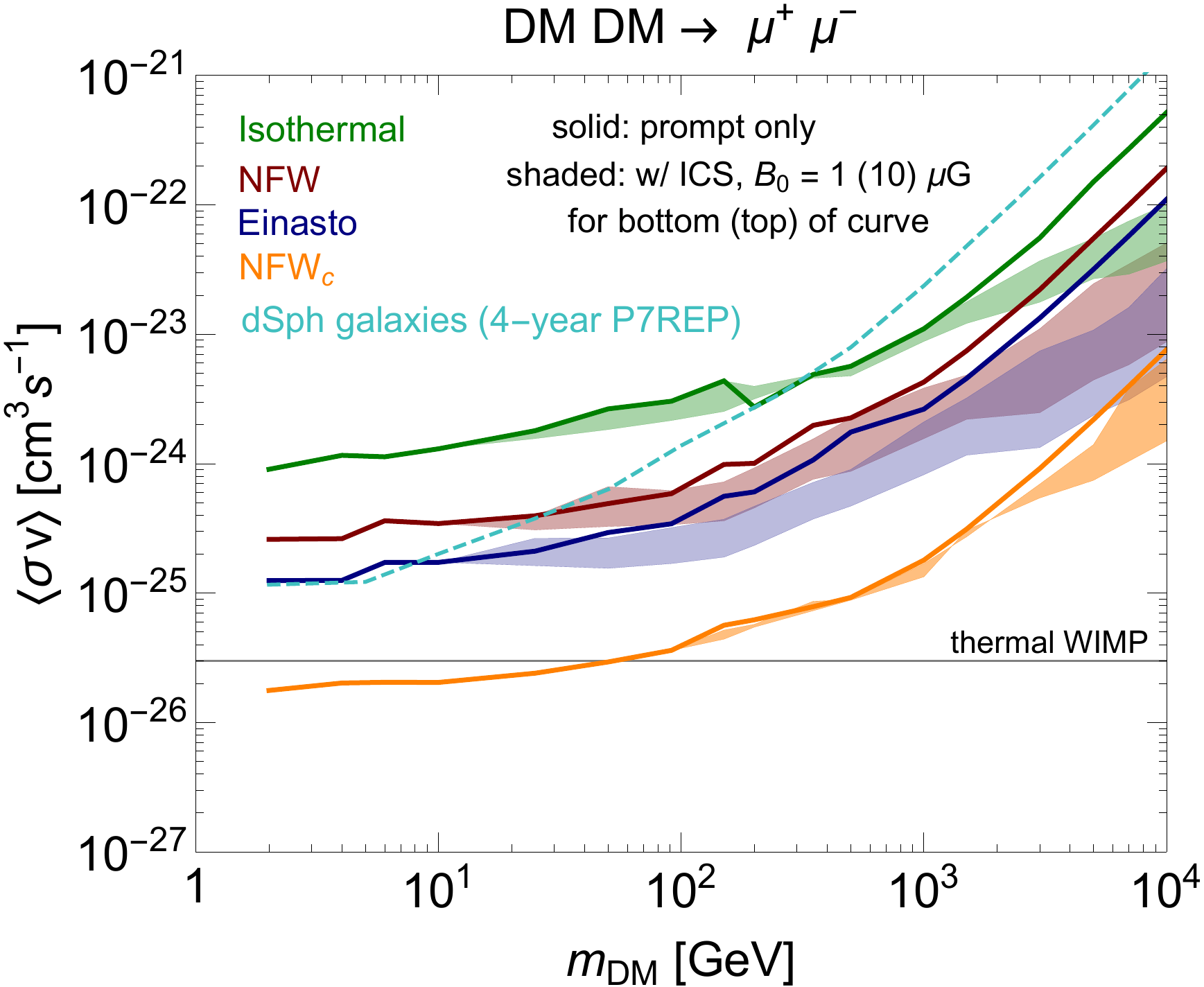}\\
\vskip 2mm
\includegraphics[width=\mywidth\textwidth]{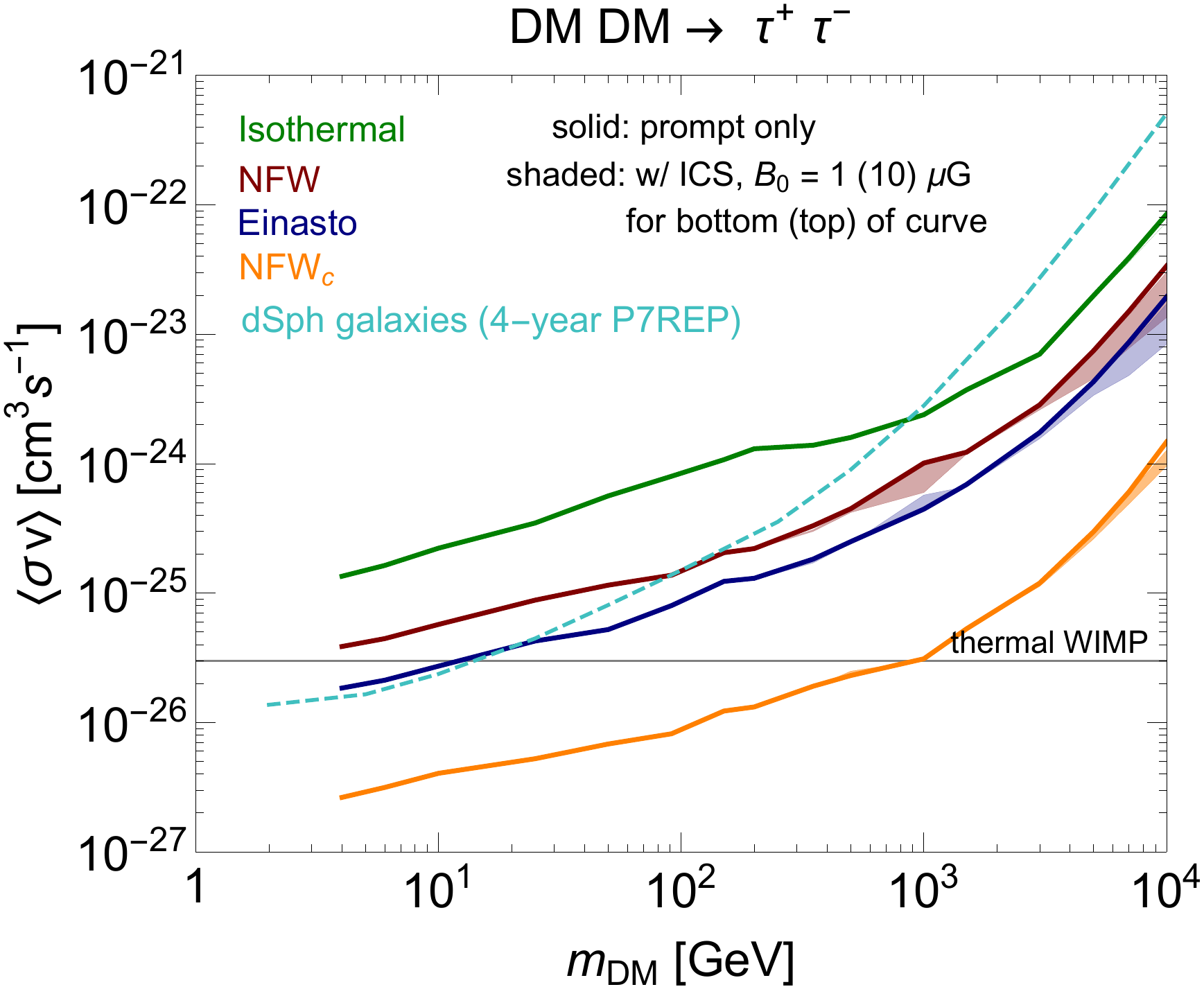}\;\;
\includegraphics[width=\mywidth\textwidth]{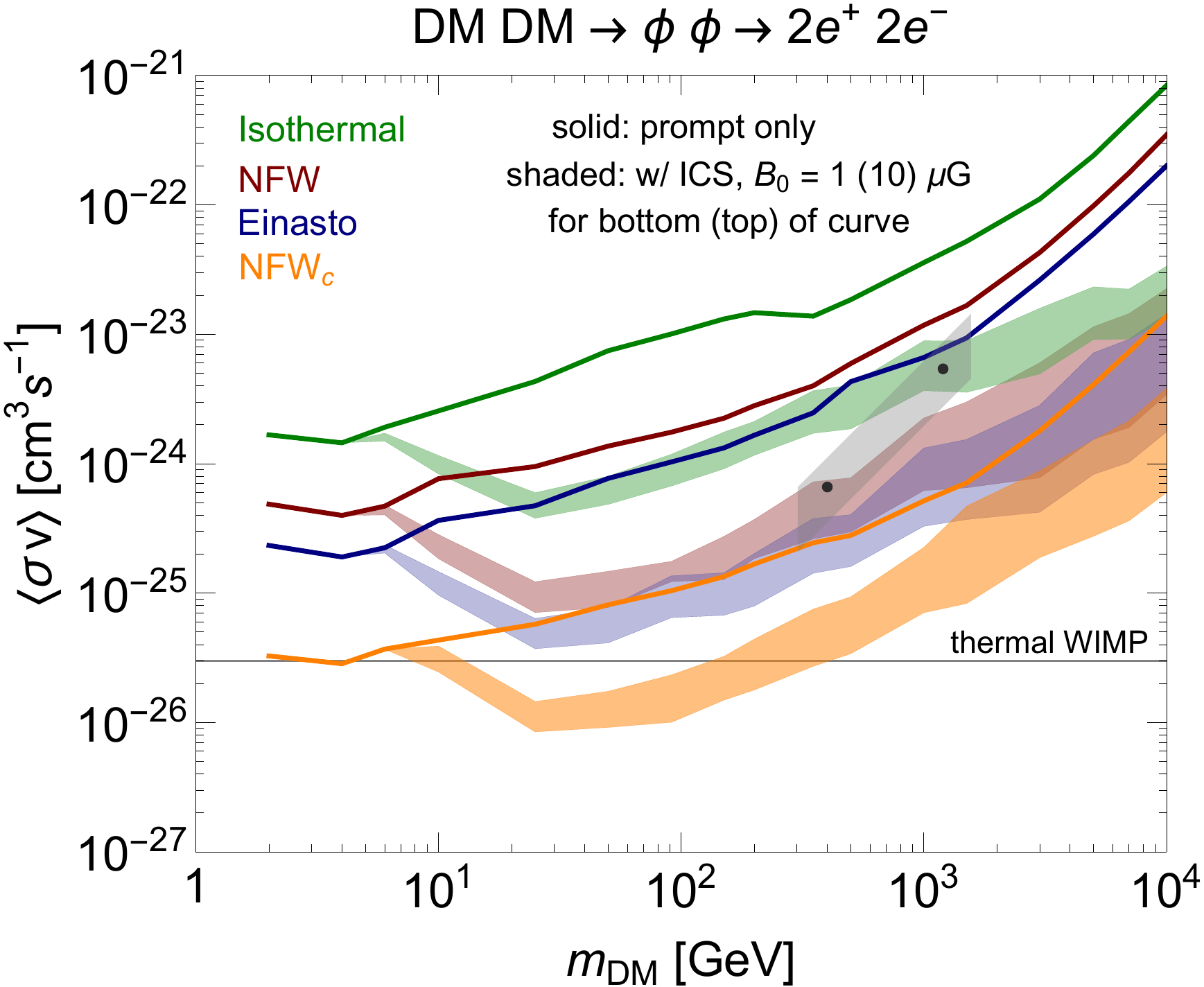}\\
\vskip 2mm
\includegraphics[width=\mywidth\textwidth]{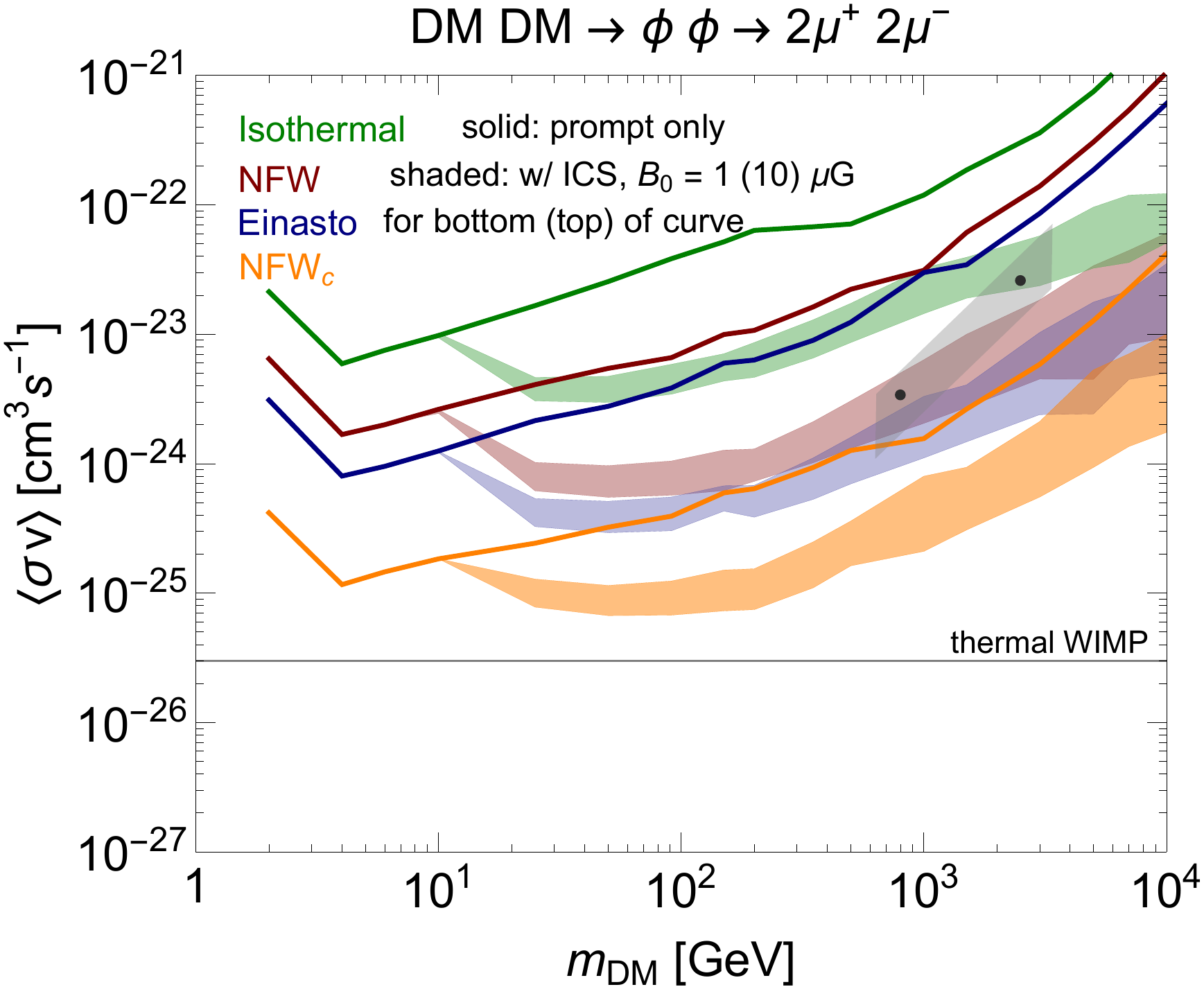}\;\;
\includegraphics[width=\mywidth\textwidth]{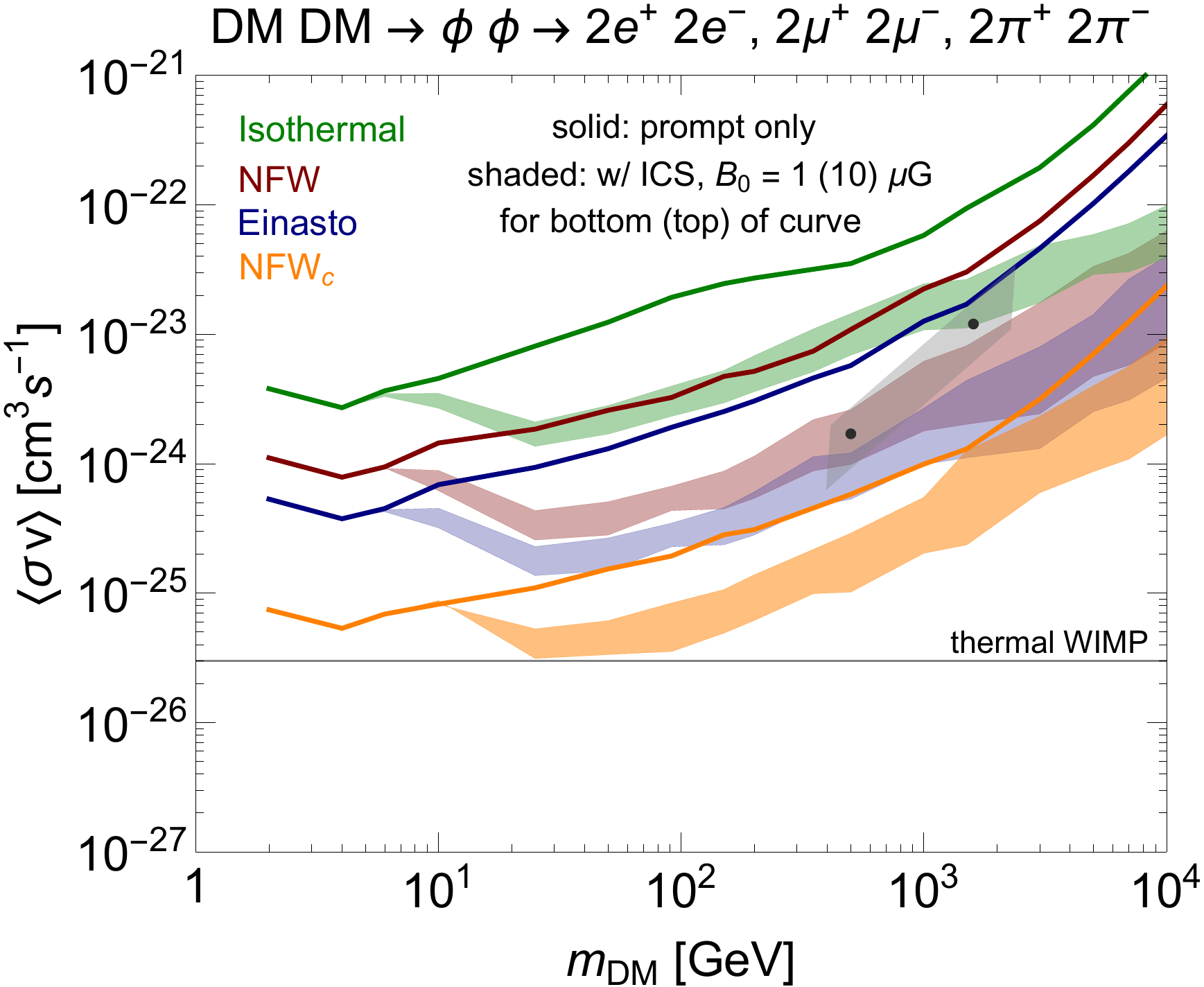}
\end{center}
\vskip -7mm
\caption{\footnotesize
95\% C.L.~upper limits on DM annihilation cross section vs. DM mass from {\it Fermi}-LAT's inclusive photon spectrum for the indicated final states.  
Each plot shows constraints for the Isothermal (green), NFW (red), Einasto (blue), and NFW$_c$ (orange) DM density profiles.   
Solid lines show constraints from the inclusion of only the prompt radiation from the annihilation, while the bands include 
the ICS off background light, with the Galactic B-field varying within $1-10~\mu$G and $D_0$ within $D_{0,\rm min} - D_{0,\rm max}$ (bottom-top of band).  
When available, we show the limits from the P7REP analysis of 15 dwarf spheroidal galaxies with a cyan dashed line~\cite{Ackermann:2013yva}.  
For the XDM models we show the approximate regions (gray) in which annihilating DM could account for the PAMELA/Fermi/AMS-02 cosmic-ray excesses. The best-fit parameters from~\cite{Cholis:2013psa} are shown as black dots.
}
\label{fig:result-Ann}
\end{figure}
\end{spacing}
\begin{spacing}{1}
\begin{figure}[t!]
\begin{center}
\includegraphics[width=\mywidth\textwidth]{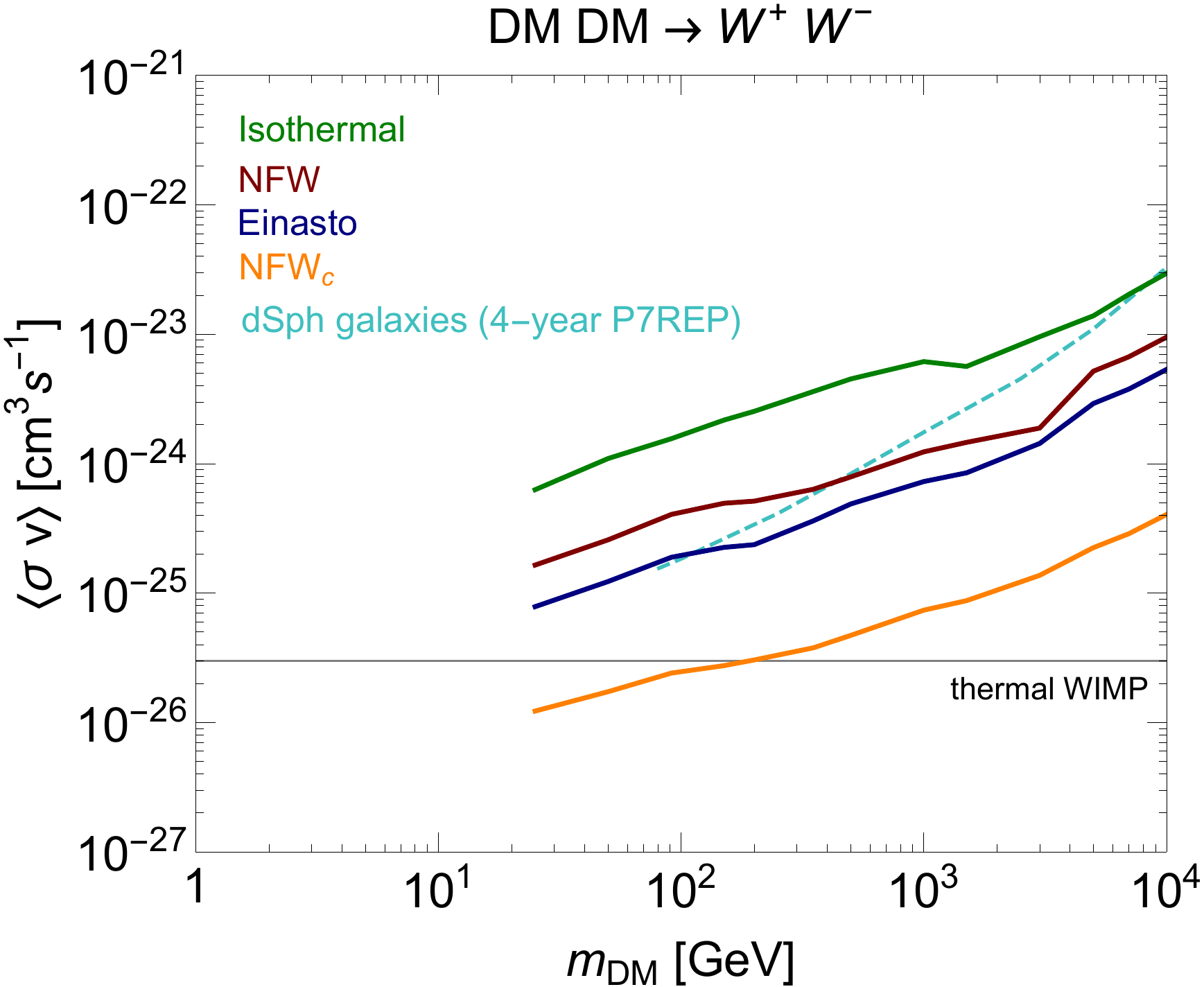}\;\;
\includegraphics[width=\mywidth\textwidth]{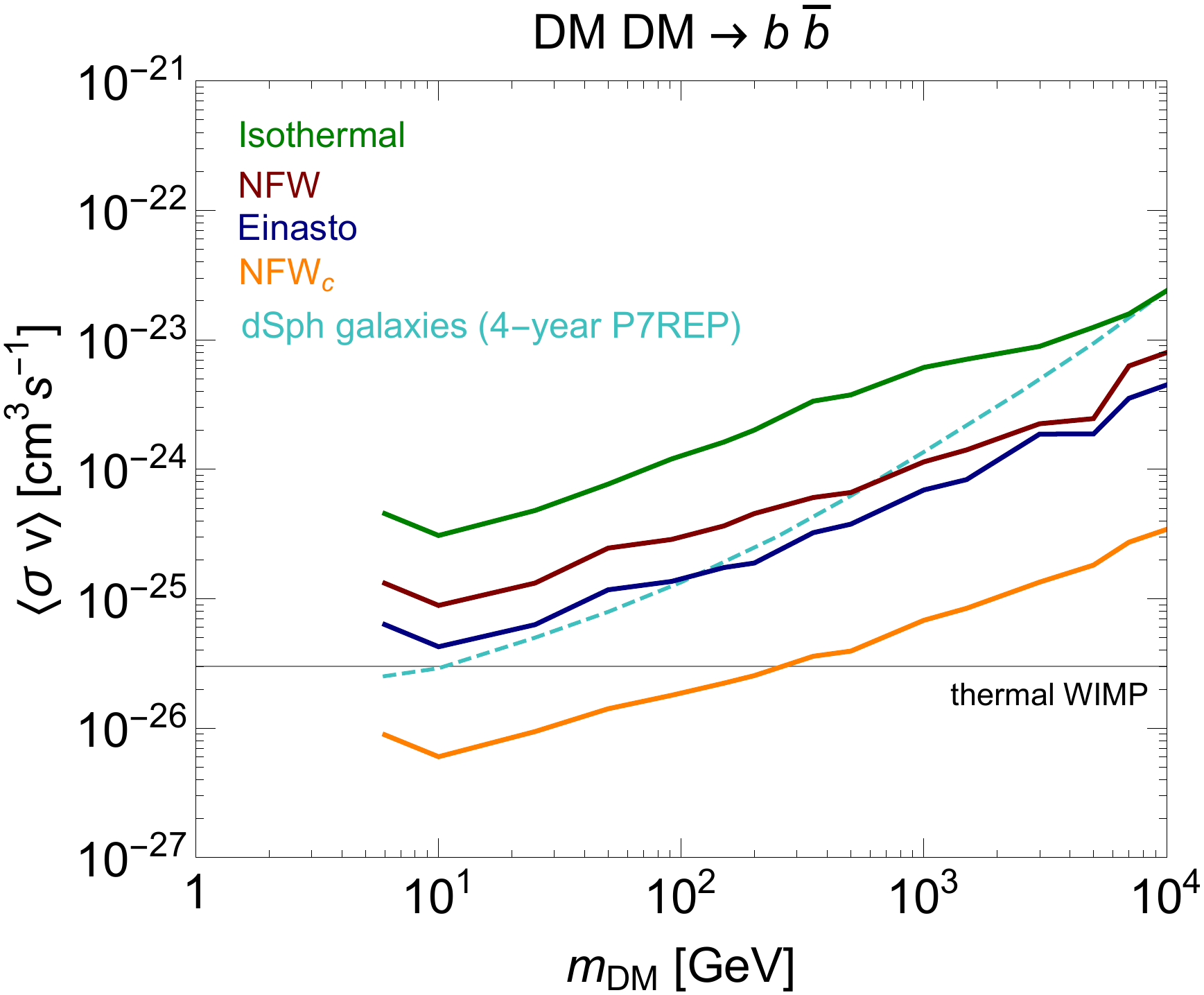}\\
\vskip 2mm
\includegraphics[width=\mywidth\textwidth]{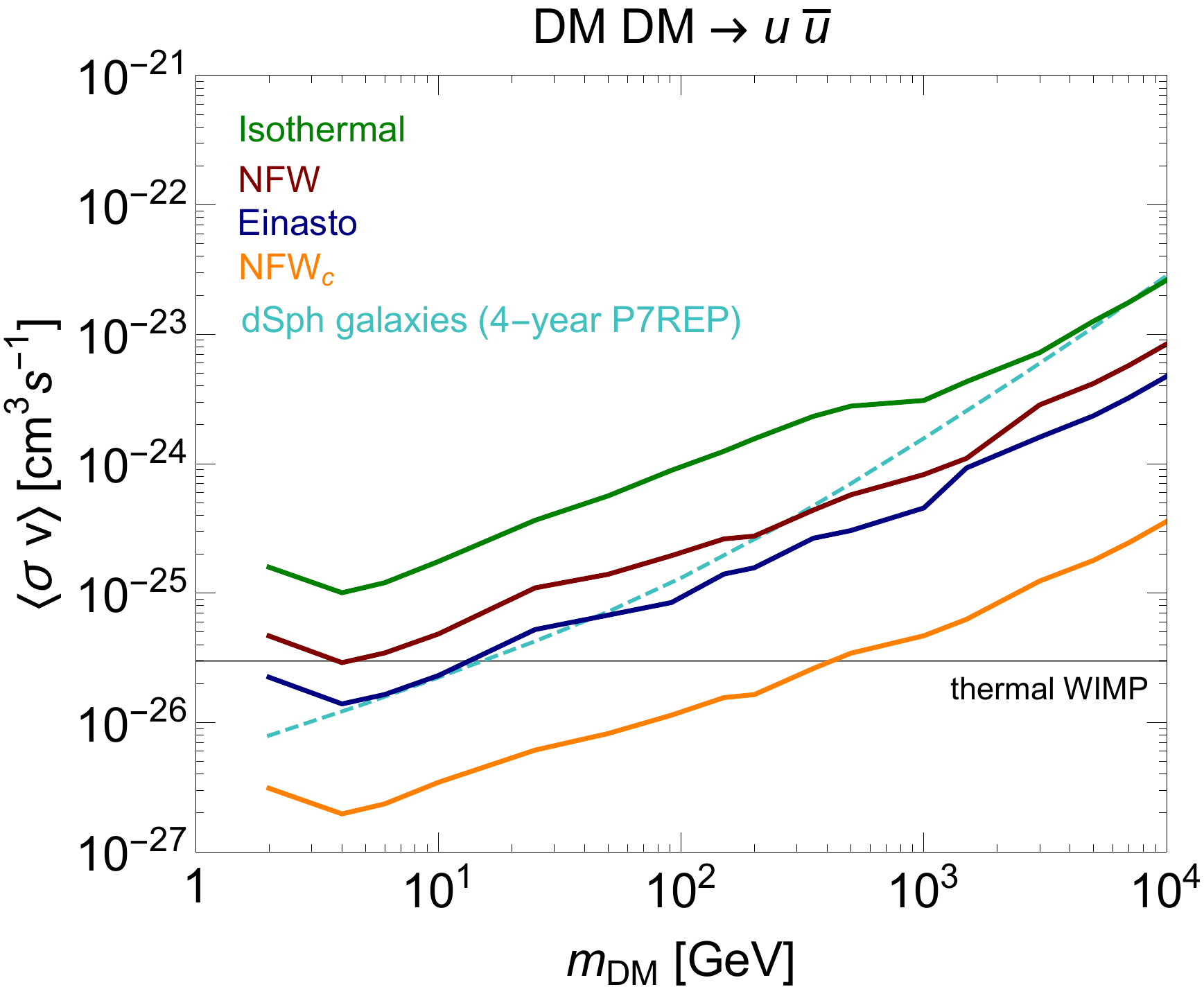}\;\;
\includegraphics[width=\mywidth\textwidth]{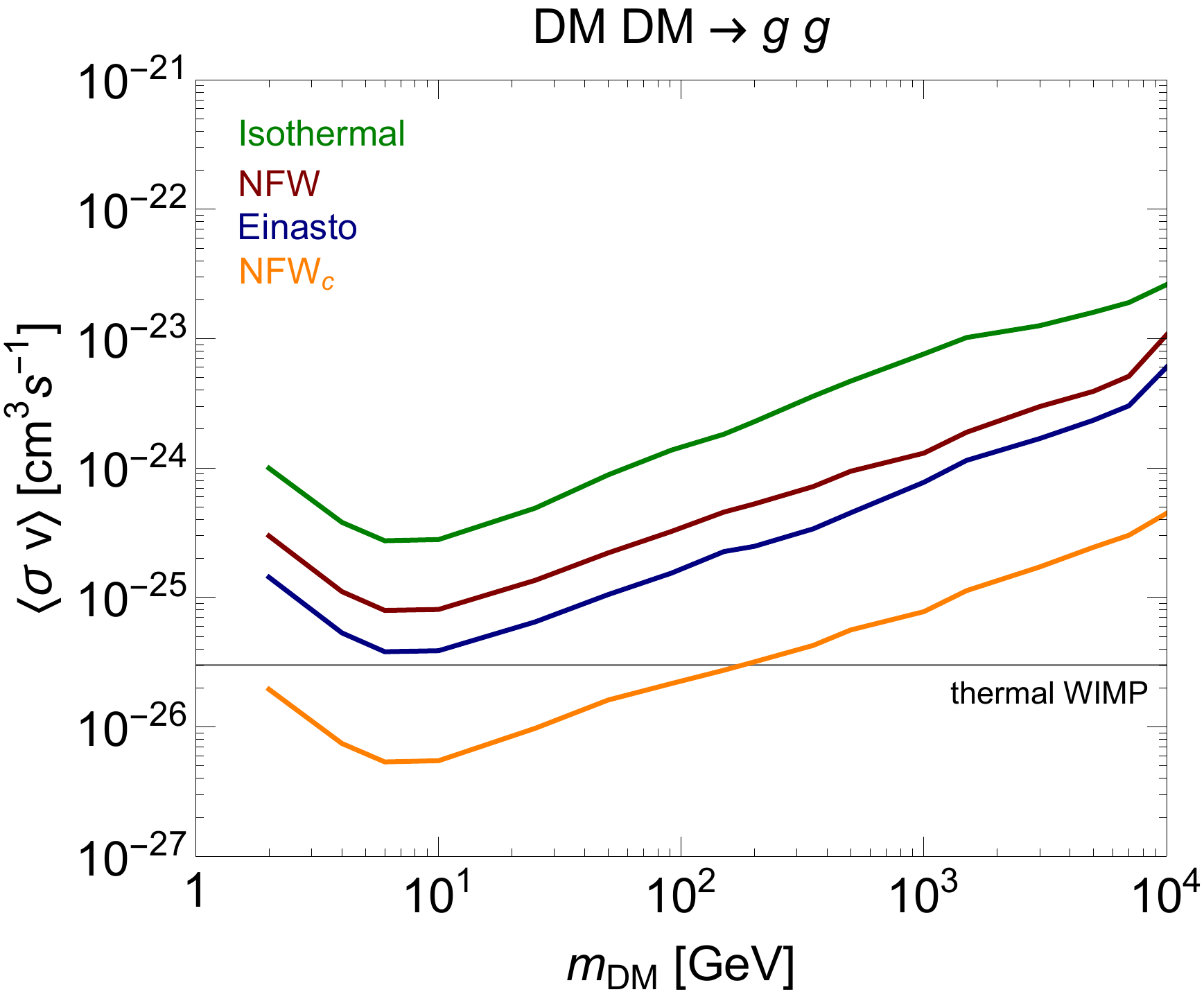}\\
\end{center}
\vskip -5mm
\caption{\footnotesize
95\% C.L.~upper limits on DM annihilation cross section vs. DM mass from {\it Fermi}-LAT's inclusive photon spectrum for the indicated final states.  
Each plot shows constraints for the Isothermal (green), NFW (red), Einasto (blue), and NFW$_c$ (orange) DM density profiles. 
Solid lines show constraints derived from including only the prompt radiation produced in the annihilation process (i.e.~final-state radiation or in the decay of hadrons).
When available, we show the limits from the 4-year P7REP analysis of 15 nearby dwarf spheroidal galaxies with a cyan dashed line~\cite{Ackermann:2013yva}.  
}
\label{fig:result-Ann-2}
\end{figure}
\end{spacing}

\vskip 8mm
The constraints disfavor the thermal WIMP cross section for low DM masses and for the cuspiest profiles (mostly the NFW$_c$ profile). 
For those cases in which the final states contain high-energy electrons, i.e.~Fig.~\ref{fig:result-Ann}, there is a contribution from prompt radiation from FSR as well as ICS. The latter, while more uncertain, considerably strengthens the bounds, especially for high DM masses. In Fig.~\ref{fig:result-Ann} the shaded band denotes the constraint from ICS as the magnitude of the Galactic magnetic field at our Solar System's location, $B_0$ is varied from 1~$\mu$G to 10~$\mu$G and correspondingly the diffusion coefficient $D_0$ from $4.797 \times10^{28}$~cm$^2$/s to $6.311 \times10^{28}$~cm$^2$/s (see \S\ref{subsec:ICS}). The propagation was performed as described in \S\ref{subsec:ICS}, i.e.~over a cylindrical geometry with radius $R_h= 20$~kpc and half-height $z_h= 4$~kpc. With ICS included and for cuspy profiles, DM annihilation to leptonic final states, particularly for electrons, can be probed well into the annihilation-cross-section regime of a thermal relic that freezes out early in the Universe, $\langle \sigma v \rangle_{\rm relic} \approx 3\times10^{-26}~\text{cm}^3$/s. The inclusion of extra particle content in DM annihilations, namely the particle $\phi$, is motivated by the best fit to the PAMELA, {\it Fermi}, and AMS-02 cosmic-ray positron and electron data~\cite{Abdo:2009zk, Adriani:2008zr, Aguilar:2013qda,Aguilar:2014mma}, if those excesses are interpreted as coming from DM annihilation. Fig.~\ref{fig:result-Ann} shows the approximate regions (shaded gray) in the cross-section--versus--mass plane, in which annihilating DM could offer an explanation for these excesses. These regions are meant to be illustrative only and chosen so that they contain the parameter choices found in~\cite{Cholis:2013psa}, shown with black dots. (See also~\cite{Boudaud:2014dta}.)
The inclusion of ICS severely constrains the favored parameter regions for all profiles except isothermal, while including only 
the prompt signal challenges the favored regions only for the cuspy NFW$_c$ profile. 

The constraints from~\cite{Ackermann:2013yva}, which, using 4 years of P7REP data, analyzed 15 dwarf spheroidal satellite galaxies (dSph) of the Milky Way to set robust constraints on DM, are shown in Fig.~\ref{fig:result-Ann} with a cyan dashed line. Due to the dSph's proximity, high DM content, and lack of astrophysical foregrounds, they are excellent targets to search for annihilating DM. Moreover, the available data on the velocity distribution of the stars in the dSph allows one to predict rather accurately the expected $\gamma$-ray flux from DM annihilation.  
This prediction is not subject to the same uncertainties as the expected flux in the Milky Way halo, which suffers from large uncertainties in the DM density profile.
Our constraints are stronger than the dSph constraints over much of the DM mass range and for several of the DM profiles that we consider, especially at high energies.
For DM masses $\lesssim 10$~GeV, our constraints are stronger than the dSph constraints for the NFW$_c$ profile, and comparable in strength for the Einasto profile, although weaker for the NFW and isothermal profiles. New results using P8 data to perform a similar analysis of the dwarf galaxies are expected soon and are somewhat more stringent than the P7REP results.

Notice that some of the ICS-inclusive limits are actually weaker than the ones with prompt radiation only. This might seem puzzling, as for a given ROI and energy range, the signal that includes prompt and ICS is obviously larger than the one with prompt only and should lead to more stringent constraints.  However, our ROI and energy range used to derive the limits from the data are dictated by the optimization of the average MC limit, such that the optimized ROIs and energy ranges for prompt+ICS and prompt-only might differ from each other. If one considers this along with the fact that the simulated data sets are not perfect representations of the real sky, the limit that includes ICS can be weaker on occasion than the prompt-only limit. 

It is useful to compare our limits with those obtained from similar analyses in the literature where no attempt was made to model the astrophysical foregrounds.  These analyses usually differ in their choice of DM-profile parameters, their procedure for constructing the limits (Gaussian error on flux versus Poisson limit on counts), their choice of propagation models for the ICS signal, and the data energy range utilized.  
Nevertheless, we can try to single out the effect of our ROI and energy-range optimization method alone by rescaling these other results to compensate for the different choices mentioned above. 
In~\cite{Cirelli:2009dv}, the limit was also constructed by scanning over a few differently shaped and located ROIs. Consequently our results are only within a factor of 1-2 stronger than theirs, across all channels.
In~\cite{Papucci:2009gd}, the construction of the bound is quite different from ours, and our results are around~2 times more stringent than theirs.
In~\cite{Ackermann:2012rg}, an optimization procedure is performed on ROIs that look very different from ours, and a less extensive optimization is done on the energy window.  
For annihilations we improve on these limits by a factor of 1--20, depending on channel and profile, and by a factor~2--4 when including ICS.
In~\cite{Gomez-Vargas:2013bea}, the ROI is optimized using the signal-over-background ratio as a figure of merit.  
For harder spectra, our improvement is between a factor of 3--8, while for softer spectra, the improvement is a factor of~1--4.
\begin{spacing}{1}
\begin{figure}[t!]
\begin{center}
\includegraphics[width=\mywidth\textwidth]{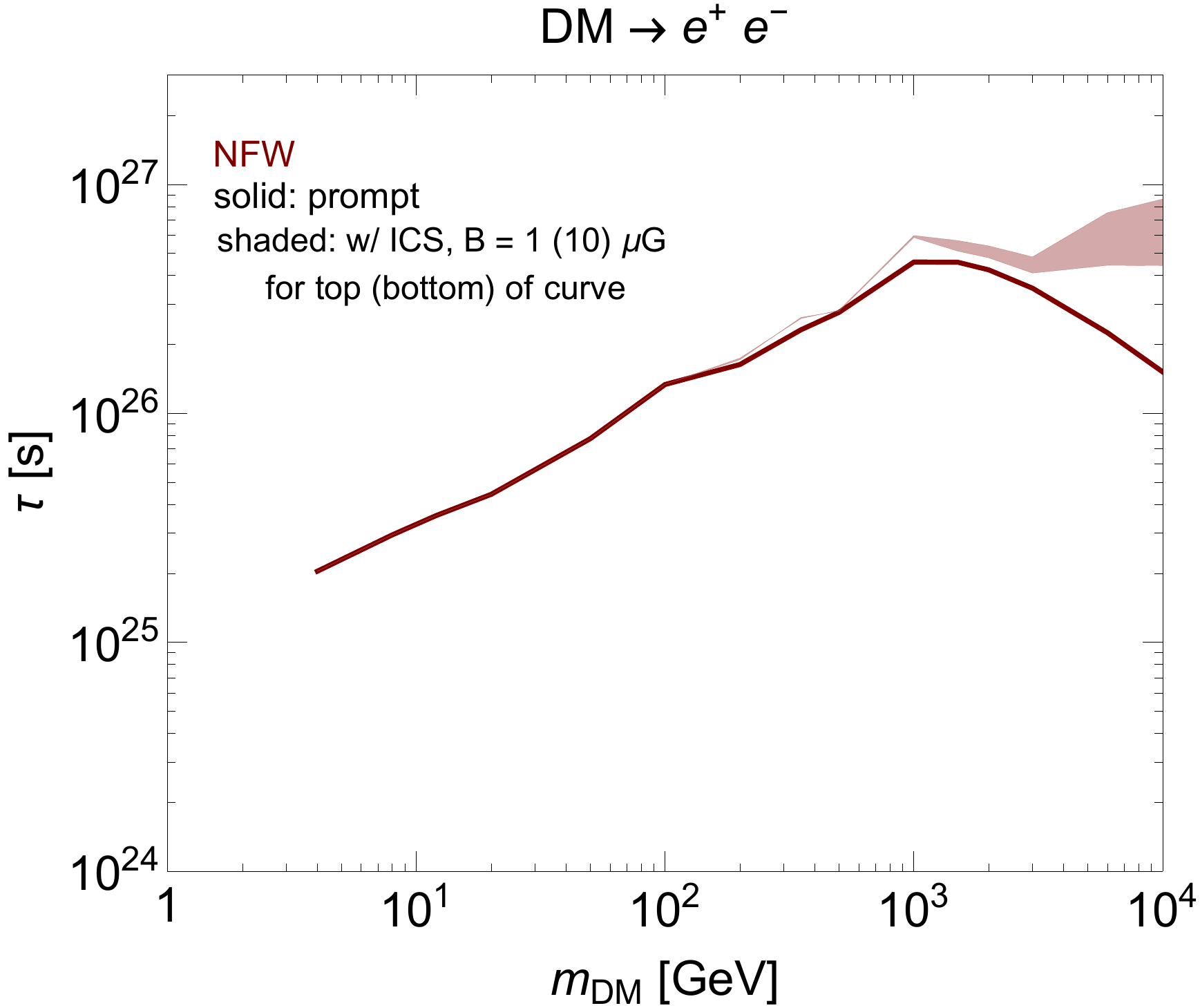}\;\;\;
\includegraphics[width=\mywidth\textwidth]{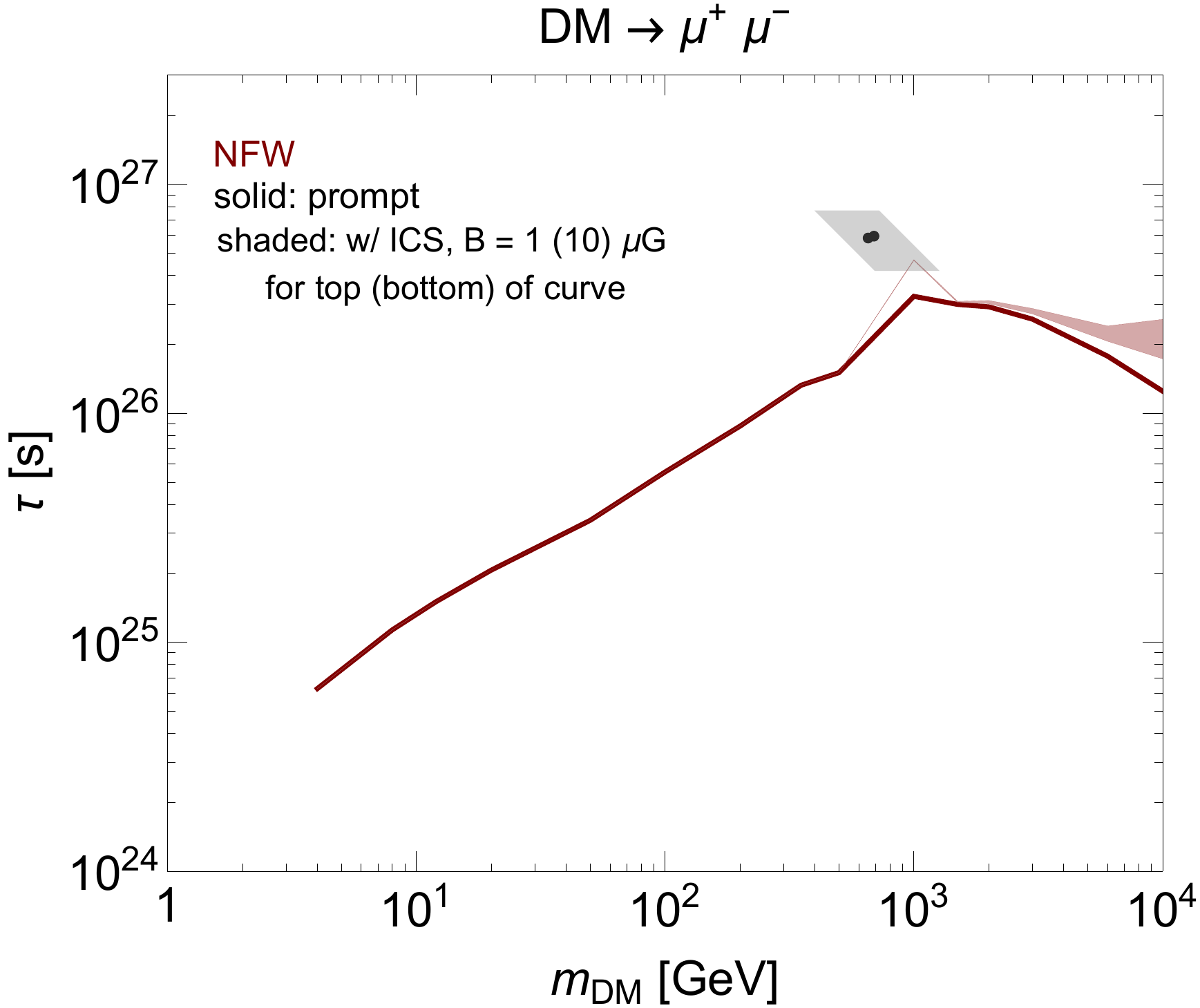}\\
\vskip 2mm
\includegraphics[width=\mywidth\textwidth]{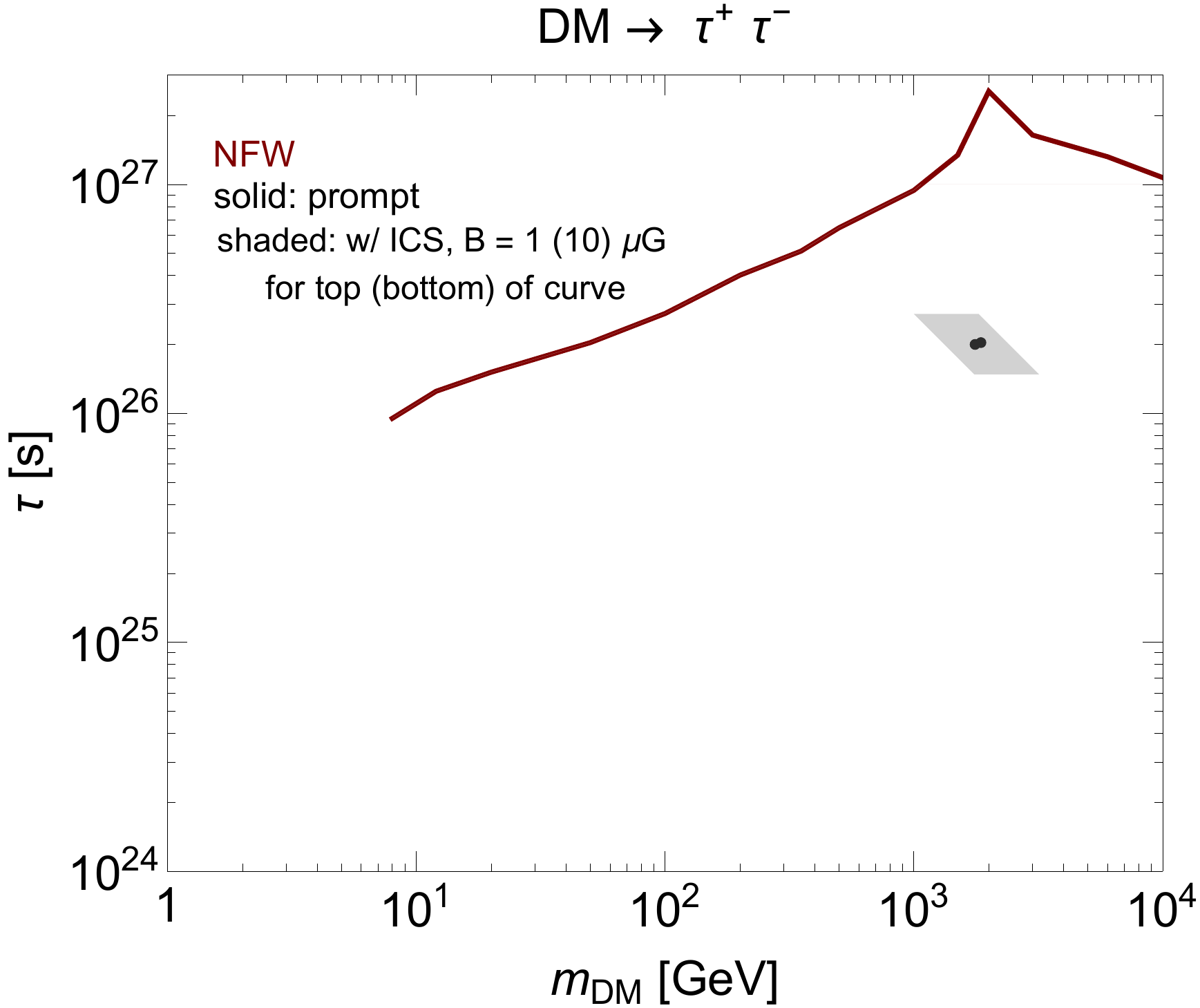}\;\;\;
\includegraphics[width=\mywidth\textwidth]{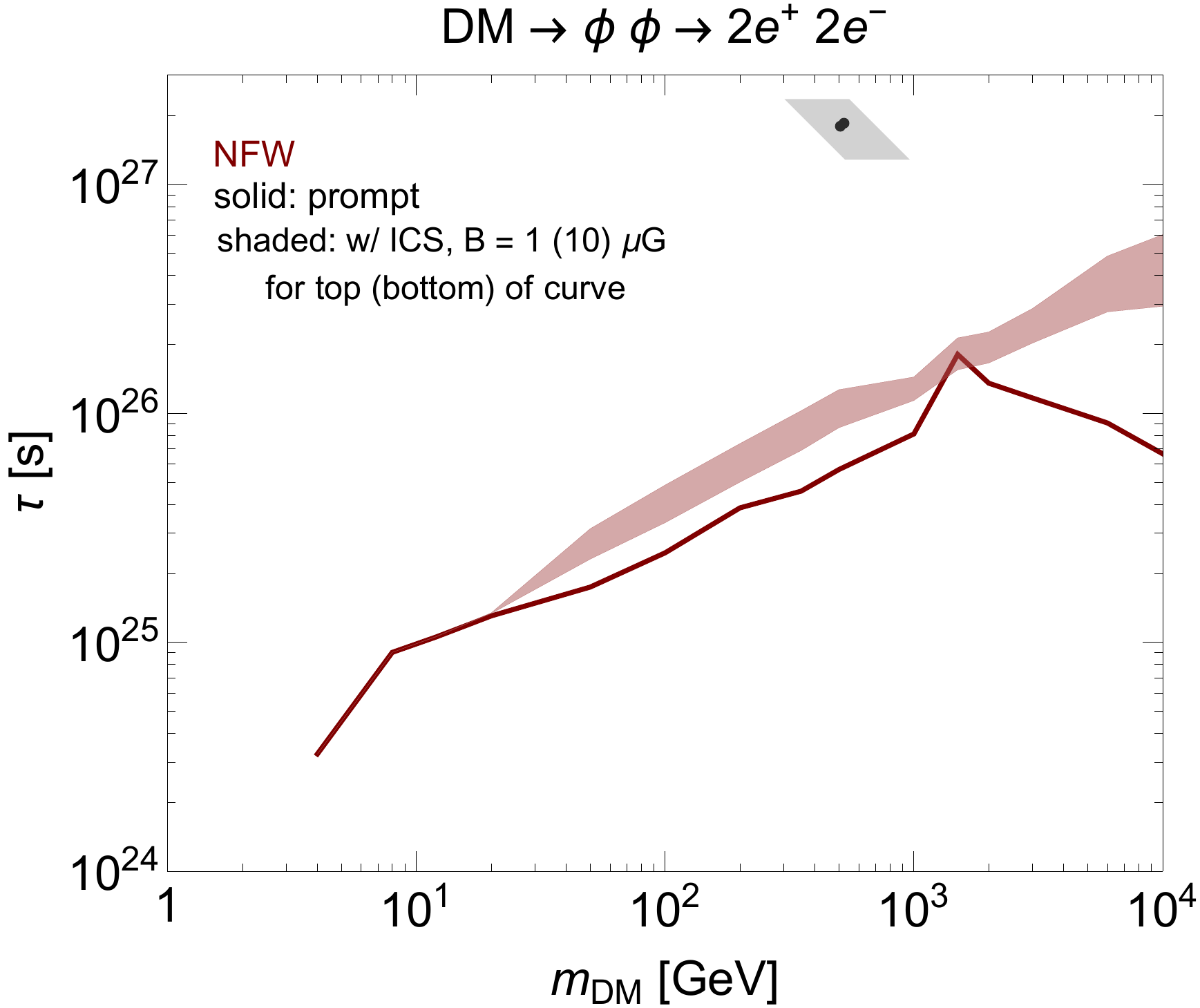}\\
\vskip 2mm
\includegraphics[width=\mywidth\textwidth]{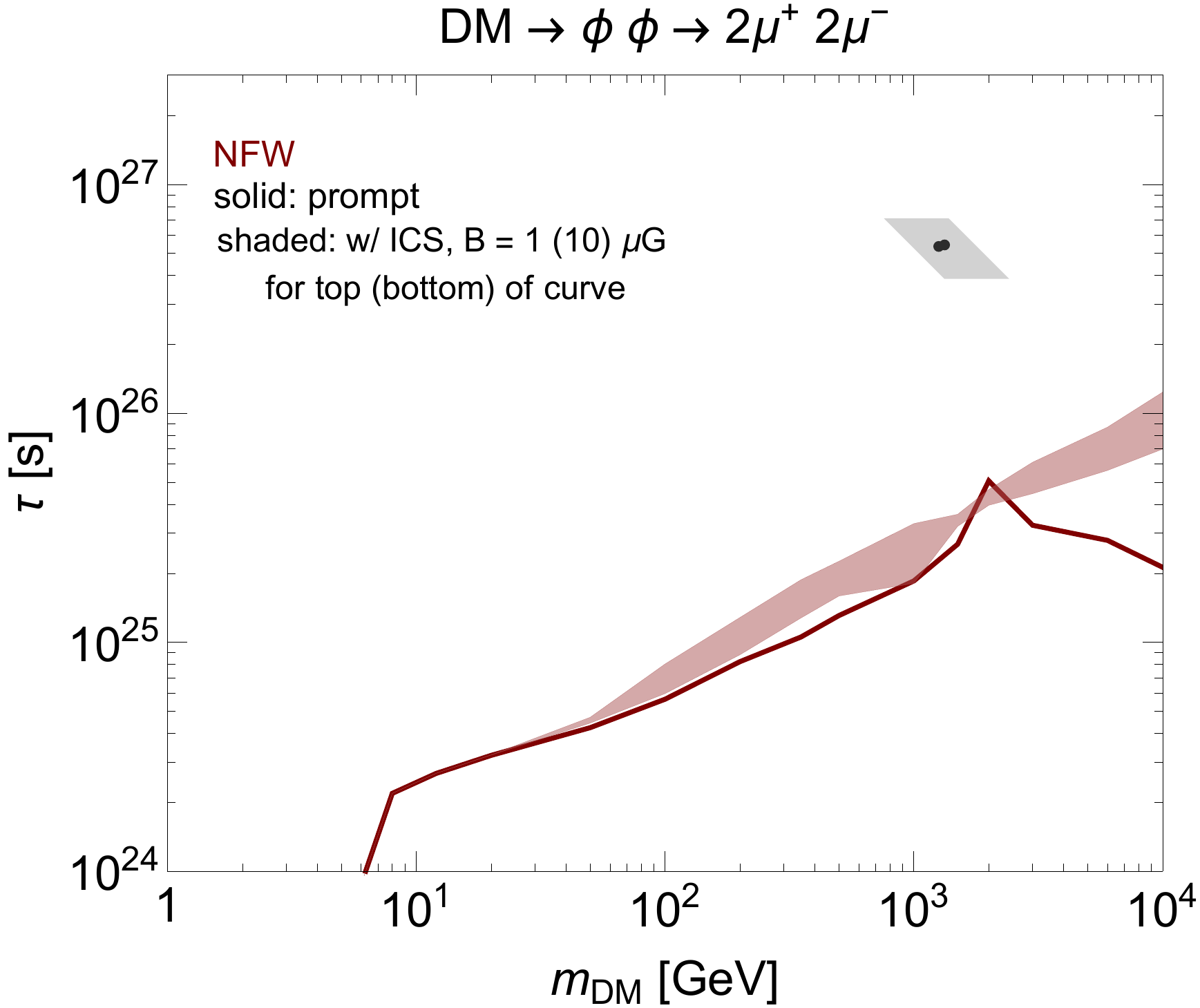}\;\;\;
\includegraphics[width=\mywidth\textwidth]{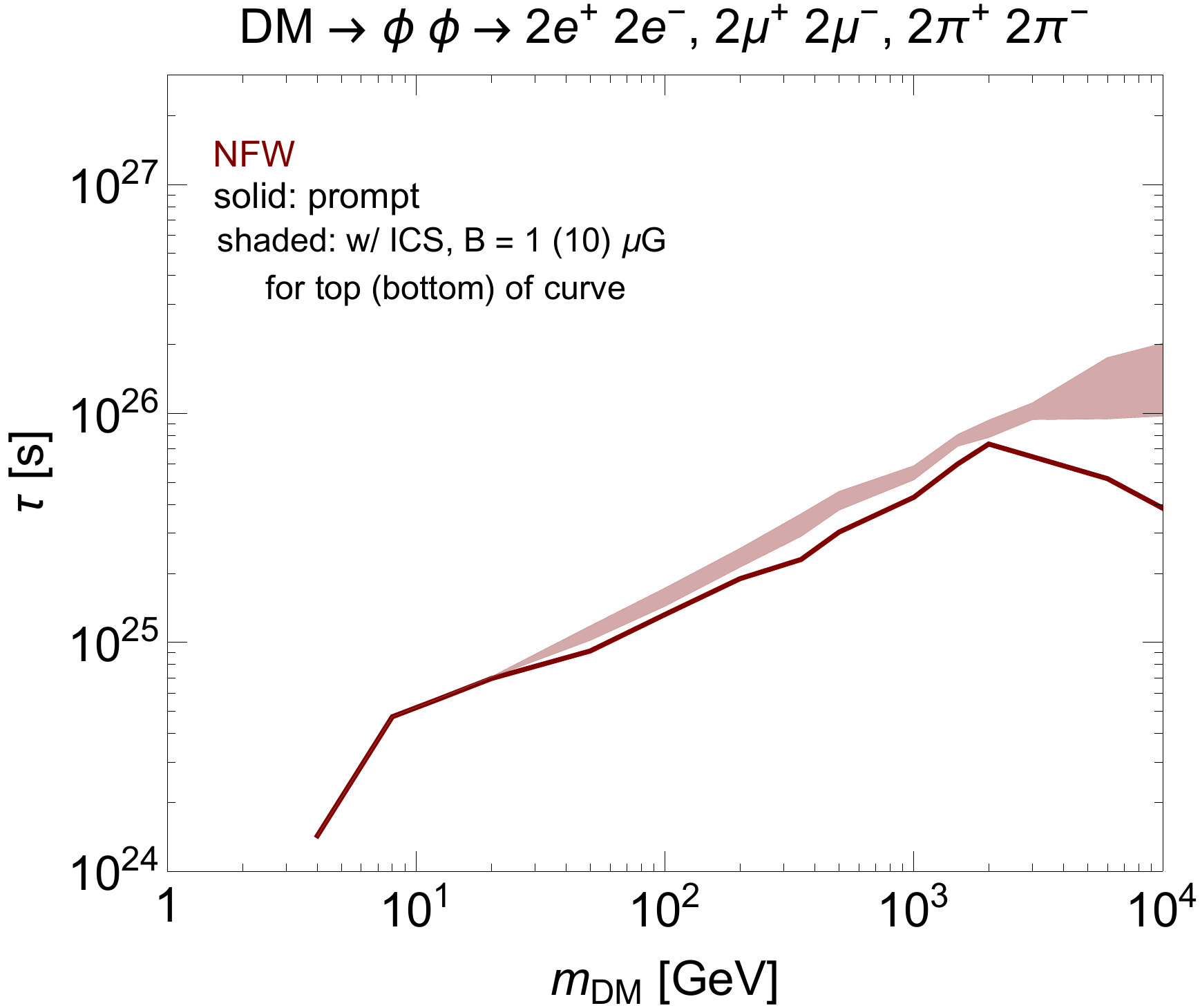}
\end{center}
\vskip -7mm
\caption{\footnotesize
95\% C.L.~lower limits on DM decay lifetime vs. DM mass from {\it Fermi}-LAT's inclusive photon spectrum for the indicated final states.  
Shown are constraints for the NFW profile (the other profiles are virtually identical). 
Solid lines show constraints derived from including only the prompt radiation produced in the annihilation process (i.e.~final-state radiation or in the decay of hadrons), while the bands include the ICS off background light, with the Galactic B-field varying within $1-10~\mu$G and $D_0$ within $D_{0,\rm min} - D_{0,\rm max}$ (bottom-top of band, when visible). For some models we show the approximate regions (gray) in which decaying DM could account for the PAMELA/Fermi/AMS-02 cosmic-ray excesses. The best-fit parameters from~\cite{Jin:2013nta} are shown as black dots.
}
\label{fig:result-Dec}
\end{figure}
\end{spacing}
\begin{spacing}{1}
\begin{figure}[t!]
\begin{center}
\includegraphics[width=\mywidth\textwidth]{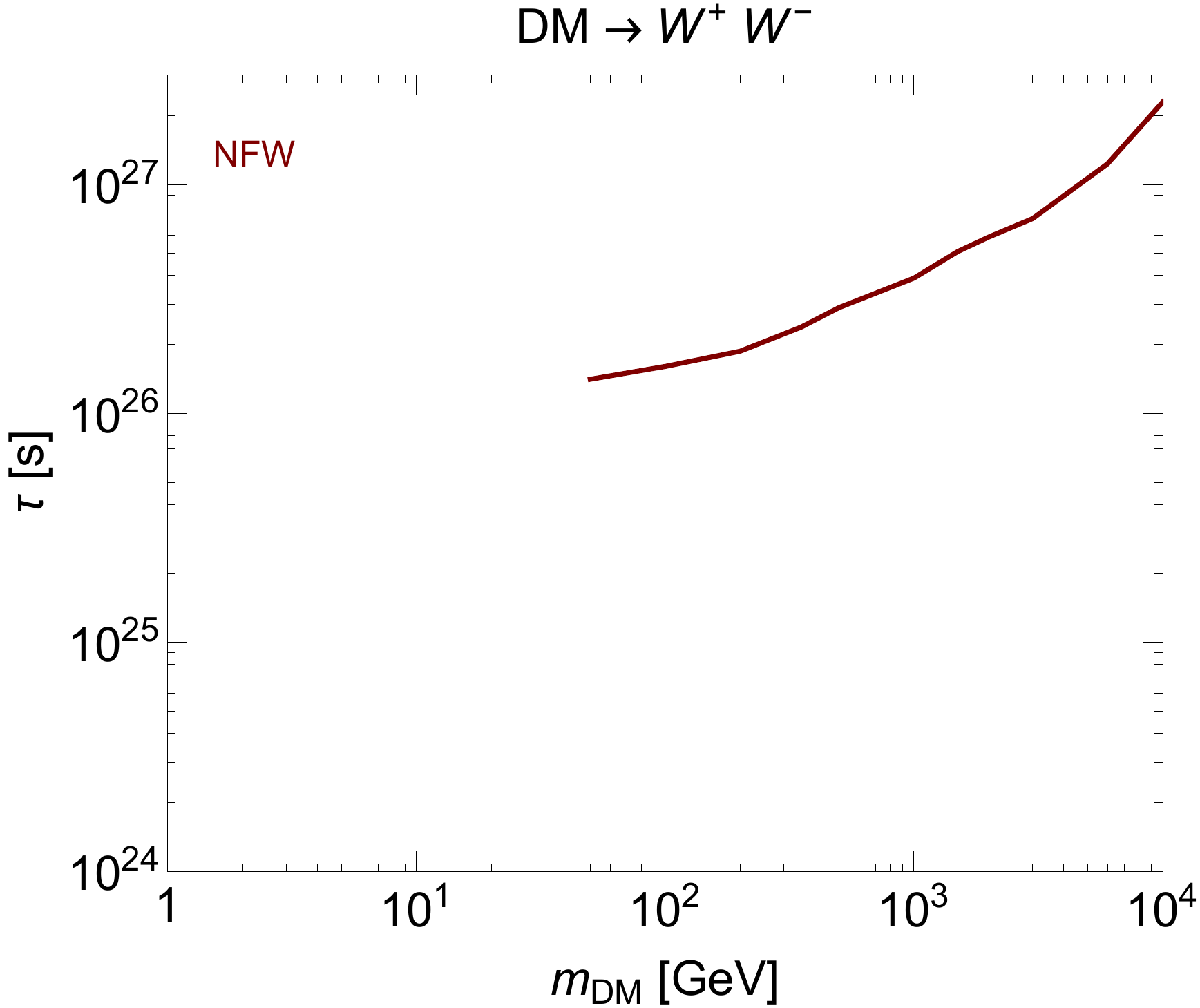}\;\;\;
\includegraphics[width=\mywidth\textwidth]{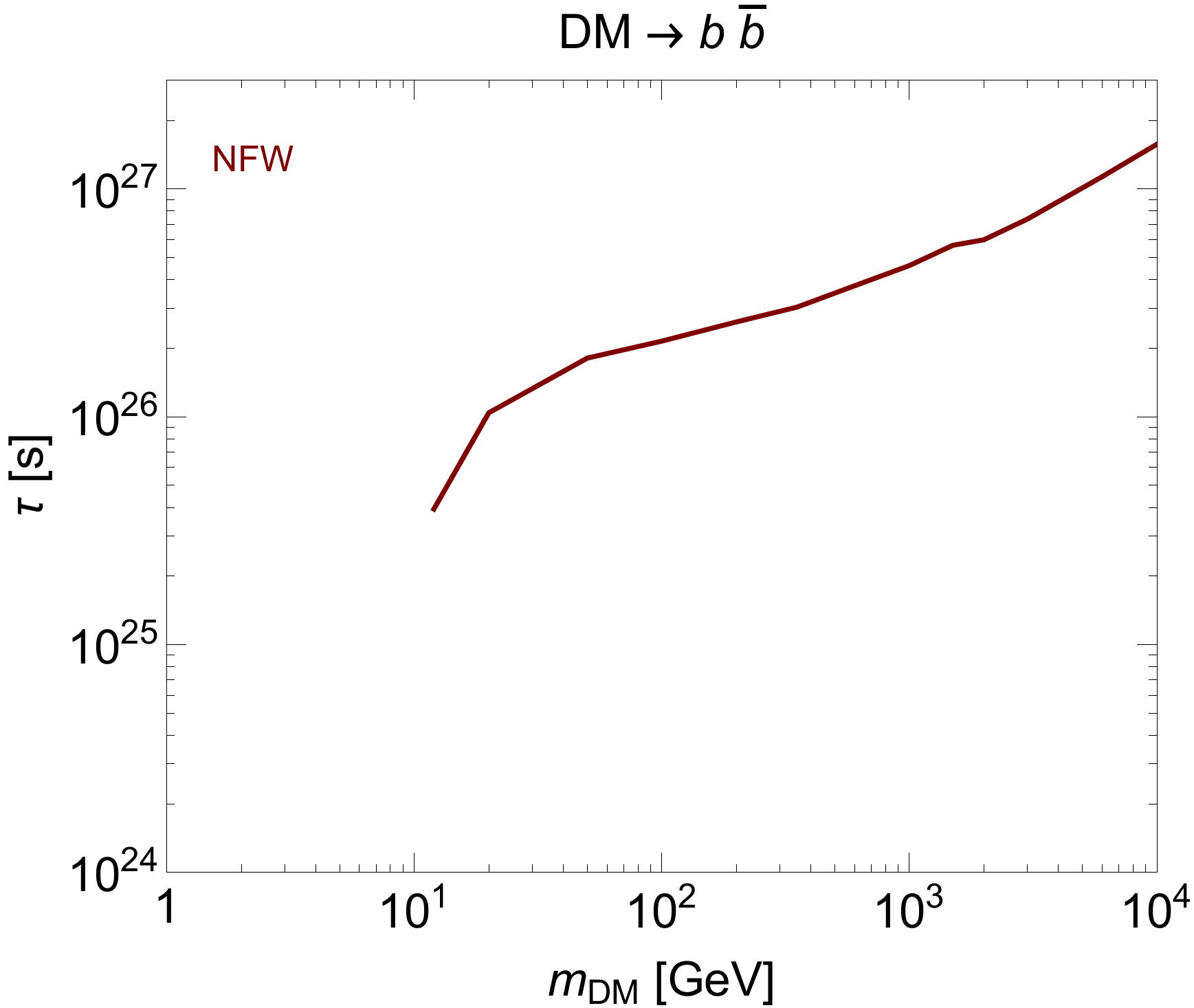}\\
\vskip 2mm
\includegraphics[width=\mywidth\textwidth]{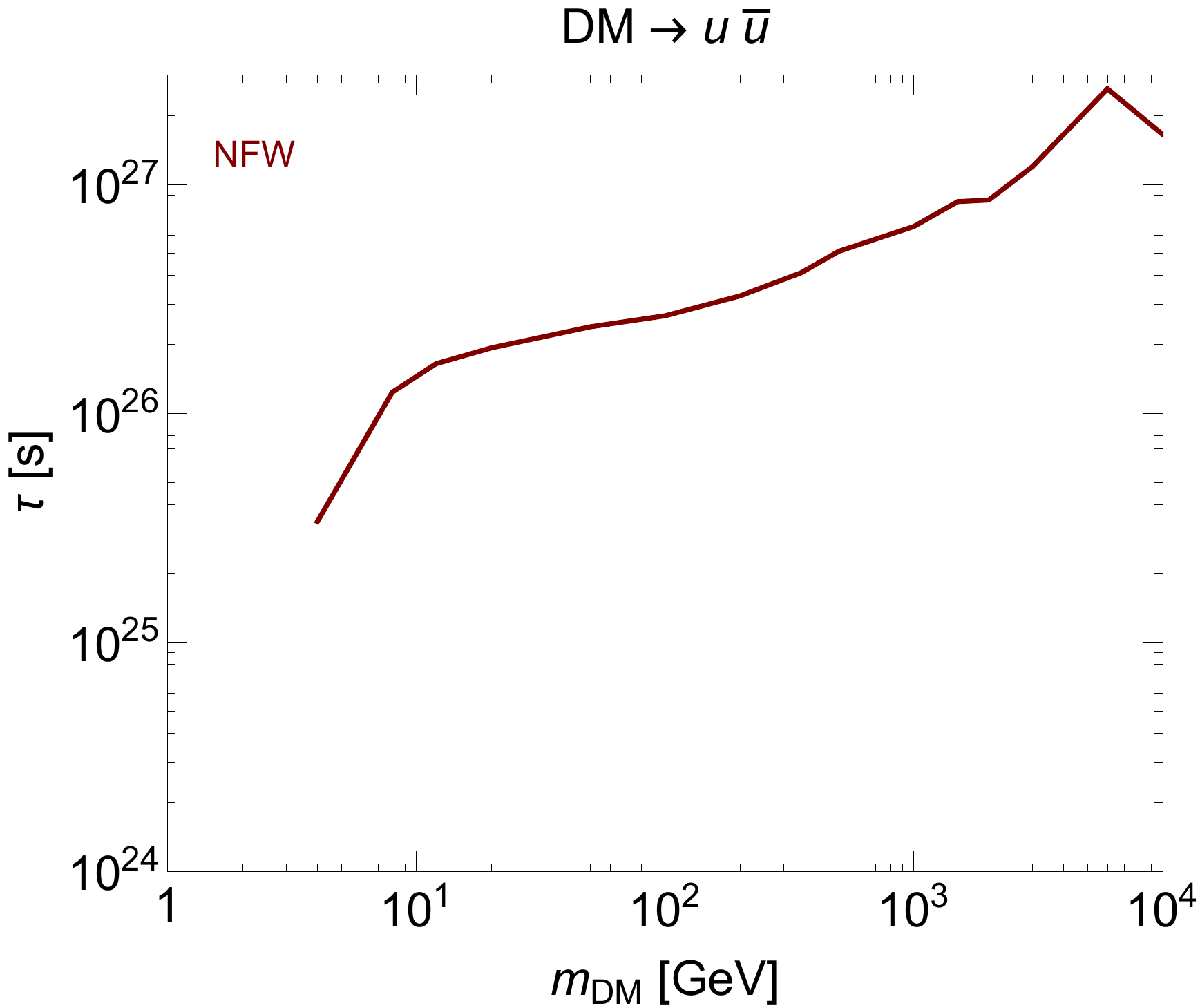}\;\;\;
\includegraphics[width=\mywidth\textwidth]{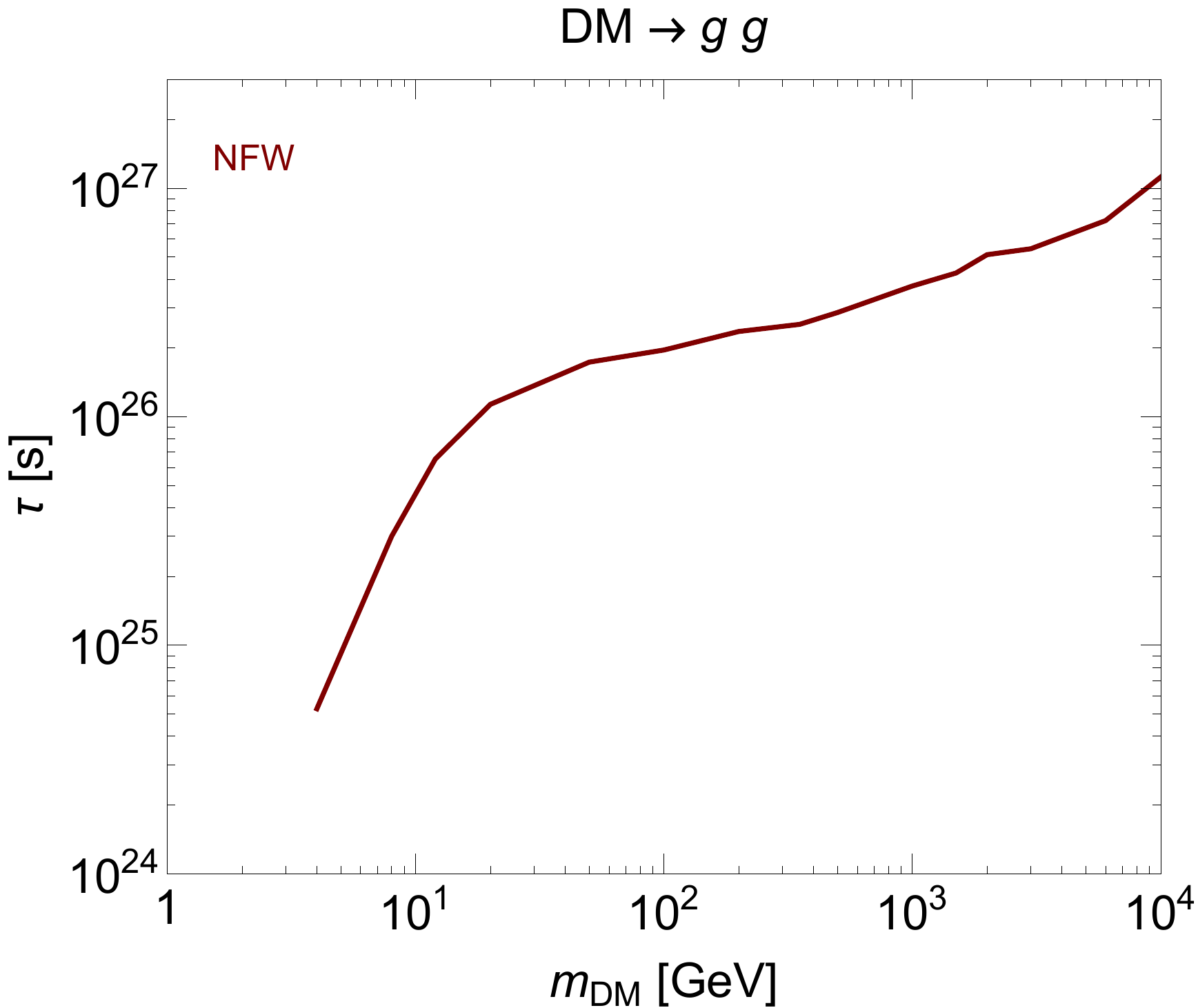}\\
\end{center}
\vskip -5mm
\caption{\footnotesize
95\% C.L.~lower limits on DM decay lifetime vs. DM mass from {\it Fermi}-LAT's inclusive photon spectrum for the indicated final states.  
Shown are constraints for the NFW profile (the other profiles are virtually identical). 
The constraints are derived from including only the prompt radiation produced in the annihilation process (i.e.~final-state radiation or in the decay of hadrons).
}
\label{fig:result-Dec-2}
\end{figure}
\end{spacing}

\subsection{Constraints on Dark Matter Decays}
\label{subsec:results-decays}

While a favorite target for the DM annihilation rate comes from the thermal freeze-out of a thermal relic, which gives the correct present-day abundance, for decaying DM no such ``favored'' lifetime exists --- the DM lifetime only has to be larger than the age of the Universe.  One possible target comes from explaining the rising fraction in the cosmic-ray positron spectrum with DM decays to final states that produce high-energy electrons and positrons, with the preferred DM lifetime 
being in the range $10^{26}-10^{27}$~s, depending on the precise final states and astrophysical assumptions~\cite{Arvanitaki:2008hq,Meade:2009iu, Nardi:2008ix, Yin:2008bs, Ibarra:2008jk, Cirelli:2009dv, Ishiwata:2008cv, Chen:2008qs,Baek:2014goa}. 
Such lifetimes do not only have a phenomenological motivation, but also arise naturally for TeV-scale DM particles that decay via a dimension-six operator generated near the scale of Grand Unified Theories (GUT's), $M \sim 10^{16}$~GeV, namely 
\bea
\tau \sim 8 \pi \, \frac{M^4}{m_{\rm DM}^5} \sim 2 \times 10^{26}~{\rm s}\, \left(\frac{1~{\rm TeV}}{m_{\rm DM}}\right)^5 \,\left(\frac{M}{10^{16}~{\rm GeV}}\right)^4\,. 
\eea
For example, in~\cite{Arvanitaki:2008hq} DM decaying via dimension-six operators in supersymmetric GUT's were posited to explain the cosmic-ray data from PAMELA.

The results for DM decays to leptonic and $\phi\phi$ final states are included in Fig.~\ref{fig:result-Dec}, whereas those decays to $b \bar{b}, u \bar{u}, g g, W^+ W^-$ are shown in Fig.~\ref{fig:result-Dec-2}. 
We only show the constraints for the NFW profile, as the other profiles lead to virtually identical constraints.  
As in the case for DM annihilation, we include ICS for decaying DM for the leptonic final states only. The additional ICS component, while very sensitive to the value of the Galactic magnetic field, can enhance the constraints significantly, as in the case for annihilating DM. Note that the bounds from prompt radiation start to deteriorate near DM masses of 1.5~TeV due to the maximum-energy selection of 750~GeV used in this study.
Our constraints compare favorably with existing constraints in the literature; for example, they are a factor of 2--3 stronger compared to~\cite{Cirelli:2009dv,Papucci:2009gd,Ackermann:2012rg}.  

While the DM decay lifetime can span an enormous range consistent with all astrophysical data, there are many scenarios that 
are being probed by the constraints presented in this analysis.  In particular, Fig.~\ref{fig:result-Dec} shows with a gray shaded 
parallelogram the approximate preferred regions in which decaying DM can explain the cosmic-ray positron and electron data.  
Black dots indicate the best-fit regions found in~\cite{Jin:2013nta}, although note that these results do not include the latest data release from AMS-02~\cite{Aguilar:2014mma} (a more careful analysis of the preferred regions is beyond the scope of this paper);
nevertheless, we expect that the preferred regions would not shift significantly, and our regions are meant to be 
taken as a useful but rough qualitative guide only.  
We see that decays to $\tau^+\tau^-$ are thoroughly disfavored, but our constraints for other channels are not strong enough 
to probe the relevant parameter regions.


\section{CONCLUSIONS}
\label{sec:conclusions}

This paper presented a conservative method for setting constraints on $\gamma$ rays originating from DM annihilation and decay, which does not rely on modeling of astrophysical foregrounds when setting a limit. Optimal regions in the sky and energy were obtained by using simulations of the $\gamma$-ray sky, and a constraint was found by only requiring that the DM signal does not over-predict the observed photon counts.  

For models of both annihilating and decaying DM, this method allows us to constrain theoretically-motivated parameter regions.  
For example, for cuspy enough profiles (e.g.,~contracted NFW), our method is able to disfavor the thermal-relic cross 
section for some leptonic and hadronic final states.  Also, for steep-enough profiles, our constraints disfavor various 
annihilating DM scenarios designed to explain the PAMELA/{\it Fermi}/AMS-02 cosmic-ray positron and electron data.   
For decaying DM, a wide range of lifetimes are excluded for various SM final states, including the preferred parameter regions 
for DM decaying to $\tau^+\tau^-$ to explain the PAMELA/{\it Fermi}/AMS-02 data.  
The conservative constraints obtained in this study are often competitive with, and in many cases stronger than, other available constraints in the literature.

\subsection*{Acknowledgements}

We thank Luca Baldini, Philippe Bruel, Seth Digel, Miguel S\'{a}nchez-Conde, and David Thompson for reading the manuscript and providing valuable comments, Neelima Sehgal for providing the photon spectra for the various dark matter annihilation and decay final states, Warit Mitthumsiri for his work on the MC simulations, and Eric Charles, Ilias Cholis, Tongyan Lin, Michele Papucci, and Gabrijela Zaharijas for helpful correspondence or discussion.
We also thank all the members of the {\it Fermi}-LAT collaboration who provided valuable comments and assistance, including Alessandro Cuoco, Alex Drlica-Wagner, Gu{\dh}laugur J\'{o}hannesson, Philipp Mertsch, Igor Moskalenko, and Matthew Wood.

RE is supported by the Department of Energy (DoE) Early Career research program DESC0008061 and by a Sloan Foundation Research Fellowship.
Research at Perimeter Institute is supported by the Government of Canada through Industry Canada and by the Province of Ontario through the Ministry of Research and Innovation. EI is partly supported by the Ministry of Research and Innovation - ERA (Early Research Awards) program.
AM is supported by the C.N. Yang Institute for Theoretical Physics (Stony Brook University) and NSF grant PHY1316617.
The work of GAGV was supported by Conicyt Anillo grant ACT1102. GAGV thanks for the support of the Spanish MINECO's Consolider-Ingenio 2010 Programme under grant MultiDark CSD2009-00064 also the partial support by MINECO under grant FPA2012-34694. 

The \textit{Fermi}-LAT Collaboration acknowledges generous ongoing support from a number of agencies and institutes that have supported both the development and the operation of the LAT as well as scientific data analysis. These include the National Aeronautics and Space Administration and the Department of Energy in the United States, the Commissariat \`a l'Energie Atomique and the Centre National de la Recherche Scientifique / Institut National de Physique Nucl\'eaire et de Physique des Particules in France, the Agenzia Spaziale Italiana and the Istituto Nazionale di Fisica Nucleare in Italy, the Ministry of Education, Culture, Sports, Science and Technology (MEXT), High Energy Accelerator Research Organization (KEK) and Japan Aerospace Exploration Agency (JAXA) in Japan, and the K.~A.~Wallenberg Foundation, the Swedish Research Council and the Swedish National Space Board in Sweden. Additional support for science analysis during the operations phase is gratefully acknowledged from the Istituto Nazionale di Astrofisica in Italy and the Centre National d'\'Etudes Spatiales in France.
\appendix

\renewcommand\thesection{\Alph{section}}
\renewcommand\thesubsection{\Alph{section}.\arabic{subsection}}
\renewcommand\thesubsubsection{\Alph{section}.\arabic{subsection}.\arabic{subsubsection}}
\makeatletter
\def\p@subsection{}
\def\p@subsubsection{}
\makeatother

\section{Constraints on DM Models invoked to explain $\gamma$ Rays from Inner Galaxy}
\label{sec:gamma-ray-excess}

In this appendix we address claims made by several groups in recent years regarding a $\gamma$-ray excess from $\sim 300$ MeV to $\sim 5$ GeV, peaking in the 1--3 GeV window, in the Inner Galaxy~\cite{Goodenough:2009gk, Hooper:2010mq, Boyarsky:2010dr, Hooper:2011ti, Abazajian:2012pn, Gordon:2013vta, Abazajian:2014fta, Hooper:2013rwa, Daylan:2014rsa, Huang:2013pda, Macias:2013vya,Calore:2014xka}. While modeling uncertainties are large and the excess may very well have a non-DM origin, we use our method to set constraints on DM scenarios that have been invoked to explain the excess. 
Since we perform no foreground subtraction, {\it a priori} we do not expect the limits derived with our method to disfavor the best-fit DM scenarios found in the literature; nevertheless, it is worthwhile to perform a careful check.  

The best fit for WIMP DM found in~\cite{Hooper:2013rwa, Daylan:2014rsa} is for $\sim 30-40$~GeV DM annihilating predominantly to $b\bar b$.  
Furthermore, the spatial distribution of the putative signal is best fit by a generalized NFW profile,
\bea
\label{eq:gen-NFW}
\rho_{\mbox{\footnotesize NFW, }\gamma}(r)&=& {\rho_0 \over \pt{r / r_s}^{\gamma} \pt{1+{r / r_s}}^{3-\gamma}}\,,
\eea
with a $\chi^2$ best fit obtained for $\gamma \approx 1.26$, although any $\gamma$ in the range $\sim 1.1-1.4$ allows for a reasonable fit. 
Analyses by other groups give results that are broadly consistent with the findings in~\cite{Hooper:2013rwa, Daylan:2014rsa}.  
In~\cite{Macias:2013vya}, it was found that DM annihilating dominantly to $b\bar{b}$ but with some admixture of $\tau^+\tau^-$ also 
provides a good fit.  Other annihilation channels may also be possible~\cite{Hooper:2012cw}.  

In Fig.~\ref{fig:GC-excess} we show the results of our optimization procedure applied to generalized NFW, Eq.~(\ref{eq:gen-NFW}), with parameters chosen from best fits found in~\cite{Daylan:2014rsa, Calore:2014xka, Abazajian:2014fta}
(which differ in part from the assumptions made in~\S\ref{subsec:results-annihilation}). 
The authors of~\cite{Daylan:2014rsa} (\cite{Calore:2014xka}) exclude from their analysis a band around the GP defined by 
$\left\vert b \right\vert < 1^\circ \,(2^\circ)$, thus not specifying a specific DM distribution within this latitude.  We therefore 
use our usual ROIs shown in Fig.~\ref{fig:ROI}, but mask a square centered on the GC of side $2^\circ \, (4^\circ)$.
We show DM annihilating to $b\bar{b}$ ({\bf left plot}) and $\tau^+\tau^-$ ({\bf right plot}). 
Unsurprisingly, the bounds that we obtain on the annihilation cross section are still a factor of $\sim 3$ or more from probing the best-fit regions shown with open or closed contours in Fig.~\ref{fig:GC-excess}. As a reference for the reader, adopting all the assumptions in~\cite{Daylan:2014rsa}, for annihilation into $b\bar b$, and choosing $m_{DM}=25$~GeV, the optimal ROI found with our method is determined by the following parameters: $R=2^\circ$, $\Delta b= 1.98^\circ$, $\Delta \ell = 0.54^\circ$, while the optimal energy range is $1.9$~GeV$\lesssim E\lesssim4.0$~GeV.
\begin{figure}[t!]
\includegraphics[width=0.48\textwidth]{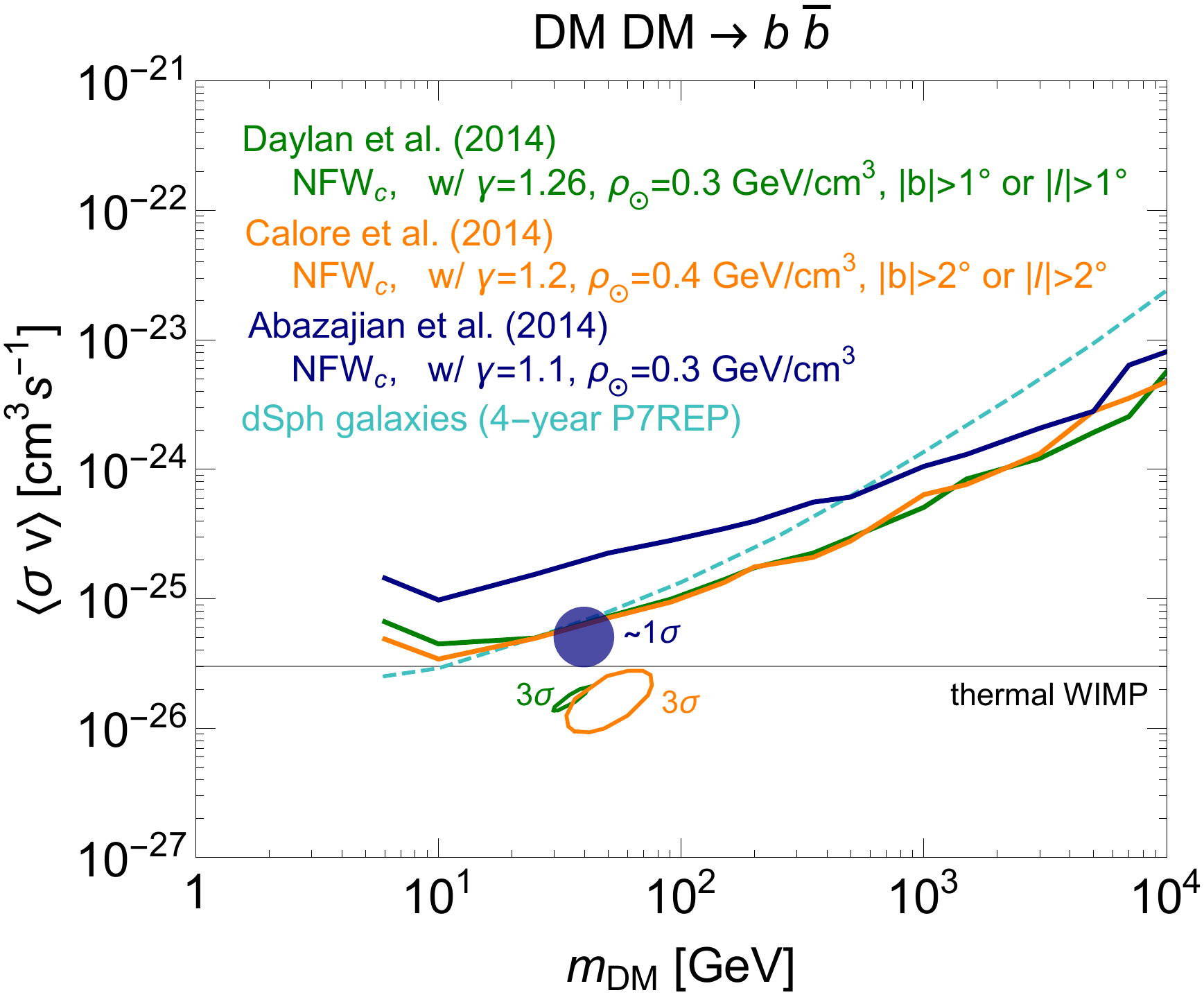}
\includegraphics[width=0.48\textwidth]{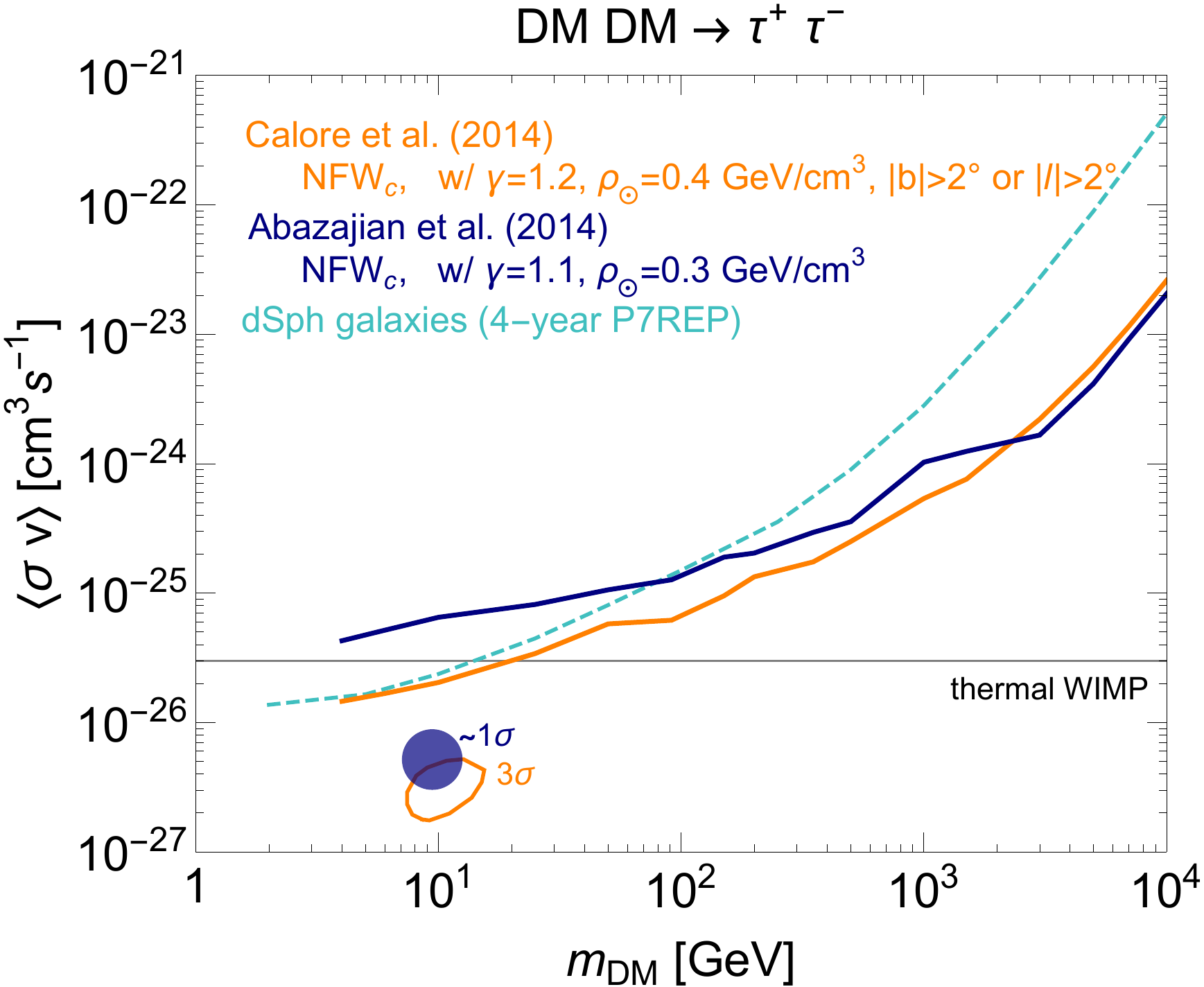}
\caption{95\% C.L.~annihilation cross section upper limits on DM annihilating to $b\bar{b}$ ({\bf left}) and $\tau^+ \tau^-$ ({\bf right}) for an NFW$_c$ profile with various inner slopes and local DM densities (note that the assumptions made in deriving these limits differ in part from those made in~\S\ref{subsec:results-annihilation}).
Also shown are the preferred regions from~\cite{Daylan:2014rsa, Calore:2014xka, Abazajian:2014fta} for DM to fit the claimed Galactic-Center $\gamma$-ray ``excess''. The constraints have been computed with the same model assumptions as the best-fit regions (including masking a square centered on the GC of side $2^\circ$ or $4^\circ$ for analyses that excluded a band around the GP with the same thickness -- see text for details).   We also show with a cyan dashed line the limit obtained from the 4-year P7REP analysis of 15 nearby dwarf spheroidal galaxies~\cite{Ackermann:2013yva}. 
}
\label{fig:GC-excess}
\end{figure}

\begin{spacing}{1}
\begin{figure}[t!]
\begin{center}
\includegraphics[width=\mywidth\textwidth]{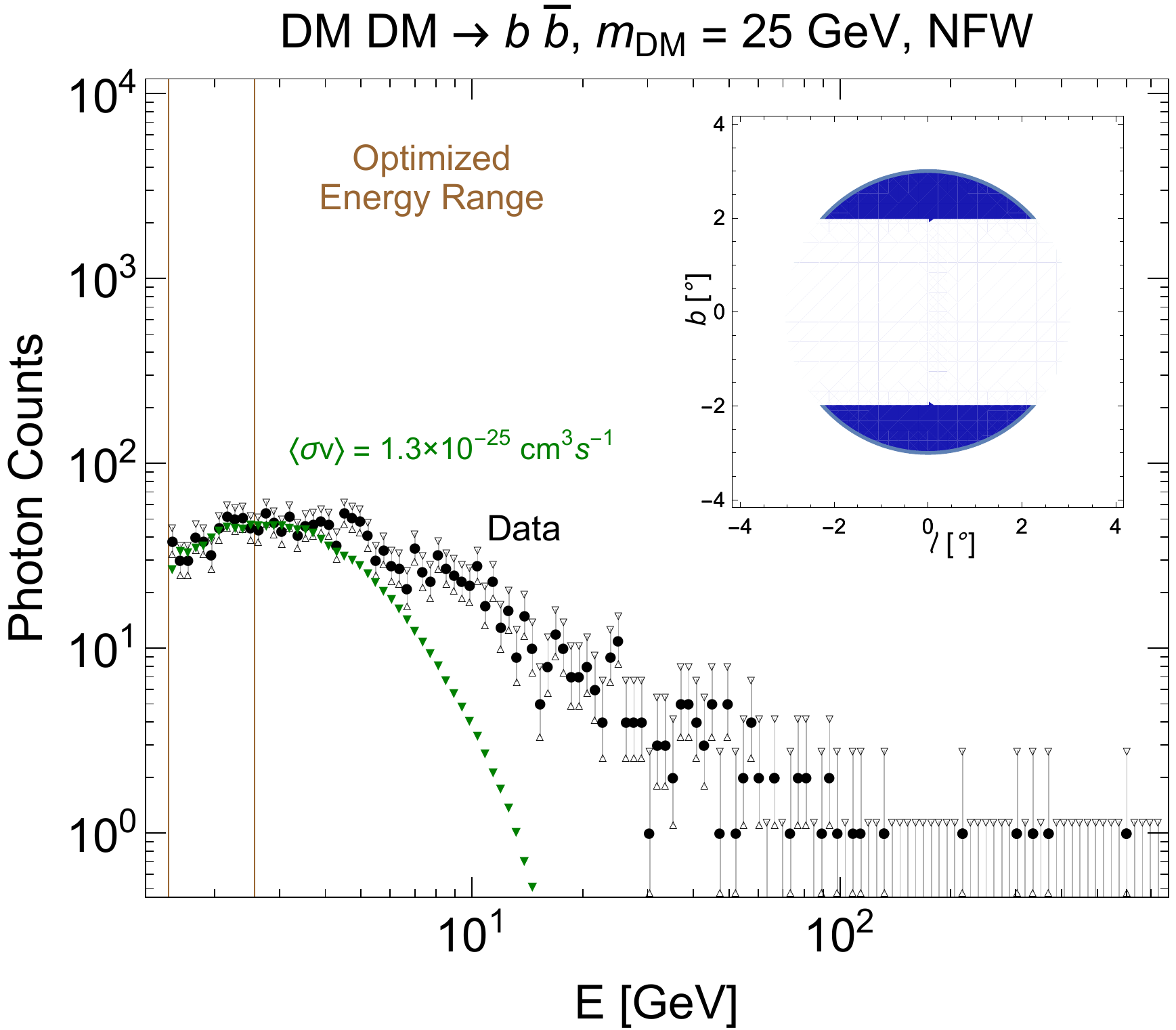}\;\;
\includegraphics[width=\mywidth\textwidth]{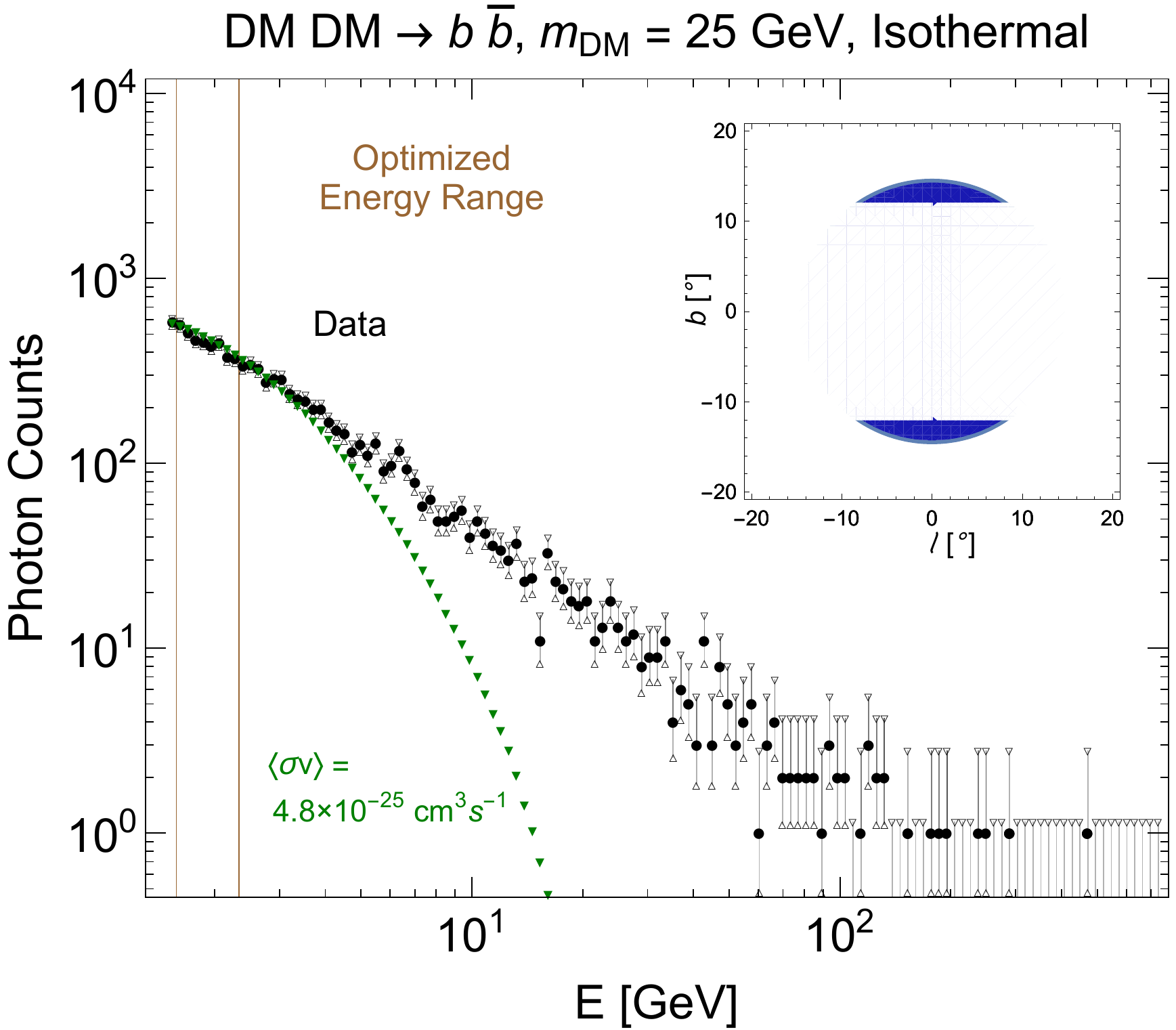}
\vskip 2mm
\includegraphics[width=\mywidth\textwidth]{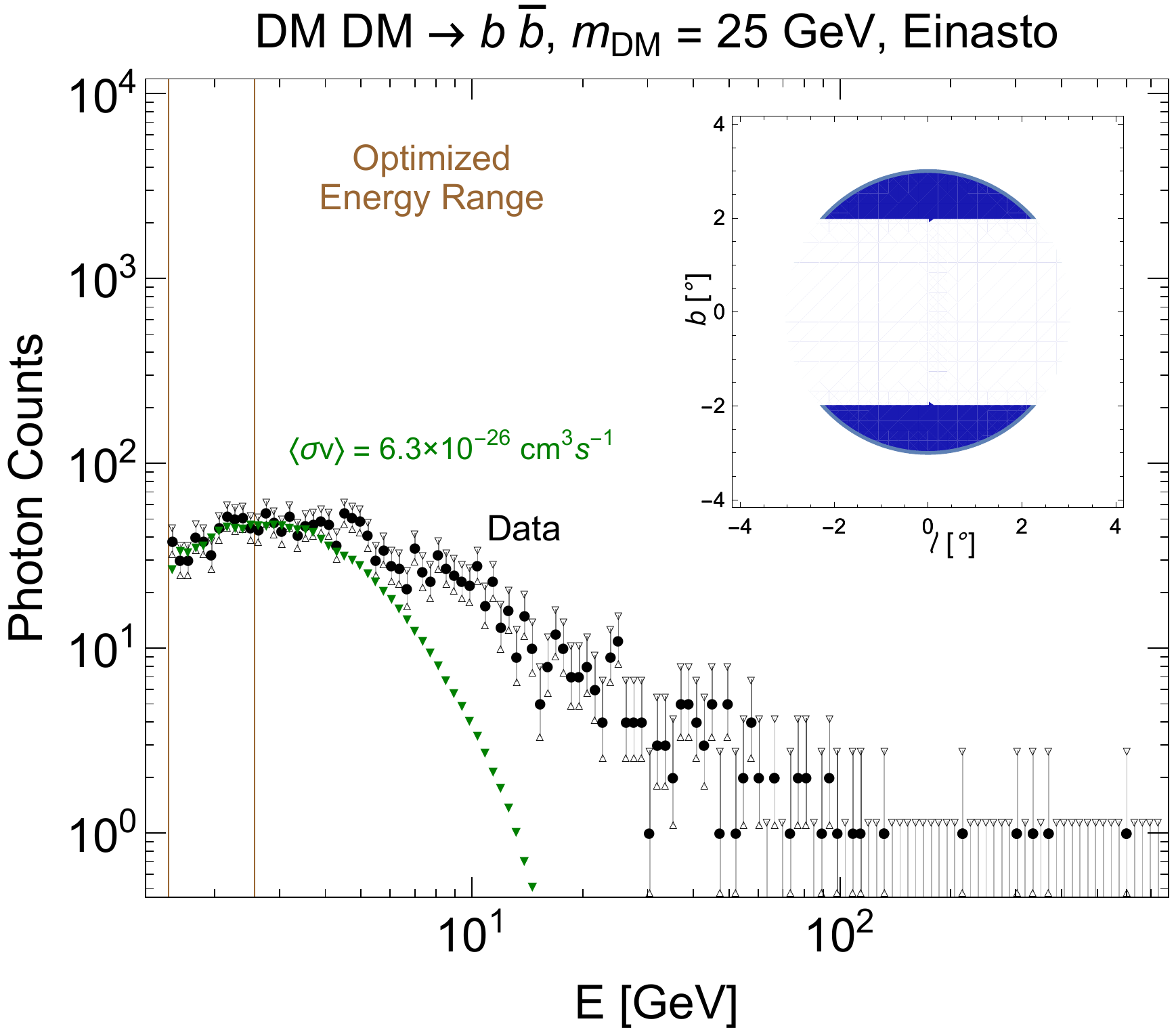}\;\;
\includegraphics[width=\mywidth\textwidth]{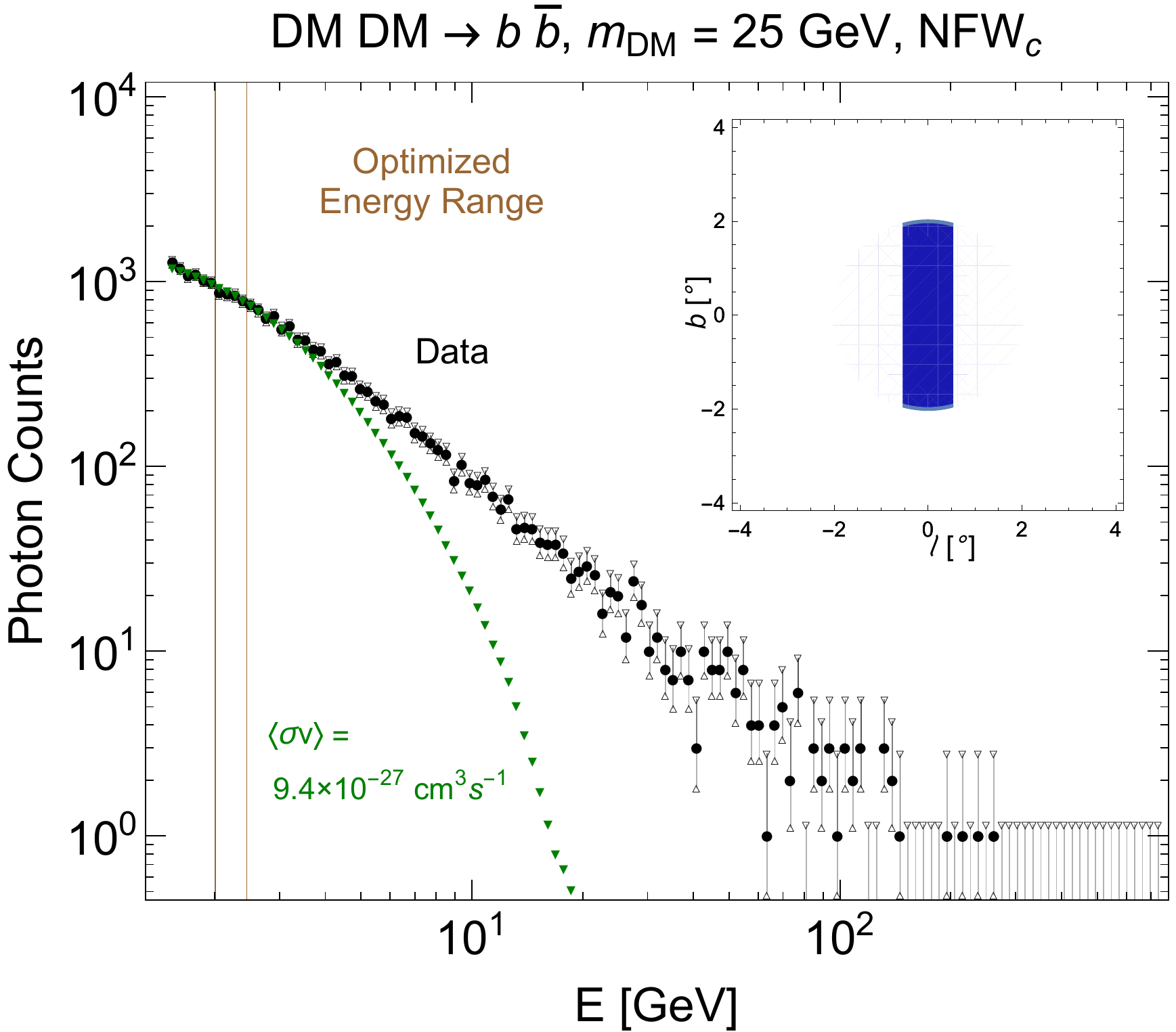}
\end{center}
\vskip -5mm
\caption{\footnotesize
Count spectrum for 25 GeV DM annihilating to $b\bar b$ for various DM density profiles. 
The vertical (brown) lines show the optimal energy range for each DM model assumption. The inset shows the optimal ROI. 
Note that PSF-convolution effects were included for the DM signal. 
The quoted $\langle\sigma v\rangle$ is the annihilation cross section that saturates the 95\% C.L.~from the data.
}
\label{fig:result-ROIs}
\end{figure}
\end{spacing}
\begin{spacing}{1}
\begin{figure}[t!]
\begin{center}
\includegraphics[width=\mywidth\textwidth]{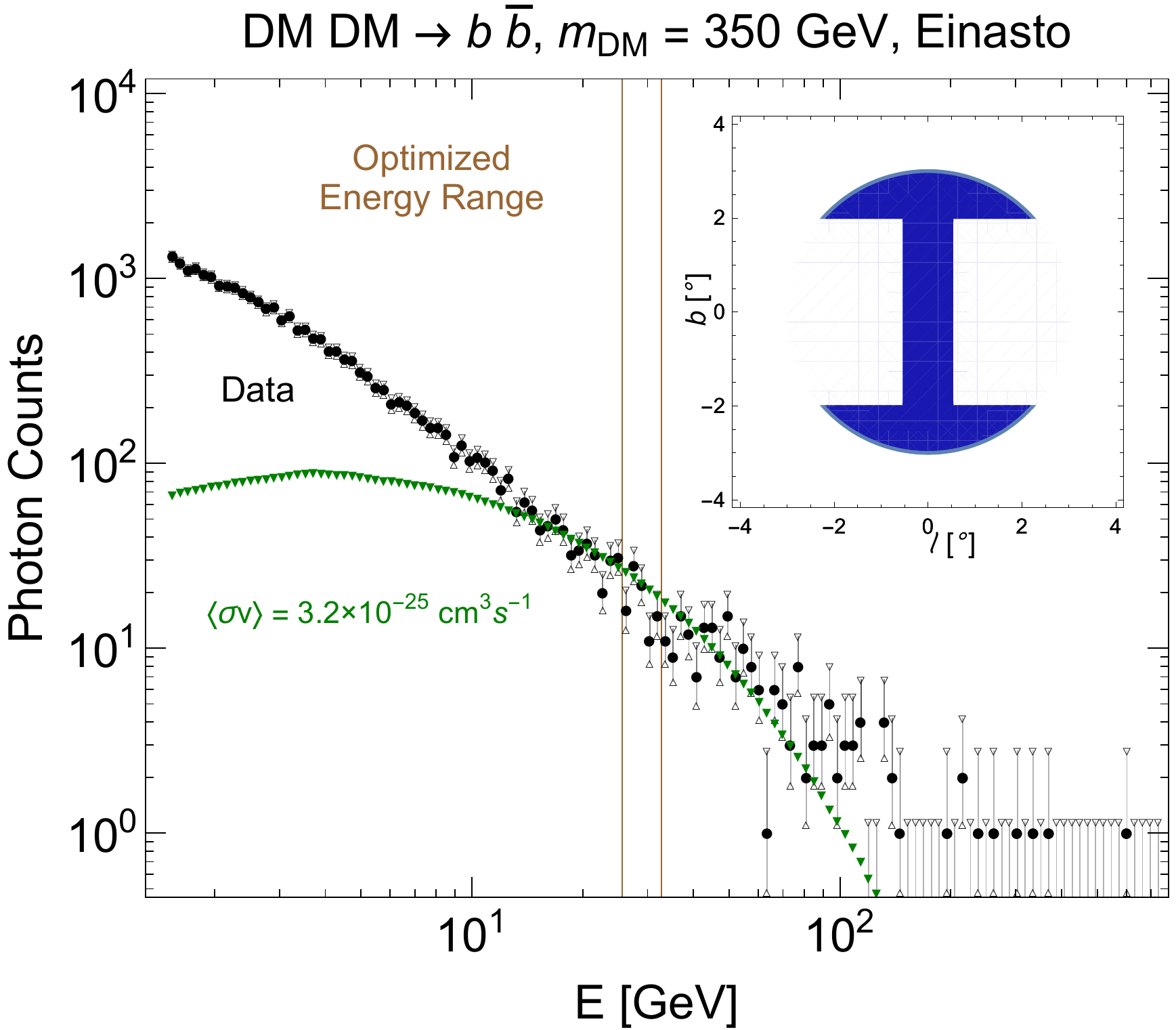}\;\;
\includegraphics[width=\mywidth\textwidth]{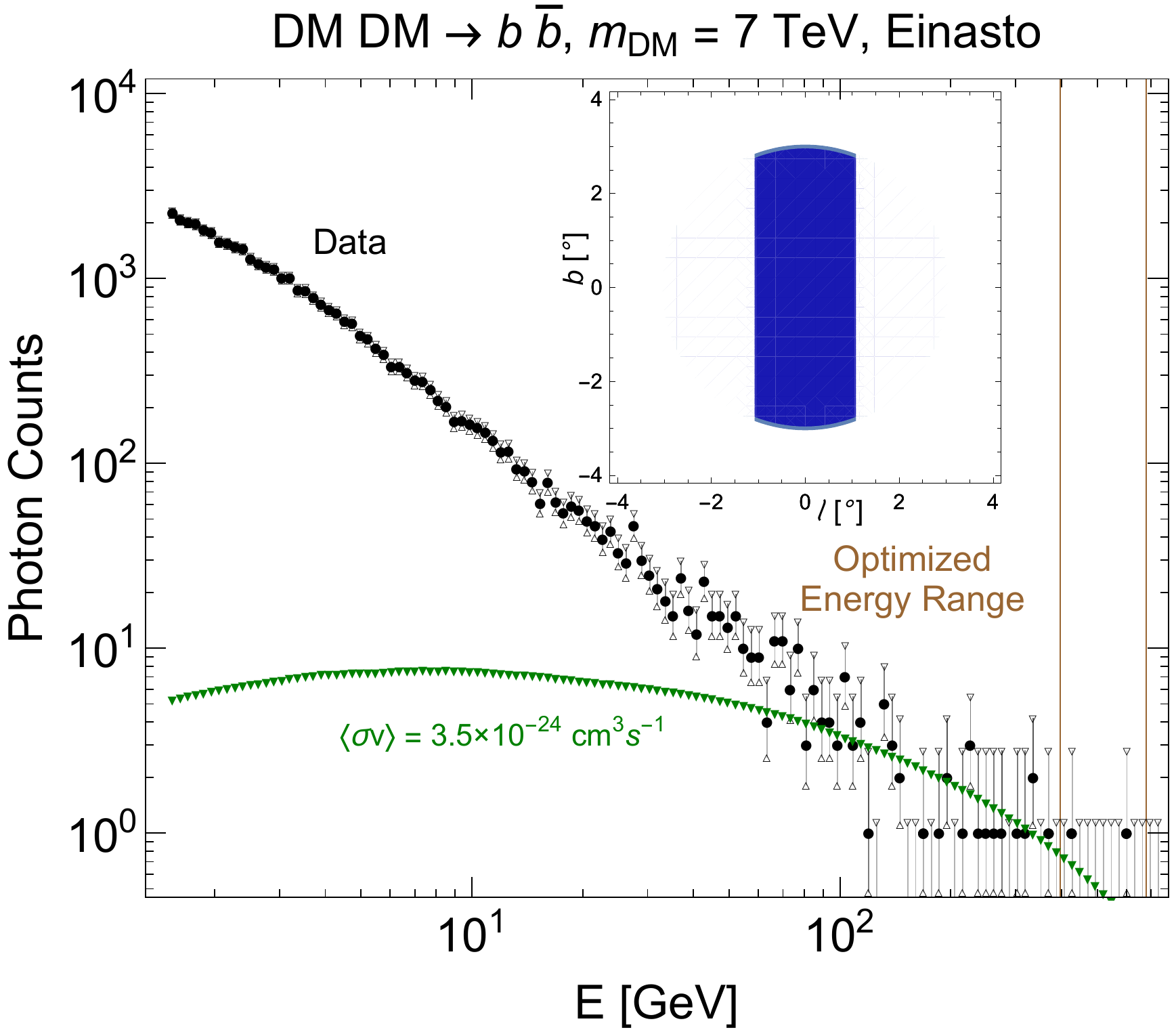}
\end{center}
\vskip -5mm
\caption{\footnotesize
Count spectrum for 350 GeV ({\bf left}) and 7 TeV ({\bf right}) DM annihilating to $b\bar b$, assuming an Einasto profile. 
The vertical (brown) lines show the optimal energy range for each DM-model assumption. The inset shows the optimal ROI. 
Note that PSF-convolution effects were included for the DM signal. 
The quoted $\langle\sigma v\rangle$ is the annihilation cross section that saturates the 95\% C.L. from the data.
}
\label{fig:result-ROIs2}
\end{figure}
\end{spacing}

\section{Dependence of Optimal ROI and Energy Range on DM Profile and DM Mass}
\label{sec:results-roispectra}

The optimal ROI and photon-energy range are found separately for each choice of DM spatial distribution, mass, and final state. In this section, we briefly illustrate the generic features of the optimal search region and its dependence on the theory hypothesis. Fig.~\ref{fig:result-ROIs} shows the obtained ROI and energy range for DM annihilation to $b\bar b$ for each of the four spatial distributions studied, and for a fixed DM mass of 25 GeV. For this final state, with the exception of NFW$_c$, where it is beneficial to look near the GC, the optimal regions in the sky involve semi-circular regions, symmetric in latitude $b$, with the GC removed. Furthermore, we find narrower optimal energy ranges for NFW$_c$-distributed DM.   

For the $b\bar b$ final state, the effect of varying the DM mass is addressed in Fig.~\ref{fig:result-ROIs2}, where the optimal regions are shown for two different masses: 350~GeV and 7~TeV, assuming NFW$_c$-distributed DM. As the DM mass is increased, the strongest optimal regions are obtained by including semi-circular regions in latitude, in addition to a rectangular area around the GC. We note that finite-resolution effects were included, by convolving the instrument's PSF with the J-factors, in the DM signal for all of the results in Fig.~\ref{fig:result-ROIs} and Fig.~\ref{fig:result-ROIs2}.

%

\section{Effect of Source Masking and Choice of Front-/Back-converting events on Limits}
\label{subsec:SS-FB}

In this appendix we investigate the effect on the DM-cross-section upper limits when masking known point sources and using front- and/or back-converting events.  

Masking known sources reduces the observed counts in an ROI and can strengthen the DM constraints, assuming that the masking does not also remove much of a potential DM signal.  
This is the case if the ROI is large, as it is expected to be for decaying DM, or for annihilating DM with shallow DM density profiles (e.g.,~isothermal).
Since astrophysical point sources at very large energies ($>$ 20~GeV) typically exhibit a small flux, their masking is expected to improve the limits for lower DM masses. For very cuspy profiles the ROIs tend to be small and concentrated around the GC region, where the number of known sources is also large; in this case, masking all the point sources would remove most of the DM signal as well and will thus not likely lead to stronger limits. 

The amount of sky that needs to be masked to remove a point source depends on the {\it Fermi}-LAT PSF, which depends on the energy and on where the photon converts in the detector.  In particular, photons that convert to an $e^+e^-$ pair in the front part of the {\it Fermi}-LAT (consisting of the first 12 layers of thin tungsten foil) have a better angular resolution (smaller PSF) than those photons that convert in the back (next 4 layers of thick tungsten foils).  
For very cuspy profiles the choice of including only front- or only back-converting events, or both, could potentially have important effects on the constraints.

We obtain the point-source coordinates from the 3FGL catalog \cite{TheFermi-LAT:2015hja} and exclude all the photons contained in pixels whose center lies within an angular radius of $2\,\gt_{68}(E)$ from any point source; here $\theta_{68}$ is an approximation of the energy-dependent P7REP 68\% point-source containment angle, 
\bea
\gt_{68}(E)[^{\circ}]=\sqrt{c_0^2\pt{E/1\, {\rm GeV}}^{-2\gb}+c_1^2}\,,
\eea
and the parameters for (front-, back-) converting events are $c_0=\pt{0.645,1.103}, c_1=\pt{0.0821,0.166}$, and $\gb=\pt{0.762,0.750}$.  
No source masking is performed within the inner  $2^\circ \times 2^\circ$ square at the GC, where the density of sources is very high and the expected DM signal peaks.  

\begin{figure}[t!]
\includegraphics[width=0.48 \textwidth]{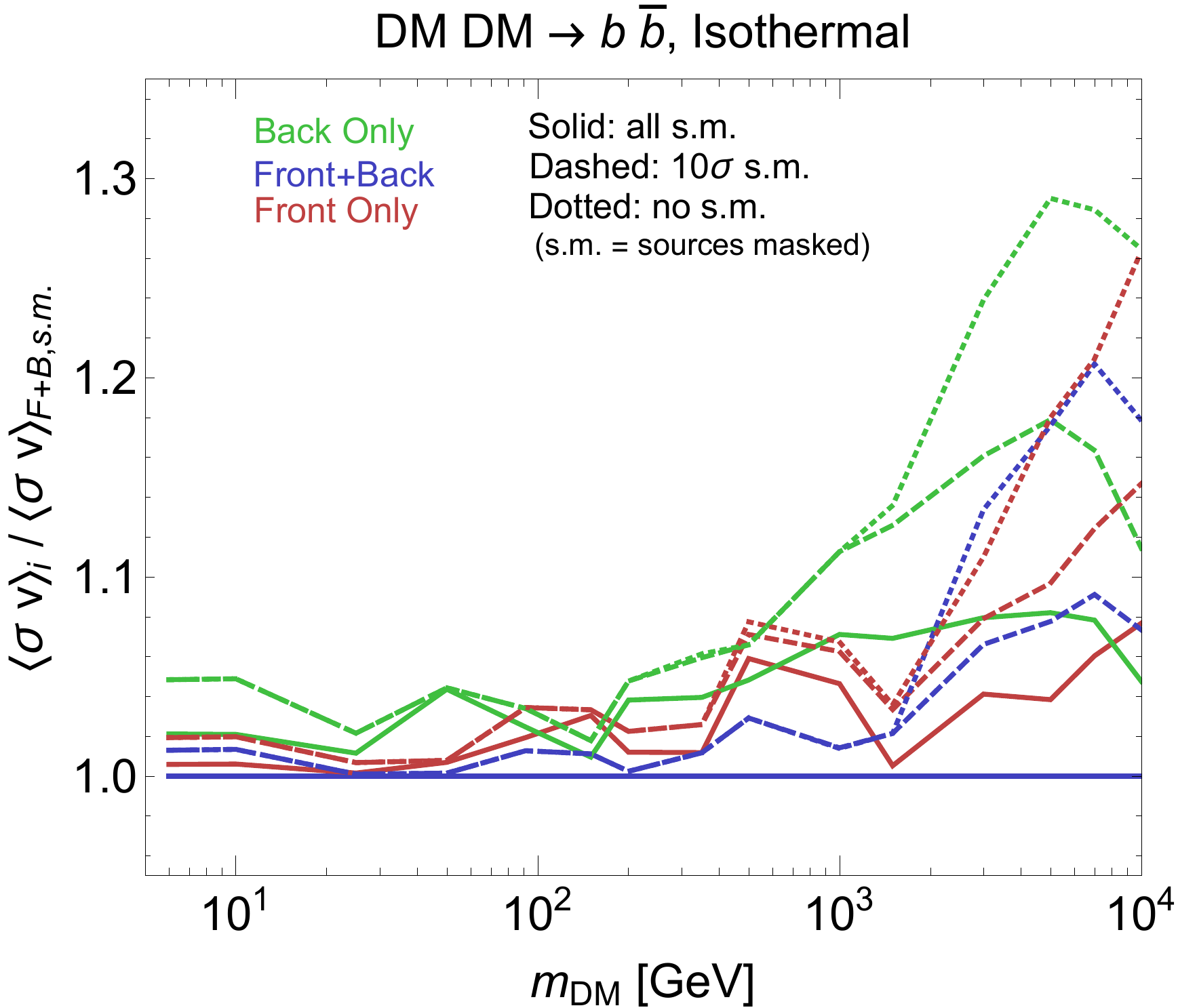}\;\;\;
\includegraphics[width=0.48 \textwidth]{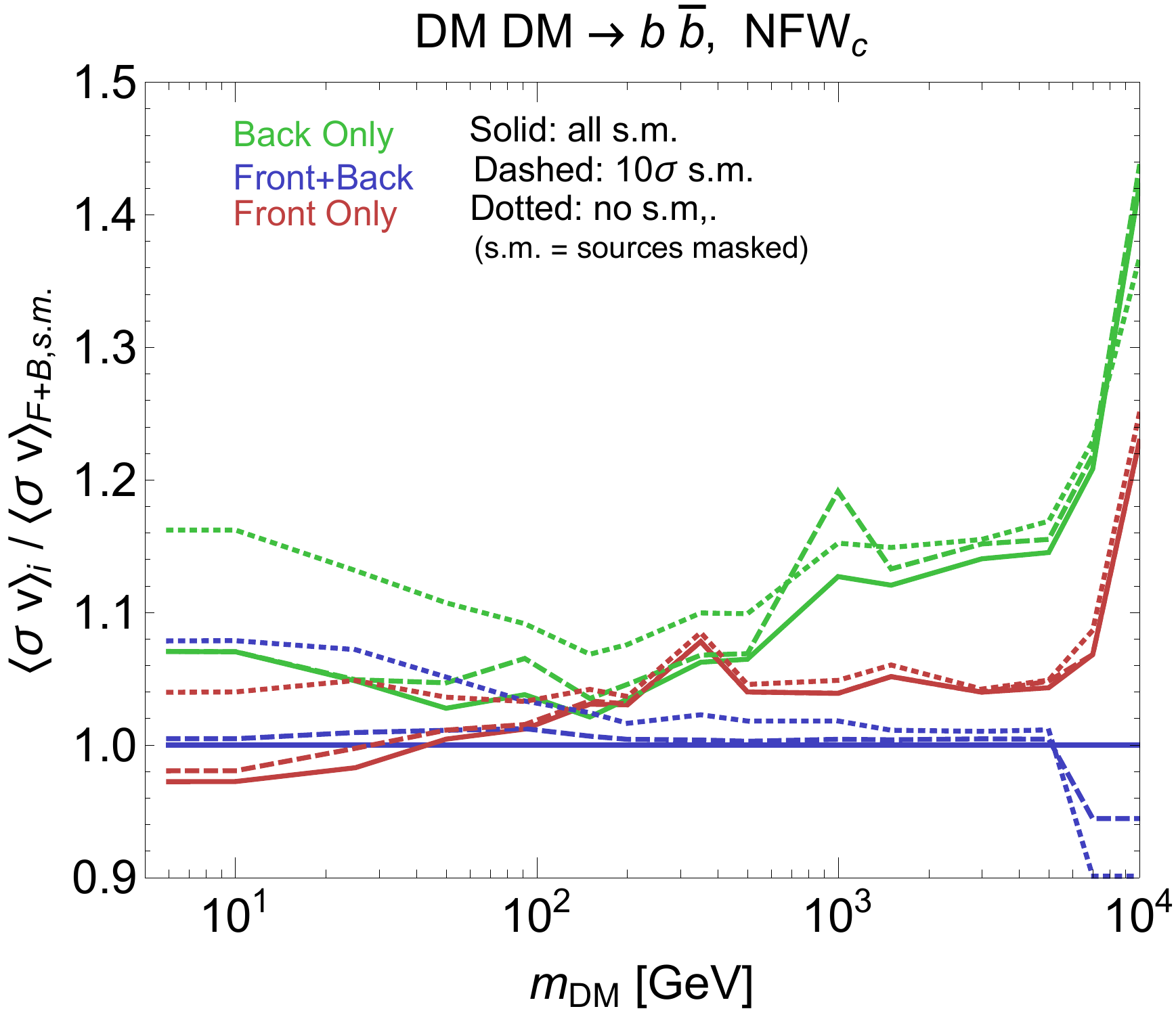}
\caption{
Ratio of expected cross section upper limits vs. DM mass from simulated MC data for DM annihilation to $b\bar{b}$ for isothermal ({\bf left}) and NFW$_c$ ({\bf right}) profiles.  
The denominator of the ratio, $\langle\sigma v\rangle_{\rm F+B, s.m.}$, is the cross section upper limit obtained when masking all known point sources in the 5-year {\it Fermi}-LAT point-source catalog, outside a $2^\circ \times 2^\circ$ square centered at the GC and including front- and back-converting events. 
The numerators of the ratios, $\langle\sigma v\rangle_i$, are the cross section upper limits obtained when masking all known point sources outside the $2^\circ \times 2^\circ$ GC square (solid lines), masking only those sources detected at more than $10\sigma$ (outside the same $2^\circ \times 2^\circ$ GC square) (dashed lines), and masking no sources (dotted lines). In each case we either include both front- and back-converting events (blue lines), only front-converting events (red lines), and only back-converting events (green lines).
\label{fig:SS-FB}
}
\end{figure}

The effect on the cross-section upper limits versus DM mass, when masking known point sources, and when including front- and/or back-converting events, is shown in Fig.~\ref{fig:SS-FB} on simulated data sets.  
The left (right) plot assumes DM annihilation to $b\bar{b}$ for our choice of an isothermal (NFW-contracted) density profile.  
We choose a shallow and cuspy profile to see how the results depend on having either large or small optimized ROIs, respectively.  
For each DM mass, and for each choice of source masking and inclusion of front-/back-converting events, we optimize the ROI choice and derive the average limit obtained from the ten simulated MC data sets.  
In Fig.~\ref{fig:SS-FB}, we show a ratio of expected cross section upper limits versus DM mass: the denominator of the ratios, $\langle\sigma v\rangle_{\rm F+B,\, s.m.}$, is the cross section upper limit obtained when masking all known point sources as described above and including front- and back-converting events; the numerators of the ratios, $\langle\sigma v\rangle_i$, are the cross section upper limits obtained when masking all known point sources (outside the $2^\circ \times 2^\circ$ GC square) (solid lines), masking only those sources detected at more than $10\sigma$ (outside the same $2^\circ \times 2^\circ$ GC square) (dashed lines), and masking no sources (dotted lines). 
In each case we either include both front- and back-converting events (blue lines), only front-converting events (red lines), or only back-converting events (green lines).  

We see from Fig.~\ref{fig:SS-FB} that, at least for the two annihilation models considered in this section, the expected limits are the same within $\mathcal{O}(10-30\%)$. Moreover, the strongest constraints are generically obtained when masking all point sources. For DM masses below $\sim 50$~GeV and cuspy profiles, the inclusion of only front-converting events is expected to provide the strongest constraints, but only marginally so.  Above $\sim 50$~GeV, the inclusion of both front- and back-converting events is best, since the photons produced in the annihilation of DM have such high energies that the PSF effects are negligible, and the inclusion of as much data as possible leads to stronger expected limits.  

Based on this, we conclude that the effect of source masking and choice of front-/back-converting events is not large on our results. We also note that the inclusion of both event-conversion types and the masking of point sources (blue solid line in Fig.~\ref{fig:SS-FB}) is expected to give constraints that are among the best. We thus make this our standard choice when showing the results in \S\ref{sec:results}.

\section{Inverse Compton Scattering}
\label{app:ICS}

In this appendix we discuss how the results from \S\ref{sec:results} depend on the parameters in the ICS computation performed in \texttt{GALPROP}.
The amount of ICS radiation depends sensitively on various key propagation parameters whose values are not known to a satisfactory degree. Here we describe the effect on our constraints from varying these parameters in order to capture some of the systematic uncertainties associated with the DM-generated ICS signal.  

\begin{figure}[t!]
\includegraphics[width=.48\textwidth]{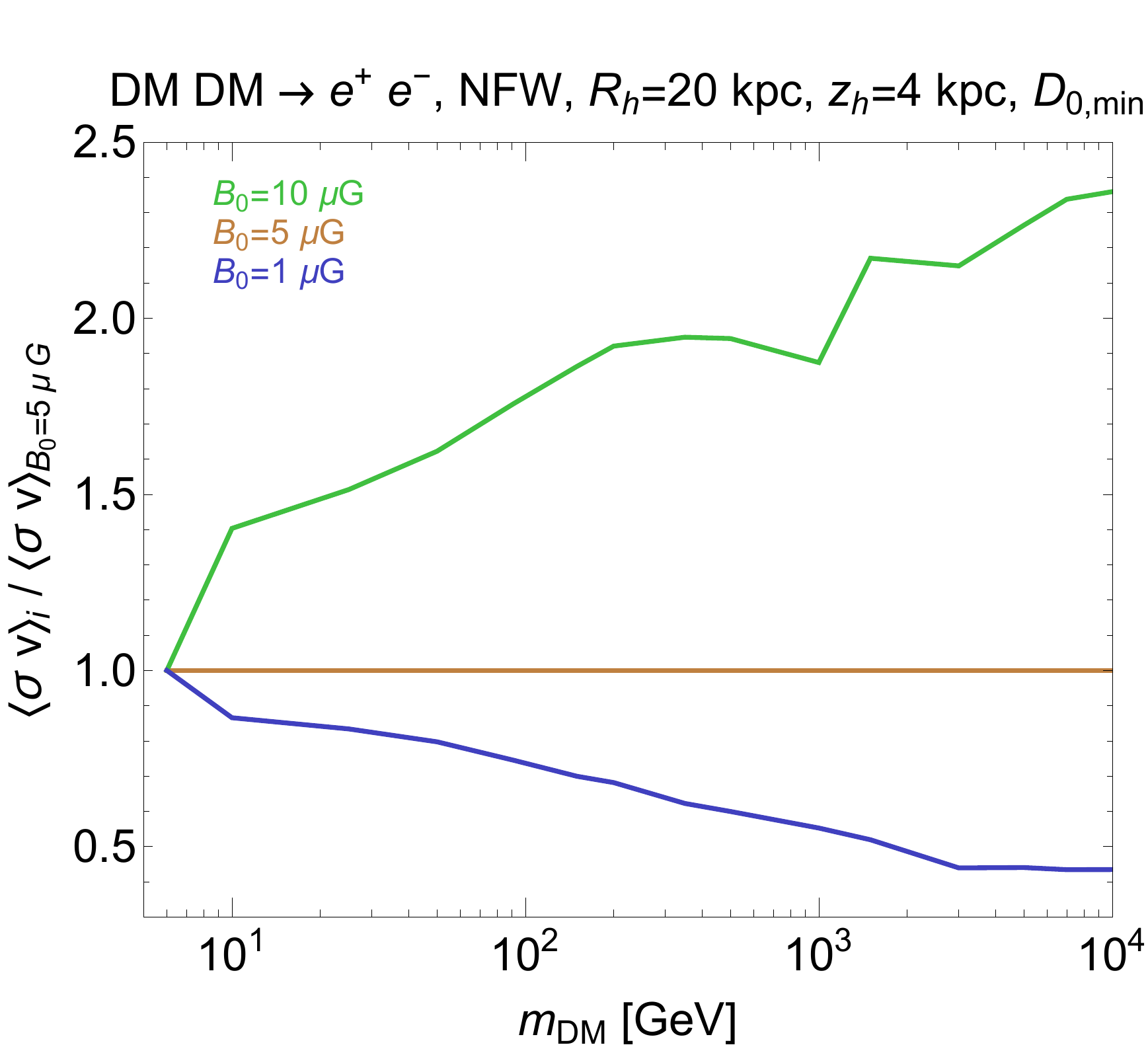}
\includegraphics[width=.48\textwidth]{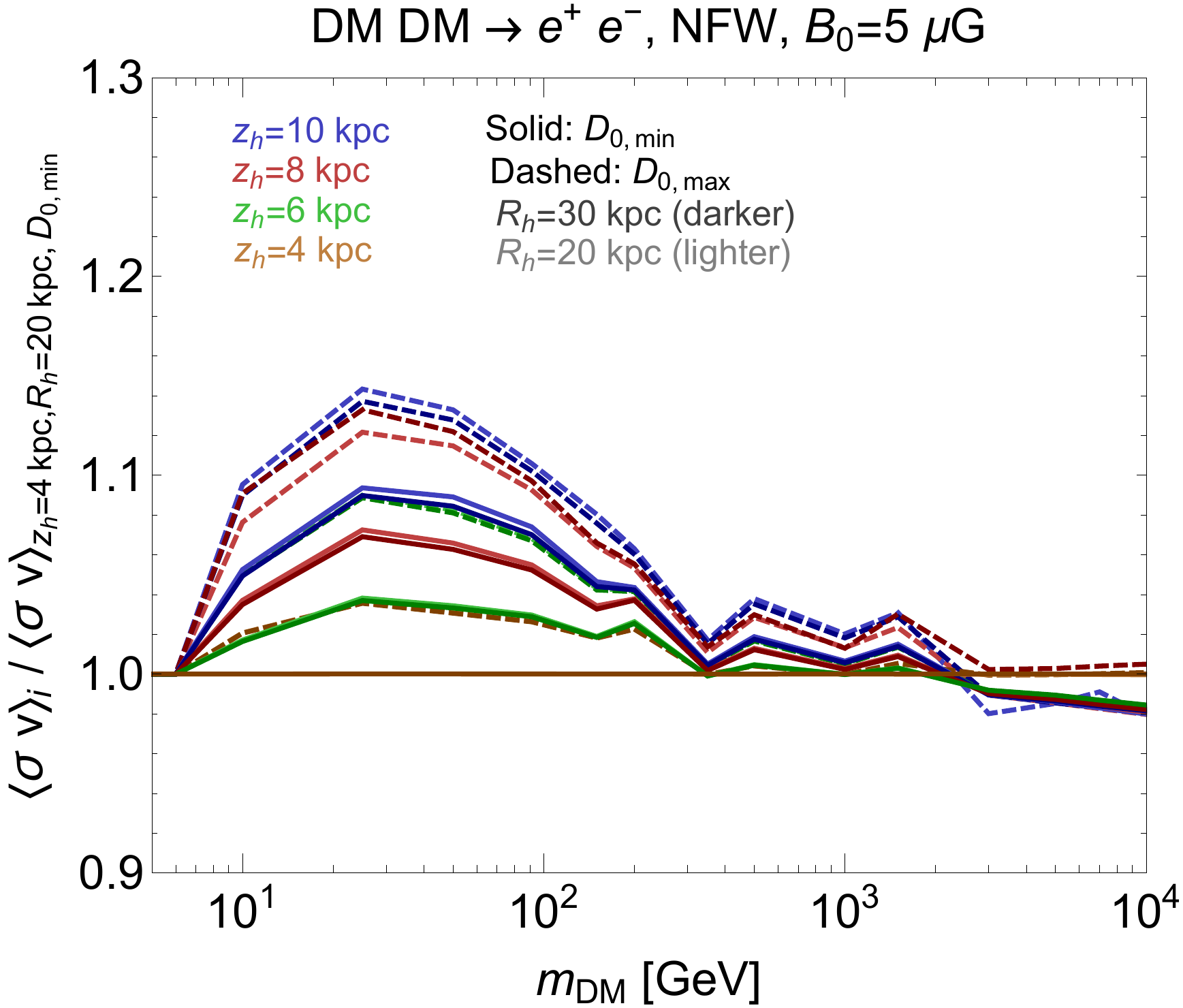}
\caption{Ratios of cross-section upper limits from simulated data on DM annihilation to $e^+e^-$ for an NFW profile, including prompt and ICS radiations, for different values of the Galactic magnetic field ({\bf left}) and different combinations of other propagation model parameters ({\bf right}). The magnetic field has the largest effect on our analysis.}
\label{fig:ICS-study}
\end{figure}

We study how different models of propagation impact our results. We use, as a starting point, the {\it Fermi}-LAT results from \cite{FermiLAT:2012aa}, in which various propagation models are fit to cosmic-ray spectra for various choices of the region of containment of the cosmic rays (parametrized with a cylindrical geometry of half-height $z_h$ and radius $R_h$). 
In our study, we vary $z_h$ and $R_h$, and two other important parameters that have a big effect on the DM ICS signal, namely the Galactic magnetic field value in the Solar System, $B_0$, and the spatial diffusion coefficient $D_0$. The values used in our study are:
\begin{enumerate}
\item $z_h= 4, 6, 8, 10 \, \rm kpc$
\item $R_h=20, 30 \,\rm kpc$
\item $D_0 = D_{0,\rm min}, D_{0,\rm max}$, where $D_{0,\rm min}$ and $D_{0,\rm max}$ are the minimum and maximum values of 
$D_0$ spanned by the various \texttt{GALPROP} models studied in~\cite{FermiLAT:2012aa} for a given ($z_h,R_h$).
\item $B_0 = 1, 5, 10~\mu$G
\end{enumerate}
As an illustration of the dependence of the DM ICS signal on these parameters, Fig.~\ref{fig:ICS-study} shows the constraint on DM annihilation to $e^+e^-$, assuming an NFW$_c$ DM profile. The greatest effect on the uncertainty of the DM ICS signal originates from the variation in the magnitude of $B_0$, as clearly shown in the left plot. Varying the other parameters (right plot) has less of an effect on the DM ICS signal. We are therefore allowed, when showing the results in \S\ref{sec:results}, to fix $z_h= 4 \,\rm kpc$ and $R_h= 20 \,\rm kpc$; whereas we show the variation of our results with $B_0$ and correspondingly $D_0=D_{0,\rm min}=4.797 \times10^{28}$~cm$^2$/s for $B_0 = 1~\mu$G (parameters yielding the strongest constraints) and $D_0=D_{0,\rm max}=6.311 \times10^{28}$~cm$^2$/s for $B_0 = 10~\mu$G (parameters yielding the weakest constraints). 

\section{Details on the Simulated data sets}
\label{app:MC}
The optimization procedure described in \S\ref{subsec:optimization} to find the optimal ROIs and energy ranges, [ROI,~$\Delta E$]$_O$, is performed on ten simulated data sets, each a 5.84-year representation of the $\gamma$-ray sky.
Here we provide a few more details on the simulations.

The generation of mock {\it Fermi}-LAT observations was carried out with the \texttt{gtobssim} routine, part of the Fermi
Science Tools package v9r34p1. Its output is a list of MC-simulated $\gamma$-ray events with relative spatial direction,
arrival time and energy, distributed according to an input source model and IRFs.

A number of model elements were put into \texttt{gtobssim} (see~\cite{Gleam1}). These include the {\it Fermi}-LAT Collaboration's model of the diffuse Galactic component,\footnote{{\texttt gll\_iem\_v05\_rev1.fit}} the isotropic component (derived for Pass 7 Reprocessed Clean front and back IRFs),\footnote{{\texttt iso\_clean\_front\_v05.txt} and {\texttt iso\_clean\_back\_v05.txt}} and the 3FGL source catalog for point and small extended sources~\cite{TheFermi-LAT:2015hja}. 

In addition, the full-sky simulations were calculated through \texttt{gtobssim} with the actual pointing and livetime history (FT2 file) of the {\it Fermi}-LAT for the first 5.84 years of the scientific phase of the mission. The source model simulated did not contain the Earth's Limb emission, which is negligible at energies above 1 GeV, compared to the celestial $\gamma$-ray signal, when a zenith angle $< 100^{\circ}$ cut is applied. The \texttt{gtobssim} tool convolves the flux components mentioned with the {\it Fermi}-LAT's response, i.e.~PSF, energy dispersion, and effective area.

Ten instances of the MC gtobssim-generated data were run, each with an independent starting seed and the same source model; thus obtaining ten statistically independent instances of the $\gamma$-ray sky. The same event selection criteria were used for the MC data sets as for the real data. One important difference between the simulated data sets and the real data is the energy range. Each simulated data set was calculated in an energy range of 0.5~GeV to 500~GeV (as opposed to 1.5~GeV to 750~GeV for the actual data). The upper bound of 500 GeV in the \texttt{gtobssim} simulations is the upper limit in the energy map of the interstellar diffuse model~\cite{Gleam1}. To deal with this mismatch, we simply fit a power-law curve to each of the ten simulated data spectra for $6.2$~GeV$<E<460$~GeV that we obtain for each ROI, and extrapolate it to 750~GeV.  (The lower value of 6.2~GeV is low enough to have enough photons to perform a meaningful fit even for small ROIs, and high enough for a single featureless power law to provide a reasonable fit to the spectra. The upper value of 460~GeV is low enough to avoid count leakages due to finite energy resolution on the sharp 500~GeV input-energy cutoff.)
We then populate each bin above 460~GeV with a random number of events chosen from a Poisson distribution whose expectation value equals the extrapolated value in a given bin. 
The subsequent optimal ROI and energy range for each theory hypothesis $T_H$ is found using the original plus extrapolated spectra.


\section{Comparison of limits between simulated and real data}\label{subsec:MC-data-comp}

\begin{figure}[t!]
\includegraphics[width=0.48\textwidth]{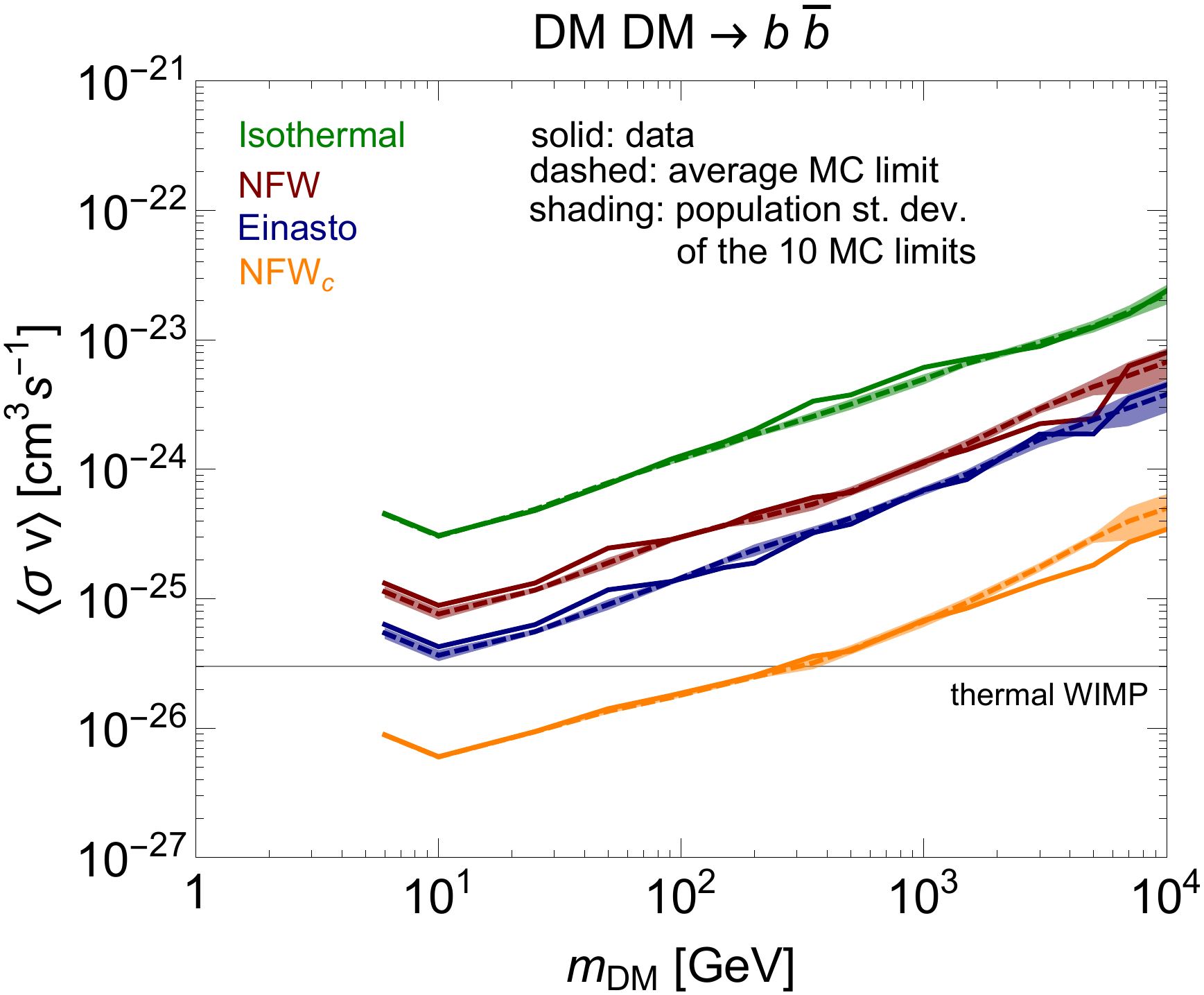}
\includegraphics[width=0.48\textwidth]{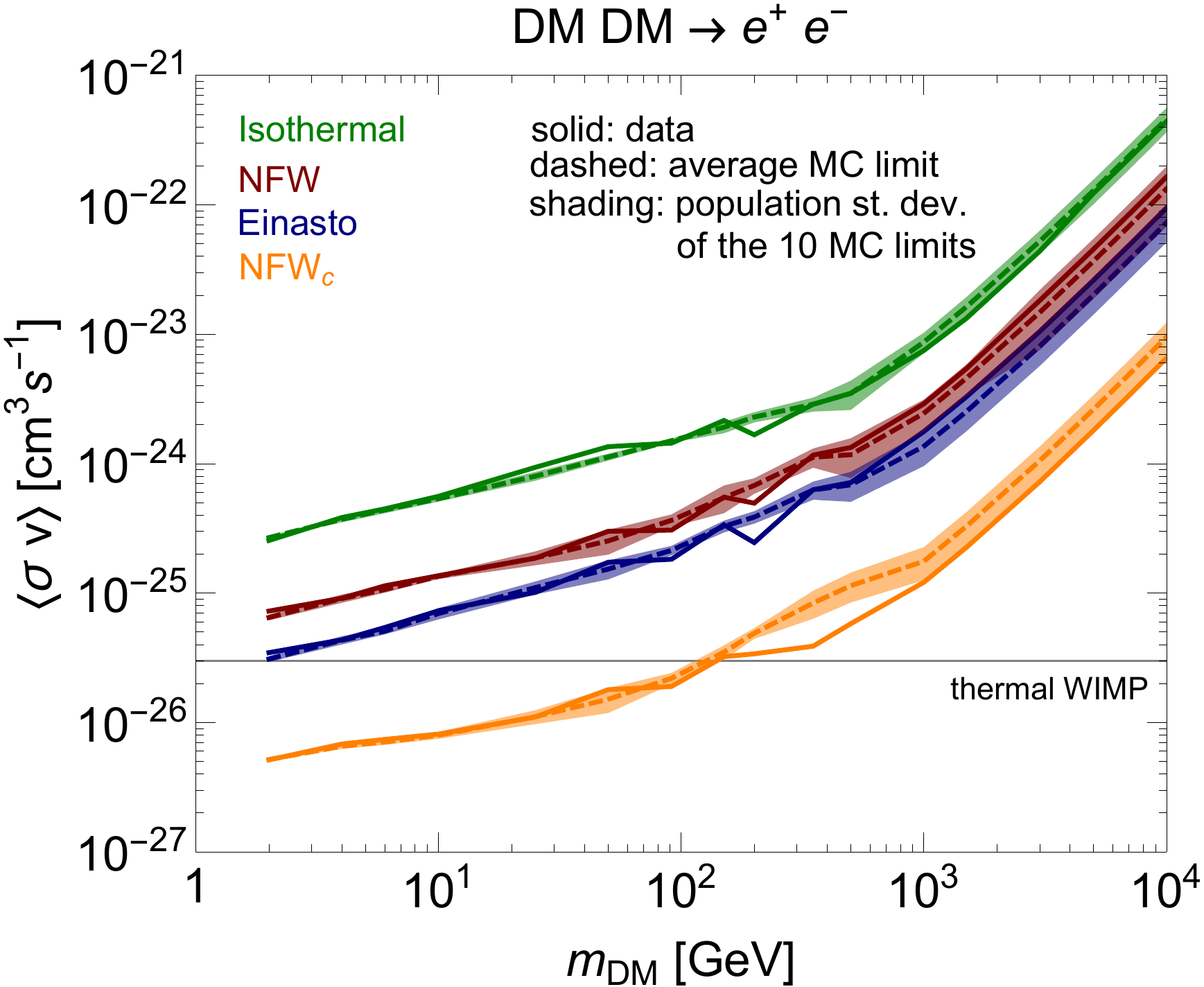}
\caption{Comparison between average MC-based expected (dashed) and real-data (solid) 95\% C.L.~annihilation cross section upper limits on DM annihilating to $b\bar{b}$ ({\bf left}) and $e^+ e^-$ ({\bf right}) for the Isothermal (green), NFW (red), Einasto (blue), and NFW$_c$ (orange) DM density profiles (we only consider prompt photons). The population standard deviations of the limits from the 10 individual MC simulations are also shown as shadings around the dashed lines.}
\label{fig:MCvsData}
\end{figure}

In this appendix we compare the results derived from the real data with those derived from simulated data. Since our simulated data is of course not a perfect representation of the real data, we do not expect that the limits derived on the real data will agree perfectly with the limits derived on simulated data.  

Fig.~\ref{fig:MCvsData} compares the simulated and observed limit on DM annihilation to $b\bar b$ ({\bf left plot}) and $e^+e^-$, including only prompt photons, ({\bf right plot}), for the four different DM density profiles introduced in \S\ref{subsec:Prompt}. Since the simulated data used in this study consists of 10 statistically independent realizations of the $\gamma$-ray sky, we present the arithmetic mean of the 10 limits (dashed lines) and the standard deviation of the population (shaded bands), as well as the observed limits (solid lines). 
We see that the limits derived using real versus simulated data agree over a wide range of masses and profiles. 


\bibliography{bibo}

\providecommand{\href}[2]{#2}\begingroup\raggedright\begin{thebibliography}{100}

\bibitem{Atwood:2009ez}
{\bf LAT Collaboration} Collaboration, W.~Atwood {\em et.~al.}, {\it {The Large
  Area Telescope on the Fermi Gamma-ray Space Telescope Mission}},  {\em
  Astrophys.J.} {\bf 697} (2009) 1071--1102,
  [\href{http://xxx.lanl.gov/abs/0902.1089}{{\tt arXiv:0902.1089}}].

\bibitem{Bertone:2004pz}
G.~Bertone, D.~Hooper, and J.~Silk, {\it {Particle dark matter: Evidence,
  candidates and constraints}},  {\em Phys.Rept.} {\bf 405} (2005) 279--390,
  [\href{http://xxx.lanl.gov/abs/hep-ph/0404175}{{\tt hep-ph/0404175}}].

\bibitem{Abdo:2010nc}
A.~Abdo, M.~Ackermann, M.~Ajello, W.~Atwood, L.~Baldini, {\em et.~al.}, {\it
  {Fermi LAT Search for Photon Lines from 30 to 200 GeV and Dark Matter
  Implications}},  {\em Phys.Rev.Lett.} {\bf 104} (2010) 091302,
  [\href{http://xxx.lanl.gov/abs/1001.4836}{{\tt arXiv:1001.4836}}].

\bibitem{Ackermann:2012qk}
{\bf LAT Collaboration} Collaboration, M.~Ackermann {\em et.~al.}, {\it {Fermi
  LAT Search for Dark Matter in Gamma-ray Lines and the Inclusive Photon
  Spectrum}},  {\em Phys.Rev.} {\bf D86} (2012) 022002,
  [\href{http://xxx.lanl.gov/abs/1205.2739}{{\tt arXiv:1205.2739}}].

\bibitem{Ackermann:2013uma}
{\bf Fermi-LAT Collaboration} Collaboration, M.~Ackermann {\em et.~al.}, {\it
  {Search for Gamma-ray Spectral Lines with the Fermi Large Area Telescope and
  Dark Matter Implications}},  {\em Phys.Rev.} {\bf D88} (2013) 082002,
  [\href{http://xxx.lanl.gov/abs/1305.5597}{{\tt arXiv:1305.5597}}].

\bibitem{Albert:2014hwa}
{\bf Fermi-LAT Collaboration} Collaboration, A.~Albert {\em et.~al.}, {\it
  {Search for 100 MeV to 10 GeV $\gamma$-ray lines in the Fermi-LAT data and
  implications for gravitino dark matter in $\mu\nu$SSM}},  {\em JCAP} {\bf
  1410} (2014), no.~10 023, [\href{http://xxx.lanl.gov/abs/1406.3430}{{\tt
  arXiv:1406.3430}}].

\bibitem{Bringmann:2012vr}
T.~Bringmann, X.~Huang, A.~Ibarra, S.~Vogl, and C.~Weniger, {\it {Fermi LAT
  Search for Internal Bremsstrahlung Signatures from Dark Matter
  Annihilation}},  {\em JCAP} {\bf 1207} (2012) 054,
  [\href{http://xxx.lanl.gov/abs/1203.1312}{{\tt arXiv:1203.1312}}].

\bibitem{Weniger:2012tx}
C.~Weniger, {\it {A Tentative Gamma-Ray Line from Dark Matter Annihilation at
  the Fermi Large Area Telescope}},  {\em JCAP} {\bf 1208} (2012) 007,
  [\href{http://xxx.lanl.gov/abs/1204.2797}{{\tt arXiv:1204.2797}}].

\bibitem{Su:2012ft}
M.~Su and D.~P. Finkbeiner, {\it {Strong Evidence for Gamma-ray Line Emission
  from the Inner Galaxy}},  \href{http://xxx.lanl.gov/abs/1206.1616}{{\tt
  arXiv:1206.1616}}.

\bibitem{Ackermann:2011wa}
{\bf Fermi-LAT collaboration} Collaboration, M.~Ackermann {\em et.~al.}, {\it
  {Constraining Dark Matter Models from a Combined Analysis of Milky Way
  Satellites with the Fermi Large Area Telescope}},  {\em Phys.Rev.Lett.} {\bf
  107} (2011) 241302, [\href{http://xxx.lanl.gov/abs/1108.3546}{{\tt
  arXiv:1108.3546}}].

\bibitem{Abdo:2010ex}
{\bf Fermi-LAT Collaboration} Collaboration, A.~Abdo {\em et.~al.}, {\it
  {Observations of Milky Way Dwarf Spheroidal galaxies with the Fermi-LAT
  detector and constraints on Dark Matter models}},  {\em Astrophys.J.} {\bf
  712} (2010) 147--158, [\href{http://xxx.lanl.gov/abs/1001.4531}{{\tt
  arXiv:1001.4531}}].

\bibitem{Ackermann:2013yva}
{\bf Fermi-LAT Collaboration} Collaboration, M.~Ackermann {\em et.~al.}, {\it
  {Dark Matter Constraints from Observations of 25 Milky Way Satellite Galaxies
  with the Fermi Large Area Telescope}},  {\em Phys.Rev.} {\bf D89} (2014)
  042001, [\href{http://xxx.lanl.gov/abs/1310.0828}{{\tt arXiv:1310.0828}}].

\bibitem{Mazziotta:2012ux}
M.~Mazziotta, F.~Loparco, F.~de~Palma, and N.~Giglietto, {\it {A
  model-independent analysis of the Fermi Large Area Telescope gamma-ray data
  from the Milky Way dwarf galaxies and halo to constrain dark matter
  scenarios}},  {\em Astropart.Phys.} {\bf 37} (2012) 26--39,
  [\href{http://xxx.lanl.gov/abs/1203.6731}{{\tt arXiv:1203.6731}}].

\bibitem{GeringerSameth:2011iw}
A.~Geringer-Sameth and S.~M. Koushiappas, {\it {Exclusion of canonical WIMPs by
  the joint analysis of Milky Way dwarfs with Fermi}},  {\em Phys.Rev.Lett.}
  {\bf 107} (2011) 241303, [\href{http://xxx.lanl.gov/abs/1108.2914}{{\tt
  arXiv:1108.2914}}].

\bibitem{Cholis:2012am}
I.~Cholis and P.~Salucci, {\it {Extracting limits on Dark Matter annihilation
  from gamma-ray observations towards dwarf spheroidal galaxies}},  {\em
  Phys.Rev.} {\bf D86} (2012) 023528,
  [\href{http://xxx.lanl.gov/abs/1203.2954}{{\tt arXiv:1203.2954}}].

\bibitem{Geringer-Sameth:2014qqa}
A.~Geringer-Sameth, S.~M. Koushiappas, and M.~G. Walker, {\it {A Comprehensive
  Search for Dark Matter Annihilation in Dwarf Galaxies}},
  \href{http://xxx.lanl.gov/abs/1410.2242}{{\tt arXiv:1410.2242}}.

\bibitem{Ackermann:2010rg}
M.~Ackermann, M.~Ajello, A.~Allafort, L.~Baldini, J.~Ballet, {\em et.~al.},
  {\it {Constraints on Dark Matter Annihilation in Clusters of Galaxies with
  the Fermi Large Area Telescope}},  {\em JCAP} {\bf 1005} (2010) 025,
  [\href{http://xxx.lanl.gov/abs/1002.2239}{{\tt arXiv:1002.2239}}].

\bibitem{Han:2012uw}
J.~Han, C.~S. Frenk, V.~R. Eke, L.~Gao, S.~D. White, {\em et.~al.}, {\it
  {Constraining Extended Gamma-ray Emission from Galaxy Clusters}},  {\em
  Mon.Not.Roy.Astron.Soc.} {\bf 427} (2012) 1651--1665,
  [\href{http://xxx.lanl.gov/abs/1207.6749}{{\tt arXiv:1207.6749}}].

\bibitem{MaciasRamirez:2012mk}
O.~Macias-Ramirez, C.~Gordon, A.~M. Brown, and J.~Adams, {\it {Evaluating the
  Gamma-Ray Evidence for Self-Annihilating Dark Matter from the Virgo
  Cluster}},  {\em Phys.Rev.} {\bf D86} (2012) 076004,
  [\href{http://xxx.lanl.gov/abs/1207.6257}{{\tt arXiv:1207.6257}}].

\bibitem{Ackermann:2012rg}
{\bf LAT collaboration} Collaboration, M.~Ackermann {\em et.~al.}, {\it
  {Constraints on the Galactic Halo Dark Matter from Fermi-LAT Diffuse
  Measurements}},  {\em Astrophys.J.} {\bf 761} (2012) 91,
  [\href{http://xxx.lanl.gov/abs/1205.6474}{{\tt arXiv:1205.6474}}].

\bibitem{Cirelli:2009dv}
M.~Cirelli, P.~Panci, and P.~D. Serpico, {\it {Diffuse gamma ray constraints on
  annihilating or decaying Dark Matter after Fermi}},  {\em Nucl.Phys.} {\bf
  B840} (2010) 284--303, [\href{http://xxx.lanl.gov/abs/0912.0663}{{\tt
  arXiv:0912.0663}}].

\bibitem{Papucci:2009gd}
M.~Papucci and A.~Strumia, {\it {Robust implications on Dark Matter from the
  first FERMI sky gamma map}},  {\em JCAP} {\bf 1003} (2010) 014,
  [\href{http://xxx.lanl.gov/abs/0912.0742}{{\tt arXiv:0912.0742}}].

\bibitem{Gomez-Vargas:2013bea}
G.~A. Gomez-Vargas, M.~A. Sanchez-Conde, J.-H. Huh, M.~Peiro, F.~Prada, {\em
  et.~al.}, {\it {Constraints on WIMP Annihilation for Contracted Dark Matter
  in the Inner Galaxy with the Fermi-LAT}},
  \href{http://xxx.lanl.gov/abs/1308.3515}{{\tt arXiv:1308.3515}}.

\bibitem{Cholis:2009gv}
I.~Cholis, G.~Dobler, D.~P. Finkbeiner, L.~Goodenough, T.~R. Slatyer, {\em
  et.~al.}, {\it {The Fermi gamma-ray spectrum of the inner galaxy:
  Implications for annihilating dark matter}},
  \href{http://xxx.lanl.gov/abs/0907.3953}{{\tt arXiv:0907.3953}}.

\bibitem{Goodenough:2009gk}
L.~Goodenough and D.~Hooper, {\it {Possible Evidence For Dark Matter
  Annihilation In The Inner Milky Way From The Fermi Gamma Ray Space
  Telescope}},  \href{http://xxx.lanl.gov/abs/0910.2998}{{\tt
  arXiv:0910.2998}}.

\bibitem{Hooper:2010mq}
D.~Hooper and L.~Goodenough, {\it {Dark Matter Annihilation in The Galactic
  Center As Seen by the Fermi Gamma Ray Space Telescope}},  {\em Phys.Lett.}
  {\bf B697} (2011) 412--428, [\href{http://xxx.lanl.gov/abs/1010.2752}{{\tt
  arXiv:1010.2752}}].

\bibitem{Boyarsky:2010dr}
A.~Boyarsky, D.~Malyshev, and O.~Ruchayskiy, {\it {A comment on the emission
  from the Galactic Center as seen by the Fermi telescope}},  {\em Phys.Lett.}
  {\bf B705} (2011) 165--169, [\href{http://xxx.lanl.gov/abs/1012.5839}{{\tt
  arXiv:1012.5839}}].

\bibitem{Hooper:2011ti}
D.~Hooper and T.~Linden, {\it {On The Origin Of The Gamma Rays From The
  Galactic Center}},  {\em Phys.Rev.} {\bf D84} (2011) 123005,
  [\href{http://xxx.lanl.gov/abs/1110.0006}{{\tt arXiv:1110.0006}}].

\bibitem{Abazajian:2012pn}
K.~N. Abazajian and M.~Kaplinghat, {\it {Detection of a Gamma-Ray Source in the
  Galactic Center Consistent with Extended Emission from Dark Matter
  Annihilation and Concentrated Astrophysical Emission}},  {\em Phys.Rev.} {\bf
  D86} (2012) 083511, [\href{http://xxx.lanl.gov/abs/1207.6047}{{\tt
  arXiv:1207.6047}}].

\bibitem{Gordon:2013vta}
C.~Gordon and O.~Macias, {\it {Dark Matter and Pulsar Model Constraints from
  Galactic Center Fermi-LAT Gamma Ray Observations}},  {\em Phys.Rev.} {\bf
  D88} (2013) 083521, [\href{http://xxx.lanl.gov/abs/1306.5725}{{\tt
  arXiv:1306.5725}}].

\bibitem{Abazajian:2014fta}
K.~N. Abazajian, N.~Canac, S.~Horiuchi, and M.~Kaplinghat, {\it {Astrophysical
  and Dark Matter Interpretations of Extended Gamma-Ray Emission from the
  Galactic Center}},  {\em Phys.Rev.} {\bf D90} (2014) 023526,
  [\href{http://xxx.lanl.gov/abs/1402.4090}{{\tt arXiv:1402.4090}}].

\bibitem{Hooper:2013rwa}
D.~Hooper and T.~R. Slatyer, {\it {Two Emission Mechanisms in the Fermi
  Bubbles: A Possible Signal of Annihilating Dark Matter}},  {\em Phys.Dark
  Univ.} {\bf 2} (2013) 118--138,
  [\href{http://xxx.lanl.gov/abs/1302.6589}{{\tt arXiv:1302.6589}}].

\bibitem{Daylan:2014rsa}
T.~Daylan, D.~P. Finkbeiner, D.~Hooper, T.~Linden, S.~K.~N. Portillo, {\em
  et.~al.}, {\it {The Characterization of the Gamma-Ray Signal from the Central
  Milky Way: A Compelling Case for Annihilating Dark Matter}},
  \href{http://xxx.lanl.gov/abs/1402.6703}{{\tt arXiv:1402.6703}}.

\bibitem{Huang:2013pda}
W.-C. Huang, A.~Urbano, and W.~Xue, {\it {Fermi Bubbles under Dark Matter
  Scrutiny. Part I: Astrophysical Analysis}},
  \href{http://xxx.lanl.gov/abs/1307.6862}{{\tt arXiv:1307.6862}}.

\bibitem{Macias:2013vya}
O.~Macias and C.~Gordon, {\it {The Contribution of Cosmic Rays Interacting With
  Molecular Clouds to the Galactic Center Gamma-Ray Excess}},  {\em Phys.Rev.}
  {\bf D89} (2014) 063515, [\href{http://xxx.lanl.gov/abs/1312.6671}{{\tt
  arXiv:1312.6671}}].

\bibitem{Calore:2014xka}
F.~Calore, I.~Cholis, and C.~Weniger, {\it {Background model systematics for
  the Fermi GeV excess}},  \href{http://xxx.lanl.gov/abs/1409.0042}{{\tt
  arXiv:1409.0042}}.

\bibitem{Drlica-Wagner:2014yca}
A.~Drlica-Wagner, G.~A. Gomez-Vargas, J.~W. Hewitt, T.~Linden, and L.~Tibaldo,
  {\it {Searching for Dark Matter Annihilation in the Smith High-Velocity
  Cloud}},  {\em Astrophys.J.} {\bf 790} (2014) 24,
  [\href{http://xxx.lanl.gov/abs/1405.1030}{{\tt arXiv:1405.1030}}].

\bibitem{Nichols:2014qsa}
M.~Nichols, N.~Mirabal, O.~Agertz, F.~J. Lockman, and J.~Bland-Hawthorn, {\it
  {The Smith Cloud and its dark matter halo: Survival of a Galactic disc
  passage}},  {\em Mon.Not.Roy.Astron.Soc.} {\bf 442} (2014) 2883,
  [\href{http://xxx.lanl.gov/abs/1404.3209}{{\tt arXiv:1404.3209}}].

\bibitem{Abdo:2010dk}
{\bf Fermi-LAT Collaboration} Collaboration, A.~Abdo {\em et.~al.}, {\it
  {Constraints on Cosmological Dark Matter Annihilation from the Fermi-LAT
  Isotropic Diffuse Gamma-Ray Measurement}},  {\em JCAP} {\bf 1004} (2010) 014,
  [\href{http://xxx.lanl.gov/abs/1002.4415}{{\tt arXiv:1002.4415}}].

\bibitem{Fornasa:2011yb}
M.~Fornasa, J.~Zavala, M.~A. Sanchez-Conde, F.~Prada, and M.~Vogelsberger, {\it
  {Dark Matter implications of the Fermi-LAT measurement of anisotropies in the
  diffuse gamma-ray background: status report}},  {\em Nucl.Instrum.Meth.} {\bf
  A692} (2012) 132--136, [\href{http://xxx.lanl.gov/abs/1110.0324}{{\tt
  arXiv:1110.0324}}].

\bibitem{Calore:2013yia}
T.~Bringmann, F.~Calore, M.~Di~Mauro, and F.~Donato, {\it {Constraining dark
  matter annihilation with the isotropic $\gamma$-ray background: updated
  limits and future potential}},  {\em Phys.Rev.} {\bf D89} (2014) 023012,
  [\href{http://xxx.lanl.gov/abs/1303.3284}{{\tt arXiv:1303.3284}}].

\bibitem{Cholis:2013ena}
I.~Cholis, D.~Hooper, and S.~D. McDermott, {\it {Dissecting the Gamma-Ray
  Background in Search of Dark Matter}},  {\em JCAP} {\bf 1402} (2014) 014,
  [\href{http://xxx.lanl.gov/abs/1312.0608}{{\tt arXiv:1312.0608}}].

\bibitem{Hisano:2009rc}
J.~Hisano, M.~Kawasaki, K.~Kohri, T.~Moroi, and K.~Nakayama, {\it {Cosmic Rays
  from Dark Matter Annihilation and Big-Bang Nucleosynthesis}},  {\em
  Phys.Rev.} {\bf D79} (2009) 083522,
  [\href{http://xxx.lanl.gov/abs/0901.3582}{{\tt arXiv:0901.3582}}].

\bibitem{CyrRacine:2009yn}
F.-Y. Cyr-Racine, S.~Profumo, and K.~Sigurdson, {\it {Protohalo Constraints to
  the Resonant Annihilation of Dark Matter}},  {\em Phys.Rev.} {\bf D80} (2009)
  081302, [\href{http://xxx.lanl.gov/abs/0904.3933}{{\tt arXiv:0904.3933}}].

\bibitem{Belikov:2009qx}
A.~V. Belikov and D.~Hooper, {\it {How Dark Matter Reionized The Universe}},
  {\em Phys.Rev.} {\bf D80} (2009) 035007,
  [\href{http://xxx.lanl.gov/abs/0904.1210}{{\tt arXiv:0904.1210}}].

\bibitem{Galli:2009zc}
S.~Galli, F.~Iocco, G.~Bertone, and A.~Melchiorri, {\it {CMB constraints on
  Dark Matter models with large annihilation cross-section}},  {\em Phys.Rev.}
  {\bf D80} (2009) 023505, [\href{http://xxx.lanl.gov/abs/0905.0003}{{\tt
  arXiv:0905.0003}}].

\bibitem{Slatyer:2009yq}
T.~R. Slatyer, N.~Padmanabhan, and D.~P. Finkbeiner, {\it {CMB Constraints on
  WIMP Annihilation: Energy Absorption During the Recombination Epoch}},  {\em
  Phys.Rev.} {\bf D80} (2009) 043526,
  [\href{http://xxx.lanl.gov/abs/0906.1197}{{\tt arXiv:0906.1197}}].

\bibitem{Huetsi:2009ex}
G.~Huetsi, A.~Hektor, and M.~Raidal, {\it {Constraints on leptonically
  annihilating Dark Matter from reionization and extragalactic gamma
  background}},  {\em Astron.Astrophys.} {\bf 505} (2009) 999--1005,
  [\href{http://xxx.lanl.gov/abs/0906.4550}{{\tt arXiv:0906.4550}}].

\bibitem{Cirelli:2009bb}
M.~Cirelli, F.~Iocco, and P.~Panci, {\it {Constraints on Dark Matter
  annihilations from reionization and heating of the intergalactic gas}},  {\em
  JCAP} {\bf 0910} (2009) 009, [\href{http://xxx.lanl.gov/abs/0907.0719}{{\tt
  arXiv:0907.0719}}].

\bibitem{Madhavacheril:2013cna}
M.~S. Madhavacheril, N.~Sehgal, and T.~R. Slatyer, {\it {Current Dark Matter
  Annihilation Constraints from CMB and Low-Redshift Data}},  {\em Phys.Rev.}
  {\bf D89} (2014) 103508, [\href{http://xxx.lanl.gov/abs/1310.3815}{{\tt
  arXiv:1310.3815}}].

\bibitem{Cirelli:2012tf}
M.~Cirelli, {\it {Indirect Searches for Dark Matter: a status review}},  {\em
  Pramana} {\bf 79} (2012) 1021--1043,
  [\href{http://xxx.lanl.gov/abs/1202.1454}{{\tt arXiv:1202.1454}}].

\bibitem{Meade:2009iu}
P.~Meade, M.~Papucci, A.~Strumia, and T.~Volansky, {\it {Dark Matter
  Interpretations of the e+- Excesses after FERMI}},  {\em Nucl.Phys.} {\bf
  B831} (2010) 178--203, [\href{http://xxx.lanl.gov/abs/0905.0480}{{\tt
  arXiv:0905.0480}}].

\bibitem{Sanchez-Conde:2013yxa}
M.~A. Sanchez-Conde and F.~Prada, {\it {The flattening of the
  concentration-mass relation towards low halo masses and its implications for
  the annihilation signal boost}},  {\em Mon.Not.Roy.Astron.Soc.} {\bf 442}
  (2014) 2271, [\href{http://xxx.lanl.gov/abs/1312.1729}{{\tt
  arXiv:1312.1729}}].

\bibitem{Catena:2009mf}
R.~Catena and P.~Ullio, {\it {A novel determination of the local dark matter
  density}},  {\em JCAP} {\bf 1008} (2010) 004,
  [\href{http://xxx.lanl.gov/abs/0907.0018}{{\tt arXiv:0907.0018}}].

\bibitem{Salucci:2010qr}
P.~Salucci, F.~Nesti, G.~Gentile, and C.~Martins, {\it {The dark matter density
  at the Sun's location}},  {\em Astron.Astrophys.} {\bf 523} (2010) A83,
  [\href{http://xxx.lanl.gov/abs/1003.3101}{{\tt arXiv:1003.3101}}].

\bibitem{Prada:2004pi}
F.~Prada, A.~Klypin, J.~Flix~Molina, M.~Martinez, and E.~Simonneau, {\it {Dark
  Matter Annihilation in the Milky Way Galaxy: Effects of Baryonic
  Compression}},  {\em Phys.Rev.Lett.} {\bf 93} (2004) 241301,
  [\href{http://xxx.lanl.gov/abs/astro-ph/0401512}{{\tt astro-ph/0401512}}].

\bibitem{Pato:2010yq}
M.~Pato, O.~Agertz, G.~Bertone, B.~Moore, and R.~Teyssier, {\it {Systematic
  uncertainties in the determination of the local dark matter density}},  {\em
  Phys.Rev.} {\bf D82} (2010) 023531,
  [\href{http://xxx.lanl.gov/abs/1006.1322}{{\tt arXiv:1006.1322}}].

\bibitem{Garbari:2012ff}
S.~Garbari, C.~Liu, J.~I. Read, and G.~Lake, {\it {A new determination of the
  local dark matter density from the kinematics of K dwarfs}},  {\em
  Mon.Not.Roy.Astron.Soc.} {\bf 425} (2012) 1445,
  [\href{http://xxx.lanl.gov/abs/1206.0015}{{\tt arXiv:1206.0015}}].

\bibitem{Ghez:2008ms}
A.~Ghez, S.~Salim, N.~Weinberg, J.~Lu, T.~Do, {\em et.~al.}, {\it {Measuring
  Distance and Properties of the Milky Way's Central Supermassive Black Hole
  with Stellar Orbits}},  {\em Astrophys.J.} {\bf 689} (2008) 1044--1062,
  [\href{http://xxx.lanl.gov/abs/0808.2870}{{\tt arXiv:0808.2870}}].

\bibitem{Navarro:1995iw}
J.~F. Navarro, C.~S. Frenk, and S.~D. White, {\it {The Structure of cold dark
  matter halos}},  {\em Astrophys.J.} {\bf 462} (1996) 563--575,
  [\href{http://xxx.lanl.gov/abs/astro-ph/9508025}{{\tt astro-ph/9508025}}].

\bibitem{Navarro:1996gj}
J.~F. Navarro, C.~S. Frenk, and S.~D. White, {\it {A Universal density profile
  from hierarchical clustering}},  {\em Astrophys.J.} {\bf 490} (1997)
  493--508, [\href{http://xxx.lanl.gov/abs/astro-ph/9611107}{{\tt
  astro-ph/9611107}}].

\bibitem{Einasto}
J.~Einasto, {\it {Kinematics and dynamics of stellar systems}},  {\em Trudy
  Inst. Astrofiz. Alma-Ata} {\bf 51} 87.

\bibitem{Navarro:2008kc}
J.~F. Navarro, A.~Ludlow, V.~Springel, J.~Wang, M.~Vogelsberger, {\em et.~al.},
  {\it {The Diversity and Similarity of Cold Dark Matter Halos}},  {\em
  Mon.Not.Roy.Astron.Soc.} {\bf 402} (2010) 21,
  [\href{http://xxx.lanl.gov/abs/0810.1522}{{\tt arXiv:0810.1522}}].

\bibitem{Bahcall:1980fb}
J.~N. Bahcall and R.~Soneira, {\it {The Universe at faint magnetidues. 2.
  Models for the predicted star counts}},  {\em Astrophys.J.Suppl.} {\bf 44}
  (1980) 73--110.

\bibitem{Burkert:1995yz}
A.~Burkert, {\it {The Structure of dark matter halos in dwarf galaxies}},  {\em
  IAU Symp.} {\bf 171} (1996) 175,
  [\href{http://xxx.lanl.gov/abs/astro-ph/9504041}{{\tt astro-ph/9504041}}].

\bibitem{2013JCAP...07..016N}
F.~{Nesti} and P.~{Salucci}, {\it {The Dark Matter halo of the Milky Way, AD
  2013}},  {\em JCAP} {\bf 7} (July, 2013) 16,
  [\href{http://xxx.lanl.gov/abs/1304.5127}{{\tt arXiv:1304.5127}}].

\bibitem{Gustafsson:2006gr}
M.~Gustafsson, M.~Fairbairn, and J.~Sommer-Larsen, {\it {Baryonic Pinching of
  Galactic Dark Matter Haloes}},  {\em Phys.Rev.} {\bf D74} (2006) 123522,
  [\href{http://xxx.lanl.gov/abs/astro-ph/0608634}{{\tt astro-ph/0608634}}].

\bibitem{deBlok:2002tg}
W.~de~Blok and A.~Bosma, {\it {High-resolution rotation curves of low surface
  brightness galaxies}},  {\em Astron.Astrophys.} {\bf 385} (2002) 816,
  [\href{http://xxx.lanl.gov/abs/astro-ph/0201276}{{\tt astro-ph/0201276}}].

\bibitem{Simon:2004sr}
J.~D. Simon, A.~D. Bolatto, A.~Leroy, L.~Blitz, and E.~L. Gates, {\it
  {High-resolution measurements of the halos of four dark matter-dominated
  galaxies: Deviations from a universal density profile}},  {\em Astrophys.J.}
  {\bf 621} (2005) 757--776,
  [\href{http://xxx.lanl.gov/abs/astro-ph/0412035}{{\tt astro-ph/0412035}}].

\bibitem{Sjostrand:2007gs}
T.~Sjostrand, S.~Mrenna, and P.~Z. Skands, {\it {A Brief Introduction to PYTHIA
  8.1}},  {\em Comput.Phys.Commun.} {\bf 178} (2008) 852--867,
  [\href{http://xxx.lanl.gov/abs/0710.3820}{{\tt arXiv:0710.3820}}].

\bibitem{Beacom:2004pe}
J.~F. Beacom, N.~F. Bell, and G.~Bertone, {\it {Gamma-ray constraint on
  Galactic positron production by MeV dark matter}},  {\em Phys.Rev.Lett.} {\bf
  94} (2005) 171301, [\href{http://xxx.lanl.gov/abs/astro-ph/0409403}{{\tt
  astro-ph/0409403}}].

\bibitem{Birkedal:2005ep}
A.~Birkedal, K.~T. Matchev, M.~Perelstein, and A.~Spray, {\it {Robust gamma ray
  signature of WIMP dark matter}},
  \href{http://xxx.lanl.gov/abs/hep-ph/0507194}{{\tt hep-ph/0507194}}.

\bibitem{Mardon:2009rc}
J.~Mardon, Y.~Nomura, D.~Stolarski, and J.~Thaler, {\it {Dark Matter Signals
  from Cascade Annihilations}},  {\em JCAP} {\bf 0905} (2009) 016,
  [\href{http://xxx.lanl.gov/abs/0901.2926}{{\tt arXiv:0901.2926}}].

\bibitem{Essig:2009jx}
R.~Essig, N.~Sehgal, and L.~E. Strigari, {\it {Bounds on Cross-sections and
  Lifetimes for Dark Matter Annihilation and Decay into Charged Leptons from
  Gamma-ray Observations of Dwarf Galaxies}},  {\em Phys.Rev.} {\bf D80} (2009)
  023506, [\href{http://xxx.lanl.gov/abs/0902.4750}{{\tt arXiv:0902.4750}}].

\bibitem{Jeltema:2008hf}
T.~E. Jeltema and S.~Profumo, {\it {Fitting the Gamma-Ray Spectrum from Dark
  Matter with DMFIT: GLAST and the Galactic Center Region}},  {\em JCAP} {\bf
  0811} (2008) 003, [\href{http://xxx.lanl.gov/abs/0808.2641}{{\tt
  arXiv:0808.2641}}].

\bibitem{ArkaniHamed:2008qn}
N.~Arkani-Hamed, D.~P. Finkbeiner, T.~R. Slatyer, and N.~Weiner, {\it {A Theory
  of Dark Matter}},  {\em Phys.Rev.} {\bf D79} (2009) 015014,
  [\href{http://xxx.lanl.gov/abs/0810.0713}{{\tt arXiv:0810.0713}}].

\bibitem{Pospelov:2008jd}
M.~Pospelov and A.~Ritz, {\it {Astrophysical Signatures of Secluded Dark
  Matter}},  {\em Phys.Lett.} {\bf B671} (2009) 391--397,
  [\href{http://xxx.lanl.gov/abs/0810.1502}{{\tt arXiv:0810.1502}}].

\bibitem{Adriani:2008zr}
{\bf PAMELA Collaboration} Collaboration, O.~Adriani {\em et.~al.}, {\it {An
  anomalous positron abundance in cosmic rays with energies 1.5-100 GeV}},
  {\em Nature} {\bf 458} (2009) 607--609,
  [\href{http://xxx.lanl.gov/abs/0810.4995}{{\tt arXiv:0810.4995}}].

\bibitem{Abdo:2009zk}
{\bf Fermi LAT Collaboration} Collaboration, A.~A. Abdo {\em et.~al.}, {\it
  {Measurement of the Cosmic Ray e+ plus e- spectrum from 20 GeV to 1 TeV with
  the Fermi Large Area Telescope}},  {\em Phys.Rev.Lett.} {\bf 102} (2009)
  181101, [\href{http://xxx.lanl.gov/abs/0905.0025}{{\tt arXiv:0905.0025}}].

\bibitem{Aguilar:2013qda}
{\bf AMS Collaboration} Collaboration, M.~Aguilar {\em et.~al.}, {\it {First
  Result from the Alpha Magnetic Spectrometer on the International Space
  Station: Precision Measurement of the Positron Fraction in Primary Cosmic
  Rays of 0.5Ð350 GeV}},  {\em Phys.Rev.Lett.} {\bf 110} (2013) 141102.

\bibitem{Aguilar:2014mma}
{\bf AMS Collaboration} Collaboration, M.~Aguilar {\em et.~al.}, {\it {Electron
  and Positron Fluxes in Primary Cosmic Rays Measured with the Alpha Magnetic
  Spectrometer on the International Space Station}},  {\em Phys.Rev.Lett.} {\bf
  113} (2014) 121102.

\bibitem{Finkbeiner:2007kk}
D.~P. Finkbeiner and N.~Weiner, {\it {Exciting Dark Matter and the INTEGRAL/SPI
  511 keV signal}},  {\em Phys.Rev.} {\bf D76} (2007) 083519,
  [\href{http://xxx.lanl.gov/abs/astro-ph/0702587}{{\tt astro-ph/0702587}}].

\bibitem{Knodlseder:2003sv}
J.~Knodlseder, V.~Lonjou, P.~Jean, M.~Allain, P.~Mandrou, {\em et.~al.}, {\it
  {Early SPI / INTEGRAL contraints on the morphology of the 511 keV line
  emission in the 4th galactic quadrant}},  {\em Astron.Astrophys.} {\bf 411}
  (2003) L457--L460, [\href{http://xxx.lanl.gov/abs/astro-ph/0309442}{{\tt
  astro-ph/0309442}}].

\bibitem{Cirelli:2010xx}
M.~Cirelli, G.~Corcella, A.~Hektor, G.~Hutsi, M.~Kadastik, {\em et.~al.}, {\it
  {PPPC 4 DM ID: A Poor Particle Physicist Cookbook for Dark Matter Indirect
  Detection}},  {\em JCAP} {\bf 1103} (2011) 051,
  [\href{http://xxx.lanl.gov/abs/1012.4515}{{\tt arXiv:1012.4515}}].

\bibitem{Essig:2013goa}
R.~Essig, E.~Kuflik, S.~D. Mcdermott, T.~Volansky, and K.~M. Zurek, {\it
  {Constraining Light Dark Matter with Diffuse X-Ray and Gamma-Ray
  Observations}},  {\em JHEP} {\bf 1311} (2013) 193,
  [\href{http://xxx.lanl.gov/abs/1309.4091}{{\tt arXiv:1309.4091}}].

\bibitem{Ade:2013zuv}
{\bf Planck Collaboration} Collaboration, P.~Ade {\em et.~al.}, {\it {Planck
  2013 results. XVI. Cosmological parameters}},  {\em Astron.Astrophys.} {\bf
  571} (2014) A16, [\href{http://xxx.lanl.gov/abs/1303.5076}{{\tt
  arXiv:1303.5076}}].

\bibitem{Su:2010qj}
M.~Su, T.~R. Slatyer, and D.~P. Finkbeiner, {\it {Giant Gamma-ray Bubbles from
  Fermi-LAT: AGN Activity or Bipolar Galactic Wind?}},  {\em Astrophys.J.} {\bf
  724} (2010) 1044--1082, [\href{http://xxx.lanl.gov/abs/1005.5480}{{\tt
  arXiv:1005.5480}}].

\bibitem{Delahaye:2007fr}
T.~Delahaye, R.~Lineros, F.~Donato, N.~Fornengo, and P.~Salati, {\it {Positrons
  from dark matter annihilation in the galactic halo: Theoretical
  uncertainties}},  {\em Phys.Rev.} {\bf D77} (2008) 063527,
  [\href{http://xxx.lanl.gov/abs/0712.2312}{{\tt arXiv:0712.2312}}].

\bibitem{Strong:1998fr}
A.~W. Strong, I.~V. Moskalenko, and O.~Reimer, {\it {Diffuse continuum
  gamma-rays from the galaxy}},  {\em Astrophys.J.} {\bf 537} (2000) 763--784,
  [\href{http://xxx.lanl.gov/abs/astro-ph/9811296}{{\tt astro-ph/9811296}}].

\bibitem{Vladimirov:2011rn}
A.~E. Vladimirov, G.~Johannesson, I.~V. Moskalenko, and T.~A. Porter, {\it
  {Testing the Origin of High-Energy Cosmic Rays}},  {\em Astrophys.J.} {\bf
  752} (2012) 68, [\href{http://xxx.lanl.gov/abs/1108.1023}{{\tt
  arXiv:1108.1023}}].

\bibitem{Cholis:2008wq}
I.~Cholis, G.~Dobler, D.~P. Finkbeiner, L.~Goodenough, and N.~Weiner, {\it {The
  Case for a 700+ GeV WIMP: Cosmic Ray Spectra from ATIC and PAMELA}},  {\em
  Phys.Rev.} {\bf D80} (2009) 123518,
  [\href{http://xxx.lanl.gov/abs/0811.3641}{{\tt arXiv:0811.3641}}].

\bibitem{Bregeon:2013qba}
{\bf Fermi-LAT Collaboration} Collaboration, J.~Bregeon, E.~Charles, and
  M.~Wood, {\it {Fermi-LAT data reprocessed with updated calibration
  constants}},  \href{http://xxx.lanl.gov/abs/1304.5456}{{\tt
  arXiv:1304.5456}}.

\bibitem{Ackermann:2012kna}
{\bf Fermi-LAT Collaboration} Collaboration, M.~Ackermann {\em et.~al.}, {\it
  {The Fermi Large Area Telescope On Orbit: Event Classification, Instrument
  Response Functions, and Calibration}},  {\em Astrophys.J.Suppl.} {\bf 203}
  (2012) 4, [\href{http://xxx.lanl.gov/abs/1206.1896}{{\tt arXiv:1206.1896}}].

\bibitem{ScienceTools}
``{Fermi Science Tools}.''
  {\url{http://www.slac.stanford.edu/exp/glast/wb/prod/sciTools_Home.htm}}.

\bibitem{TheFermi-LAT:2015hja}
{\bf The Fermi-LAT Collaboration} Collaboration, {\it {Fermi Large Area
  Telescope Third Source Catalog}},
  \href{http://xxx.lanl.gov/abs/1501.0200}{{\tt arXiv:1501.0200}}.

\bibitem{Cholis:2013psa}
I.~Cholis and D.~Hooper, {\it {Dark Matter and Pulsar Origins of the Rising
  Cosmic Ray Positron Fraction in Light of New Data From AMS}},  {\em
  Phys.Rev.} {\bf D88} (2013) 023013,
  [\href{http://xxx.lanl.gov/abs/1304.1840}{{\tt arXiv:1304.1840}}].

\bibitem{Boudaud:2014dta}
M.~Boudaud, S.~Aupetit, S.~Caroff, A.~Putze, G.~Belanger, {\em et.~al.}, {\it
  {A new look at the cosmic ray positron fraction}},
  \href{http://xxx.lanl.gov/abs/1410.3799}{{\tt arXiv:1410.3799}}.

\bibitem{Jin:2013nta}
H.-B. Jin, Y.-L. Wu, and Y.-F. Zhou, {\it {Implications of the first AMS-02
  measurement for dark matter annihilation and decay}},  {\em JCAP} {\bf 1311}
  (2013) 026, [\href{http://xxx.lanl.gov/abs/1304.1997}{{\tt
  arXiv:1304.1997}}].

\bibitem{Arvanitaki:2008hq}
A.~Arvanitaki, S.~Dimopoulos, S.~Dubovsky, P.~W. Graham, R.~Harnik, {\em
  et.~al.}, {\it {Astrophysical Probes of Unification}},  {\em Phys.Rev.} {\bf
  D79} (2009) 105022, [\href{http://xxx.lanl.gov/abs/0812.2075}{{\tt
  arXiv:0812.2075}}].

\bibitem{Nardi:2008ix}
E.~Nardi, F.~Sannino, and A.~Strumia, {\it {Decaying Dark Matter can explain
  the e+- excesses}},  {\em JCAP} {\bf 0901} (2009) 043,
  [\href{http://xxx.lanl.gov/abs/0811.4153}{{\tt arXiv:0811.4153}}].

\bibitem{Yin:2008bs}
P.-f. Yin, Q.~Yuan, J.~Liu, J.~Zhang, X.-j. Bi, {\em et.~al.}, {\it {PAMELA
  data and leptonically decaying dark matter}},  {\em Phys.Rev.} {\bf D79}
  (2009) 023512, [\href{http://xxx.lanl.gov/abs/0811.0176}{{\tt
  arXiv:0811.0176}}].

\bibitem{Ibarra:2008jk}
A.~Ibarra and D.~Tran, {\it {Decaying Dark Matter and the PAMELA Anomaly}},
  {\em JCAP} {\bf 0902} (2009) 021,
  [\href{http://xxx.lanl.gov/abs/0811.1555}{{\tt arXiv:0811.1555}}].

\bibitem{Ishiwata:2008cv}
K.~Ishiwata, S.~Matsumoto, and T.~Moroi, {\it {Cosmic-Ray Positron from
  Superparticle Dark Matter and the PAMELA Anomaly}},  {\em Phys.Lett.} {\bf
  B675} (2009) 446--449, [\href{http://xxx.lanl.gov/abs/0811.0250}{{\tt
  arXiv:0811.0250}}].

\bibitem{Chen:2008qs}
C.-R. Chen, M.~M. Nojiri, F.~Takahashi, and T.~Yanagida, {\it {Decaying Hidden
  Gauge Boson and the PAMELA and ATIC/PPB-BETS Anomalies}},  {\em
  Prog.Theor.Phys.} {\bf 122} (2009) 553--559,
  [\href{http://xxx.lanl.gov/abs/0811.3357}{{\tt arXiv:0811.3357}}].

\bibitem{Baek:2014goa}
S.~Baek, P.~Ko, W.-I. Park, and Y.~Tang, {\it {Indirect and direct signatures
  of Higgs portal decaying vector dark matter for positron excess in cosmic
  rays}},  {\em JCAP} {\bf 1406} (2014) 046,
  [\href{http://xxx.lanl.gov/abs/1402.2115}{{\tt arXiv:1402.2115}}].

\bibitem{Hooper:2012cw}
D.~Hooper, N.~Weiner, and W.~Xue, {\it {Dark Forces and Light Dark Matter}},
  {\em Phys.Rev.} {\bf D86} (2012) 056009,
  [\href{http://xxx.lanl.gov/abs/1206.2929}{{\tt arXiv:1206.2929}}].

\bibitem{FermiLAT:2012aa}
{\bf Fermi-LAT Collaboration} Collaboration, {\it {Fermi-LAT Observations of
  the Diffuse Gamma-Ray Emission: Implications for Cosmic Rays and the
  Interstellar Medium}},  {\em Astrophys.J.} {\bf 750} (2012) 3,
  [\href{http://xxx.lanl.gov/abs/1202.4039}{{\tt arXiv:1202.4039}}].

\bibitem{Gleam1}
\url{http://fermi.gsfc.nasa.gov/ssc/data/access/lat/BackgroundModels.html}.
\newblock Galactic diffuse model used for all MC is based on (with flux
  renormalization to 1 GeV lowest energy)
  \url{http://fermi.gsfc.nasa.gov/ssc/data/analysis/software/aux/gal_2yearp7v6_trim_v0.fits}.

\end{thebibliography}\endgroup
\bibliographystyle{JHEP}


\end{document}